\documentclass[onecollarge,natbib]{svjour2}
\bibpunct{[}{]}{;}{n}{}{,} 
\smartqed  
\usepackage{graphicx}
\usepackage{amsfonts}
\usepackage{amsmath}
\usepackage{amssymb}
\usepackage{latexsym}
\usepackage{color}
\usepackage{colordvi}
\usepackage{epsfig}
\usepackage{array}
\usepackage{pifont}

\usepackage{amsfonts}
\usepackage{amsmath}
\usepackage{amssymb}
\usepackage{latexsym}
\usepackage{color}
\usepackage{colordvi}
\usepackage{epsfig}
\usepackage{array}
\usepackage{pifont}
\usepackage{axodraw}

\newcommand{\beq}{\begin{eqnarray}}
\newcommand{\eeq}{\end{eqnarray}}
\newcommand{\bea}{\begin{eqnarray}}
\newcommand{\eea}{\end{eqnarray}}

\newcommand{\Lams}{\Lambda_{\overline{\rm MS}}}

\newcommand{\be}{\begin{equation}}
\newcommand{\ee}{\end{equation}}
\newcommand{\lwrsim}{\raise0.3ex\hbox{$<$\kern-0.75em\raise-1.1ex\hbox{$\sim$}}}

\newcommand{\alvp}{(\frac{\alpha_s}{4\pi})}
\newcommand{\alvps}{\frac{\alpha_s}{4\pi}}
\newcommand{\smsbar}{\overline{\mbox{{\scriptsize \sc ms}}}} 
\newcommand{\tr}{\text{Tr}}

\newcommand{\Break}{ \nonumber \\ & & }

\def\braket#1{\langle #1 \rangle}
\def\VA{\langle A^2 \rangle}

\def\C2#1#2{({\cal C}_2)_{#1}^{#2}}

\def\eq#1{Eq.~(\ref{#1})}

\journalname{Few-Body Systems}
\newcommand{\ghostSD}{\begin{picture}(150,25)(0,0)
\SetWidth{1.2}
\DashArrowLine(12.5,0)(37.5,0){5}
\DashArrowLine(37.5,0)(75,0){5}
\DashLine(75,0)(112.5,0){5}
\DashArrowLine(112.5,0)(137.5,0){5}
\SetWidth{1}
\Vertex(112.5,0){2}
\GlueArc(75,0)(37.5,0,90){-4}{6}
\GlueArc(75,0)(37.5,90,180){-4}{6}
\CCirc(75,0){5}{Black}{Yellow}
\CCirc(75,37.5){5}{Black}{Yellow}
\CCirc(37.5,0){5}{Black}{Yellow}
\Text(20,-10)[l]{a,k}
\Text(50,15)[l]{d,$\nu$}
\Text(100,-10)[l]{e}
\Text(100,15)[r]{f,$\mu$}
\Text(50,-10)[l]{c,q}
\Text(120,-10)[l]{b,k}
\Text(75,48)[c]{q-k}
\end{picture}}

\newcommand{\ghostDr}{\begin{picture}(100,25)(0,0)
\SetWidth{1.2}
\DashArrowLine(12.5,0)(50,0){5}
\DashArrowLine(50,0)(87.5,0){5}
\CCirc(50,0){5}{Black}{Yellow}
\Text(12.5,-10)[l]{a}
\Text(87.5,-10)[r]{b}
\Text(50,-10)[c]{k}
\end{picture}}
\newcommand{\ghostBr}{\begin{picture}(100,25)(0,0)
\SetWidth{1.2}
\DashArrowLine(12.5,0)(87.5,0){5}
\Text(12.5,-10)[l]{a}
\Text(87.5,-10)[r]{b}
\Text(50,-10)[c]{k}
\end{picture}}

\newcommand{\ghThreeTwo}{\begin{picture}(120,40)(0,0)
\SetWidth{1.2}
\DashArrowLine(10,20)(30,20){5}
\DashArrowLine(30,20)(90,20){5}
\DashArrowLine(90,20)(110,20){5}
\SetWidth{1}
\Vertex(30,20){2}
\Vertex(90,20){2}
\Gluon(60,20)(60,-10){4}{3}
\GlueArc(60,20)(30,0,75){-4}{4}
\GlueArc(60,20)(30,105,180){-4}{4}
\CCirc(60,50){10}{Black}{Blue}
\end{picture}}

\newcommand{\ghThreeOneRS}{\begin{picture}(150,45)(0,0)
\SetWidth{1.2}
\DashArrowLine(12.5,0)(37.5,0){5}
\DashArrowLine(37.5,0)(112.5,0){5}
\DashArrowLine(112.5,0)(137.5,0){5}
\SetWidth{1}
\Vertex(37.5,0){2}
\Vertex(112.5,0){2}
\Vertex(40,31){2}
\Gluon(37.5,0)(37.5,48){-4}{4}
\GlueArc(67.5,0)(45,70,135){-4}{4}
\GlueArc(67.5,0)(45,0,70){-4}{5}
\CCirc(78,40){10}{Black}{Blue}
\Text(135,5)[]{q}
\Text(20,5)[]{k}
\Text(25,47)[]{q-k}
\end{picture}}
\newcommand{\gluonSDi}{\begin{picture}(112.5,18.75)(0,0)
\SetScale{0.75}
\SetWidth{1.2}
\Gluon(12.5,0)(37.5,0){-4}{2}
\Gluon(37.5,0)(75,0){-4}{3}
\Gluon(75,0)(112.5,0){-4}{3}
\Gluon(112.5,0)(137.5,0){-4}{2}
\SetWidth{1}
\Vertex(112.5,0){2}
\GlueArc(75,0)(37.5,0,90){-4}{6}
\GlueArc(75,0)(37.5,90,180){-4}{6}
\CCirc(75,0){5}{Black}{Yellow}
\CCirc(75,37.5){5}{Black}{Yellow}
\CCirc(37.5,0){5}{Black}{Yellow}
\end{picture}}

\newcommand{\gluonSDii}{\begin{picture}(112.5,18.75)(0,0)
\SetScale{0.75}
\SetWidth{1.2}
\Gluon(15,-5)(75,-5){-3}{4}
\Gluon(75,-5)(135,-5){-3}{4}
\SetWidth{1}
\GlueArc(75,18.75)(18.75,-90,90){3}{4}
\GlueArc(75,18.75)(18.75,90,270){3}{4}
\CCirc(75,-2.5){5}{Black}{Yellow}
\CCirc(75,37.5){5}{Black}{Yellow}
\end{picture}}

\newcommand{\gluonSDiii}{\begin{picture}(112.5,18.75)(0,0)
\SetScale{0.75}
\SetWidth{1.2}
\Gluon(12.5,0)(37.5,0){-4}{2}
\DashLine(37.5,0)(75,0){4}
\DashLine(75,0)(112.5,0){4}
\Gluon(112.5,0)(137.5,0){-4}{2}
\SetWidth{1}
\Vertex(112.5,0){2}
\DashCArc(75,0)(37.5,0,90){4}
\DashCArc(75,0)(37.5,90,180){4}
\CCirc(75,3){5}{Black}{Yellow}
\CCirc(75,37.5){5}{Black}{Yellow}
\CCirc(37.5,3){5}{Black}{Yellow}
\end{picture}}

\newcommand{\gluonSDiv}{\begin{picture}(112.5,18.75)(0,0)
\SetScale{0.75}
\SetWidth{1.2}
\Gluon(15,-5)(75,-5){-3}{4}
\Gluon(75,-5)(135,-5){-3}{4}
\SetWidth{1}
\DashCArc(75,18.75)(18.75,-90,90){4}
\DashCArc(75,18.75)(18.75,90,270){4}
\CCirc(75,-2.5){5}{Black}{Yellow}
\CCirc(75,37.5){5}{Black}{Yellow}
\end{picture}}

\begin{document}

\title{The Infrared Behaviour of the Pure Yang-Mills Green Functions 
}


\author{Ph.~Boucaud \and J.P.~Leroy \and A.~Le~Yaouanc \and J.~Micheli \and O.~P\`ene \and J.~Rodr\'iguez-Quintero
}


\institute{Ph.~Boucaud, J.P.~Leroy, A.~Le~Yaouanc, J. Micheli, O. P\`ene \at
               Laboratoire de Physique Th\'eorique, 
CNRS et Universit\'e  Paris-Sud XI, B\^at, 210, 91405 Orsay, France\\
Report Number LPT-Orsay/11-45
           \and
           J. Rodr\'{\i}guez-Quintero \at
              Dpto. F\'isica Aplicada, Fac. Ciencias Experimentales,Univ. de Huelva, 21071 Huelva, Spain\\
Report Number UHU-FT/11-10 \\cdot
              Tel.: +34-959-219787\\
              Fax: +34-959-219777\\
              \email{jose.rodriguez@dfaie.uhu.es}           
}

\date{Received: date / Accepted: date}

\maketitle

\begin{abstract}

We review the infrared properties of the pure Yang-Mills correlators and discuss 
recent results concerning the two classes of low-momentum solutions for them reported in literature, 
{\it i.e.} decoupling and scaling solutions. We will mainly focus on the Landau gauge and 
pay special attention to the results inferred from the analysis of the Dyson-Schwinger equations of the 
theory and from ``{\it quenched}'' lattice QCD. The results obtained from properly interplaying both 
approaches are strongly emphasized. 

\end{abstract}




\section{Introduction}
\label{sec:Intro}

The whole set of correlation functions fully describes a Quantum Field Theory,
as it is related to the S-matrix elements. In  QCD or pure Yang-Mills theories
Green functions are most often gauge dependent quantities which have  no direct
relationship with physical observables, the latter being necessarily gauge
invariant. However, their indirect physical relevance is well known. In
particular, long distance (or small momentum) Green functions will hopefully
shed some light on the deepest mysteries of QCD such as confinement, spontaneous
chiral symmetry breaking, etc. 

Indeed, more than thirty five years after the discovery
of QCD and notwithstanding its numerous successes, notwithstanding either the fact that
everybody is convinced that QCD implies confinement, a real proof of it from 
first principles  has not yet been achieved. This is doubtless one
of the major scientific challenges of this century. In fact, in the case of QCD,
one is not even provided with a precise mathematical formulation of confinement. 
In the case of the pure Yang-Mills theory, such a mathematical formulation at least exists: it 
is the area-law for Wilson loops which unluckily does not hold for QCD due to
the string breaking by the sea quarks. 
Furthermore, the Polyakov loop (the product of link variables along a curve in time direction 
closed by periodic boundary conditions), probing the screening properties of a static colour 
triplet test charge, appears to be the order parameter for the deconfinement phase transition. 
Here we will restrict ourselves to the
pure Yang-Mills theory. We will not enumerate all the tracks which have been
followed to understand confinement by means of the peculiarities of QCD in 
the infrared domain, there are many of them 
(see, for instance, the classical Wilson's~\cite{Wilson:1974sk} or Cornwall's~\cite{Cornwall:1979hz} works,  
or the very recent introduction to the confinement problem in ref.~\cite{Greensite:2011zz} 
and references therein). 
In  this review we will only mention the Kugo-Ojima and the Gribov-Zwanziger approaches. 
The latter aimed in principle to deal with the problem of Gribov copies, but it is also 
thought to be connected with confinement scenarios.  
Both approaches are related to the behaviour of the Green functions of the gluons and ghosts 
in the deep infrared. 

It is now well established that the vacuum of a quantum field theory is never
trivial, and especially not in the case of a non-Abelian gauge theory. It is
believed that the low modes of the vacuum, the condensates, are the keys to 
understand its non-perturbative properties. The vacuum is a gauge invariant
state. But the configurations of the fields in the vacuum have different 
features in different gauges. This allows for different, complementary and
rewarding views into its properties. We will address this issue (sec.~\ref{sec:A2lat}). 

We will concentrate our efforts on a
review  of the  gluon and ghost Green functions properties at small momentum in
a pure Yang-Mills theory.  We will invoke results from the analysis of
Dyson-Schwinger equations (DSE), Slavnov-Taylor  identities (ST) and from
lattice QCD,  paying special attention to the interplay of  all these
techniques and particularly to the study of Yang-Mills solutions when lattice 
results happen to be applied as DSE inputs.

We work hereafter mainly in the Landau gauge, 
but some of the results we will present are actually valid in any covariant gauge. 
We will also discuss some results in Coulomb gauge and focus especially on the 
similarities of the general properties of the solutions in both Landau and Coulomb 
gauges.

Our notations will be the following :  
\beq
(F^{(2)})^{ab}(k) &=& -\delta^{ab} \frac{F(k^2)}{k^2} \label{dressghost}\\
(G^{(2)}_{\mu\nu})^{ab}(k) &=& \delta^{ab} \frac{G(k^2)}{k^2} \left(\delta_{\mu\nu} -
\frac{k_\mu k_\nu}{k^2} \right)\label{dressglue} \\
\Gamma^{abc}_{\mu\nu\rho}(p,q,r) &=& f^{abc} \Gamma_{\mu\nu\rho}(p,q,r)  
\\
\rule[0cm]{0cm}{0.6cm} \widetilde{\Gamma}_{\mu}^{abc}(q,k;p) &=& f^{abc} (-i q_\nu) g_0
\widetilde{\Gamma}_{\nu\mu}(q,k;p) 
\nonumber  \\
&=&  i g_0 f^{abc} \left( \ -q_\nu H_1(-q,k) + p_\nu H_2(-q,k) \ \right) \ ,
\label{vertghost} 
\eeq
respectively for the ghost propagator, the gluon propagator, the three-gluons vertex and the
ghost-gluon vertex\footnote{We stick to the decomposition given in ref.  \cite{Chetyrkin:2000dq} {\sl except for  the arguments of the scalar functions, for which we keep the same order as in $\Gamma$ itself }}. All momenta 
are taken as entering. In eq.~(\ref{vertghost}) 
 $-q$ is the momentum of the outgoing ghost, $k$ 
 the momentum of the incoming one and $p=-q-k$ the momentum of the gluon. 
 $H_{1(2)}$ corresponds to the gluon-transverse (longitudinal) form factor that will be extensively invoked in the following. 
 $g_0$ is the bare coupling that should be properly renormalized~\footnote{By $Z_g^{-1}=\widetilde{Z}_3 Z_3^{1/2} \widetilde{Z}_1^{-1}$, where 
 $Z_3$($\widetilde{Z}_3$) is the gluon (ghost) propagator renormalization constant and $\widetilde{Z}_1$ is the proper ghost-gluon 
 vertex renormalization constant.} and, in the limit of vanishing incoming ghost momentum ($k \to 0$), one has
 \beq\label{alphaTfirst}
 \alpha_T(q^2) = \frac{g_0^2} {4 \pi^2} F^2(q^2) G(q^2) \ ,
 \eeq
which gives the running coupling in the Taylor renormalization scheme, 
where $F(p^2)$ and $G(p^2)$, defined by Eqs.(\ref{dressghost},\ref{dressglue}), are the dressing functions
of the ghost and  gluon propagators respectively. 
Up to logarithms, we parametrise the propagators in the infrared by setting at leading order
\begin{equation}
\label{param}
\begin{split}
& G(p^2)=\left(\frac{p^2}{\lambda^2}\right)^{\alpha_G}  
\\ & F(p^2)=\left(\frac{p^2}{\eta^2}\right)^{\alpha_F}, \quad\text{when }p^2\text{ is small},
\end{split}
\end{equation}
where $\lambda,\eta$ are some dimensional parameters.  Finally we shall set $ D(q^2)= G(q^2) / q^2$

As we mentioned, our goal will be to describe the current state-of-the-art concerning the 
low-momentum properties of the Green functions in gluo-dynamics, mainly by focussing on the  
results obtained with two main approaches: DSEs and Lattice QCD. 
The paper is organized as follows: a first section (sec.~\ref{sec:analytic}) will be devoted 
to provide the reader with a first insight on the problem of the Yang-Mills Green functions low-momentum 
behaviour through reviewing the main analytical properties recently derived in the literature for 
the solutions; we will review then, as exhaustively as possible, the numerical lattice QCD results on 
the subject in sec.~\ref{sec:lattice}; the results obtained from the numerical resolution of DSEs, with 
different truncation approaches, will be discussed and put in connection with lattice results in 
sec.~\ref{sec:num}; and we finally conclude in sec.~\ref{sec:conclu}.
However, we will start now by ``saying'' a few words to introduce those two main approaches and 
their results, but also about other approaches as the ones based on redefining the 
QCD lagrangian to properly correct the Gribov ambiguity. 

\subsection{Lattice results}

The lattice technique in gauge field theories was initiated by Wilson in 1974. It has now been used  
for the field  of Yang Mills Green's functions for nearly 25 years. 
Let us begin by recalling what it consists in. The general idea is to rewrite the lagrangian on a 
discretised space-time (the lattice) instead of the ordinary continuous one. Accordingly  all 
derivatives (including the covariant derivatives) are replaced by finite differences. If in addition 
one limits oneself to a finite volume, the path integral is now over a finite (but huge) number of 
field variables and can be estimated as a sum over a stochastically determined set of configurations.  
The advantages  are well known :
\begin{itemize}
\item The only ingredient is the original lagrangian, excluding any additional hypothesis. 
So, in some sense, the results obtained in this way can be considered as exact from the theoretical 
point of view.
\item  The technique is essentially non-perturbative.
 \item The lattice spacing $a$ acts as an ultraviolet cutoff : no divergence occurs. 
The continuum limit can be recovered by letting $a$ go to zero, at the price of an appropriate renormalisation.
\end{itemize}
 
There are some drawbacks to the method, however. 
\begin{itemize}
\item It is very demanding in computing power.
\item The results are given in numerical form and, as such, they  suffer from uncertainties. 
The first one is of statistical nature and stems from the finite number of configurations which 
are used in the evaluation ; it can in principle be reduced at will by increasing this number. 
The other sources are more intrisically related to the technique, they give rise to systematic errors.  
We mention them briefly below although they will be discussed  in greater detail in dedicated sections in the following (see section~\ref{artefacts}). 
 \end{itemize}

First of all the actual calculations are necessarily performed at finite values of the lattice spacing. 
Na\"\i vely, at tree level, the discretisation procedure generates effects of order $a$ in the fermionic lagrangian and of order $a^2$ in the Yang-Mills one.
The gap between those conditions and the continuum limit can be partly reduced by using an improved lagrangian (see for instance refs.~\cite{Sheikholeslami:1985ij} for the fermionic case and \cite{Iwasaki:1984cj,Luscher:1984xn,Luscher:1985zq} in the pure gauge one). 
The remaining part has to be treated numerically.

Second, going to the lattice implies, as we have already said,  that one works in a finite volume. 
This induces, in turn, a $1/L$ spacing in $p$-space. Since we are interested in the infrared  limit 
of the  Green's functions we shall have to take special care  of potential discontinuities in the 
neighbourhood of  $0$. 

Third, the space on which we are working is actually  a  torus since the field configurations are 
usually chosen to be periodic. One can therefore wonder what kind of relationship the functions 
obtained in this way entertain with the ``real world'' ones.

As a final remark, we note that the Gribov copies problem, which is absolutely general, 
can be specifically dealt with on the lattice. 
The gauge fixing procedure consists in gauge-transforming the 
field configuration into one of its gauge equivalents which satisfy the given gauge  condition. 
Generally the different possible choices do not contribute equally to the path integral. 
  It is therefore an important issue to know how large this effect is and how one can minimize it. 
Within the numerical precision, this can be explicitely done on the lattice, as will be discussed 
in section~\ref{AmbGribov}.

The number of papers dealing with lattice simulations for Green functions in the IR is rather large, and their results  are given according to  
quite a variety  of presentations, which makes the task of comparing them not so easy. We shall try to give an 
account of those works in sec.~\ref{sec:lattice}. Note that the most recent studies have been performed in the unquenched 
theory (\cite{Bowman:2004jm,Bowman:2005vx,Parappilly:2006si,Silva:2010vx}),  
so that they fall out of the scope of this review and will not be discussed.

\subsection{The DSE picture}

In principle, the field equations for the Green functions of a quantum field theory 
can be derived from the integral representation of the theory; these are the 
Dyson-Schwinger equations (DSEs) of this 
theory which are considered to describe its non-perturbative 
dynamics~\cite{Dyson:1949ha,Schwinger:1951ex}.
Under the only assumption of the existence of a well-defined measure in a functional integral 
representation of the generating functional for the Green functions of a theory, QCD 
for instance, the corresponding DSEs can be derived (see for instance \cite{Alkofer:2000wg})
as an infinite tower of coupled integral equations. Of course, in practice,
these equations need to be truncated to be studied. Typically, one can make appeal to 
additional sources of information, for instance Slavnov-Taylor identities, to express 
higher $n$-point functions in terms of elementary two-point ones or to generate some 
general ans\"atze for vertices. Thus, a closed coupled system of DSEs is to be obtained 
and can be numerically solved, 
 subject to the validity of the truncation rules which are applied. 
This DSE approach has been extensively applied to investigate the low-momentum behaviour of the QCD        
Green functions; in particular, for the Yang-Mills gluon and ghost propagators. We will dedicate 
secs.~\ref{sec:analytic} and~\ref{sec:num}                       
to describe some of the main results contributing to the current DSE picture for this low-momentum 
behaviour of Yang-Mills Green functions. Let us however    
briefly introduce those results and how they emerged in the last few years. 

After the first pioneering works of Mandelstam \cite{Mandelstam:1979xd,Mandelstam:1980ii} and 
the further ones of Brown and Pennington \cite{Brown:1988bn}, only a few years ago,  
a new paradigm emerged when it was widely accepted (see for instance \cite{Alkofer:2000wg}) 
that a vanishing gluon propagator and a diverging ghost
dressing function at zero-momentum in Landau gauge 
made up  the unique solution of the truncated tower of DSEs. 
In contrast, alternative DSE solutions were also predicted to give a massive gluon 
propagator~\cite{Aguilar:2006gr,Aguilar:2008xm}. Lattice QCD (LQCD) estimates for those propagators appeared 
to be also in contradiction with a gluon propagator that vanishes at zero-momentum or with a 
ghost dressing function that diverges~\cite{Cucchieri:2007md,Bogolubsky:2007ud,IlgenGrib,Boucaud:2005ce,Sternbeck:2007ug}.
We addressed this issue in two recent papers~\cite{Boucaud:2008ji,Boucaud:2008ky} and tried to clarify 
the contradiction. After assuming in the vanishing momentum limit a ghost dressing function behaving as 
$F(q^2) \sim (q^2)^{\alpha_F}$ and a gluon propagator as 
$D(q^2) \sim (q^2)^{\alpha_G-1}$ (or, by following a notation commonly used,
a gluon dressing function as $G(q^2)= q^2 D(q^2) \sim (q^2)^{\alpha_G}$), 
we proved that the ghost propagator DSE (GPDSE) admits two types of solutions:

\begin{itemize}
\item If $\alpha_F \neq 0$, the low-momentum behaviour of both gluon and ghost propagators 
are related by the condition $2 \alpha_F+\alpha_G = 0$ implying that $F^2(q^2)G(q^2)$ goes to a
non-vanishing constant when $q^2 \to 0$. This solution is now called ``{\it scaling}''.
\item If $\alpha_F=0$, the low-momentum leading term of the gluon propagator is  not constrained any longer 
by the leading but instead by the next-to-leading one of the ghost propagator, and LQCD 
solutions indicating that $F^2(q^2)G(q^2) \to 0$  when $q^2 \to0$~\cite{IlgenGrib,Boucaud:2005ce} 
can be pretty well accommodated within this case. This solutions is named as ``{\it decoupling}''.
\end{itemize}

In particular, the numerical study in ref.~\cite{Boucaud:2008ji} of the GPDSE
using a LQCD  gluon input finds that two classes of solutions emerge, depending
on the value of the strong coupling constant at the renormalization point, which
is a free parameter  in this exercise. Indeed, it seems to be by now well
established that the two classes  of solutions, decoupling and  
scaling may emerge from the
tower of DSE~\cite{Aguilar:2006gr,Aguilar:2008xm,Fischer:2008uz}. Such a
nomenclature, despite being widely accepted,  can be misleading. The
perturbative running for the  coupling constant renormalized in Taylor-scheme is
given by \eq{alphaTfirst} and one can thus extend this definition, although not univocally, 
to the IR domain.  However,  a scale invariance ( a decoupling)
of the IR dynamics for the theory  cannot be inferred  from the low-momentum
behaviour of such a coupling in the scaling (decoupling) solution. 
In particular, as will be seen in next \eq{coefC2} of subsection \ref{subsec:next} 
(see also Eqs.~(2.30-2.32) in ref.~\cite{Aguilar:2009nf}),  
an effective charge can be properly defined for
phenomenological purposes such that it reaches a  constant at zero-momentum in
the ``decoupling'' case, which means the absence of decoupling. 
Leaving aside nomenclature, the two classes of solutions are definitely different and the
Taylor-scheme coupling, although maybe not appropriate for  phenomenological
purposes in the IR domain,  is a well defined and very convenient quantity to
discriminate them.

How both types of IR solutions for Landau gauge DSE emerge and how the
transition between them occurs,  in relation with the size of the coupling
(taken as an integration boundary condition at  the renormalization momentum), was
initially discussed in ref.~\cite{Boucaud:2008ji} through the analysis of a
ghost propagator DSE combined with a gluon propagator taken from lattice
computations.  It should be remembered that one needs to know the QCD mass scale
to predict the QCD coupling at any momentum.  This mass scale should be of
course supplied to get a particular solution from DSE and can be  univocally
related to the boundary condition needed, after applying a truncation scheme,
to solve  the equation.   The existence of a {\it critical} value for the
coupling at any renormalization momentum was suggested  by that partial
analysis. No solution was proven to exist for any coupling bigger than the
critical one and  the unique scaling solution\footnote{The authors of
\cite{Fischer:2006vf} proved there, once the scaling behaviour  is assumed, the
uniqueness for Yang-Mills infrared solutions}  seemed to emerge when the
coupling took this critical value. Later, the authors of
ref.~\cite{Fischer:2008uz} confirmed, by the analysis of the tower of DSE
truncated  within two different schemes and also in the framework of the
functional renormalization group, that  the boundary condition for the DSE
integration determined whether a decoupling or the scaling solution  occurs. A
similar analysis has been recently done in the Coulomb
gauge~\cite{Watson:2010cn} leading to  the same pattern as in
ref.~\cite{Boucaud:2008ji}, although the authors interpret the boundary
condition in terms  of the gauge-fixing ambiguity (see also
\cite{Epple:2007ut,Leder:2010ji}). 
Furthermore, an analytic study based on the pinch
technique (PT) in ref.~\cite{Cornwall:2009ud} shows that, within some 
approximations, there is a lower limit for the gluon mass, below which the PT coupling 
is singular in the IR, that can be also interpreted as an
upper limit to the coupling at some renormalization point.  
Very recently also, a next-to-leading low-momentum
asymptotic formula  for the decoupling ghost dressing function solutions was
obtained by studying the ghost propagator DSE under the  assumption, for the
truncation, of a constant ghost-gluon vertex and of a simple model  for a
massive gluon propagator~\cite{Boucaud:2010gr}. In this asymptotic formula, the 
ghost-propagator low-momentum behaviour appears to be regulated by  the
zero-momentum effective charge in Taylor scheme~\cite{Aguilar:2009nf}  and by
the Landau-gauge gluon mass scale.

That DSEs, being an intricated tower of coupled functional differential equations, admit multiple solutions belonging to
 different types is not  very surprising. The curent lattice results appear to be clearly compatible with only one 
type of solutions and, after the appropriate physical calibration of the simulations has been performed (and all the lattice 
artefact have been put properly under control), must help to select the 
``{\it physical}'' solutions for the QCD Green functions among the multiple DSE ones. 
Furthermore, as will be discussed in the next subsection, an effective action 
incorporating the so-called Gribov horizon  can be properly built
  to deal with the Gribov problem, and the lattice-like solutions can also appear as those minimizing this effective 
  action (the emergence of non-vanishing dimension-two condensates, that we will also discuss in 
  section~\ref{sec:A2lat}, plays a crucial role in obtaining the lattice-like solutions).

\subsection{The Gribov and the Gribov-Zwanziger approaches}
\label{subsec:GZ}

The computation of gauge dependent quantities like the correlation 
functions we are interested in  suffers from the Gribov ambiguity: imposing a  constraint like 
the Landau gauge condition $\partial_\mu A^\mu = 0 $  is not enough to pick  unambiguously
a unique representative from each gauge orbit  \cite{Gribov:1978}. 
Gribov then proposed to reinforce the gauge condition by restricting the space of gauge fields 
to representatives  minimizing 
the functional 
\beq
{\cal F}(\Phi) = Tr \int {\mathrm d}x A_\mu^\Phi(x) A_\mu^\Phi(x)
\eeq
where $ A_\mu^\Phi$ is the gauge transformed field.  
This is appealing from two points of view : 
first the Gribov problem is reduced as gauge copies which are maxima or saddle points of ${\cal F}$ are pushed  
out of the game  and second this restricted space, named the Gribov region,  is rather easy to characterize mathematically. Actually the second derivatives of  the functional ${\cal F}$ 
defines an operator, the so-called Fadeev-Popov operator, which has to be positive in the Gribov region.
For the Landau gauge this operator reads :
\beq
{\cal M}^{ab} \ =\ \partial^\mu D_\mu^{ab} \ ,
\eeq
where $D$ is the covariant derivative. But in general along a gauge orbit  the minimum for the functional  is not unique.
The ambiguity in the gauge fixing is not fully eliminated this way. It is then  natural to try to restrict further 
the space of gauge fields to configurations which are not only a minimum but an absolute minimum of ${\cal F}$. 
This region, named the fundamental modular region, in which on {\sl each} gauge orbit {\cal F} possesses a unique absolute minimum  
(cf reference~\cite{Dell'Antonio:1991xt}), would be theoretically convenient to 
eliminate the Gribov problem but unfortunately no practical explicit characterization has been found so far.

Gribov proposed a heuristic method to perform the restriction to the Gribov region. He computed 
the inverse of the Faddeev-Popov operator in perturbation theory at one loop and found an expression 
with a structure like~:
\beq
({\cal M}^{-1})^{ab} \ =\ \frac{1}{k^2} \  \frac{1}{1 - \sigma(k,A)}
\eeq
The explicit form for $\sigma$ can be found in the original paper by Gribov but is not necessary for the general discussion here.
The Gribov region is a space where the eigenvalues of the Fadeev-Popov operator are positive.
Forbidding the crossing of the boundary of this region is equivalent to forbid the appearance of a zero eigenvalue.
This is the origin of a condition proposed by Gribov to restrict the space of gauge fields to the Gribov region~:
\beq
\sigma(0,A) \ <\ 1
\eeq
With this restriction as a constraint, Gribov computed the ghost and the gluon propagators (at one loop) and found~:
\bea
G_{Gribov}(k^2) &=& \frac {k^4}{k^4 + m_G^4}\\
F_{Gribov}(k^2) &=& \frac {128 \pi^2 m_G^2}{N_c g^2k^2}
\eea
Namely, when $k^2$ goes to 0,  a vanishing gluon propagator and an enhanced ghost propagator are exhibited.
 $m_G$ is called the Gribov mass.

One step further has been accomplished by Zwanziger \cite{Zwanziger:1989} to relax the approximation 
used in the determination of the constraint and to extend it to all orders.
He found a condition to restrict the space of gauge fields \cite{Zwanziger:1989}
\beq
h(A) \ <\ d(N_C^2 -1)
\eeq
where $h(A)$, the horizon function , is a non local functional of the gauge fields. This approach has been subsequently refined and developed by Zwanziger himself (\cite{Zwanziger:1993}) and other authors (\cite{Kondo:2009ug,Dudal:2005na,Dudal:2007cw,Dudal:2008sp}). Zwanziger showed that this condition can be exponentiated to be transformed into a contribution ${\cal S}_{GZ}$ to the action~:
\beq
{\cal S}_{GZ} \ =\ \gamma \left(h(A) -  d(N_C^2 -1)\right)
\eeq
to be added to the usual contributions coming from the gauge action and the gauge fixing terms.
 $\gamma$ is defined by a relation, the horizon condition, required to implement the exponentiation~:
\beq\label{eq:horizon}
<h(A)>_{GZ}\ = \   d(N_C^2 -1)
\eeq
This action has many interesting properties : auxiliary fields can be introduced to transform ${\cal S}_{GZ}$ in a local form \cite{Zwanziger:1989ren} ; the theory has been proven to be renormalizable \cite{Zwanziger:1989ren} ; the gluon and ghost propagators have been computed and
 results similar  to the Gribov ones are found: 
a vanishing gluon propagator and an enhanced ghost propagator.
These results were in accordance with the common prejudice inferred at that time from the Dyson-Schwinger equations.

As it became clearer and  clearer that the lattice results for the propagators were not in accordance with 
these predicted behaviors, a refined version of the GZ approach was necessary  to have an answer for this discrepancy. 
The key point is the explicit introduction of the dimension two  condensates \cite{Dudal:2005na,Dudal:2008sp}.
This has the very nice feature that the renormalizability is not spoiled. With this refinement
the gluon and ghost propagators are in qualitative agreement with the lattice results and the
prediction from the ``decoupling" solution of the Dyson-Schwinger equations. In particular, the gluon 
propagator in the so-called ``refined'' Gribov-Zwanziger (RGZ) formalism is shown 
to behave as
\beq\label{eq:RGZgluon}
D(k^2) \ = \ \frac{k^2 + M^2}
{\displaystyle k^4 + k^2 \left(m^2 + M^2\right) + 2 g^2 N_C \gamma^2 + M^2 m^2}
\eeq
where $\gamma$ is the Gribov-Zwanziger parameter determined by the horizon condition, \eq{eq:horizon}, 
while $M$ and $m$ are two mass parameters related to the two above mentioned condensates . In particular, 
$m$ is related to the dimension-two gluon condensate, $\VA$, that will be  discussed in detail
below   (cf. section~\ref{sec:A2lat})  and plays a crucial role for the RGZ gluon propagator to account for lattice 
results~\cite{Dudal:2010tf}. Furthermore, the authors of ref.~\cite{Dudal:2011gd} studied the 
effective action in RGZ and provided  firm evidences that the dimension-two condensates should be 
non-vanishing (and hence the mass parameters $M$ and $m$) and  a strong indication that, 
in addition to the non-zero gluon propagator at vanishing momentum, the ghost propagator is not enhanced, 
in consistence with lattice data and with the previously discussed DSE picture for decoupling solutions.

\subsection{Other approaches}

Many other approaches have been applied to investigate the low-momentum properties of the 
Yang-Mills Green functions. We will end this introduction by indicating some of them and 
addressing the interested reader to some original works. 
Apart from DSEs or Lattice QCD, one of the most followed approaches is that of functional 
renormalization group equations (FRGs)~\cite{Ellwanger:1995qf,Ellwanger:1996wy,Fischer:2004uk,Pawlowski:2003hq,Pawlowski:2005xe}. 
Indeed, FRGs and DSEs appear to be rather interconnected and, for instance, it has been shown that the integrated flow equations 
define a set of DSEs within a particular renormalization scheme~\cite{Pawlowski:2005xe}. First, it was reported that 
FRGs analysis of the Landau-gauge low-momentum behaviour for the Yang-Mills Green functions leads to the uniqueness of 
a scaling-type solution~\cite{Pawlowski:2003hq,Fischer:2006vf}, but it has been recently proven, also within FRGs, 
that both scaling and decoupling solutions exist~\cite{Fischer:2009tn}. 
On the other hand, other approaches like the infrared mapping of $\lambda \phi^4$ and Yang-Mills 
theories in ref.~\cite{Frasca:2007uz,Frasca:2008tg,Frasca:2009yp} or the massive extension of the Fadeev-Popov 
action in ref.~\cite{Tissier:2010ts,Tissier:2011ey} appear to support a massive gluon propagator and a free ghost, 
{\it i.e.} a decoupling-type solution. In particular,  a very accurate description of lattice data for Yang-Mills 
gluon and ghost correlators in Landau gauge is obtained by means of a one-loop computation 
with this last massive extension of the Fadeev-Popov action in four dimensions, while the main features for lattice 
results in d=2,3 can be also accounted for. Furthermore, the low-momentum behaviour for the one-loop results with 
this massive action matched pretty well with the one expected for a decoupling solution within DSEs approach.
One can find in the literature still other methods, such as the application of stochastic quantization~\cite{Zwanziger:2003cf,Zwanziger:2001kw}, that have  been applied to  the subject. 

\subsection{The low-momentum correlators and the confinement problem}

Let us end this introduction with one comment about the 
Kugo-Ojima~\cite{Kugo:1979oj} confinement criterion which might be in order here. 
 In the Kugo-Ojima colour confinement picture, the physical spectrum of the theory is free of 
 coloured asymptotic states as a consequence of the so-called ``quartet mechanism''. 
 A sufficient condition for it to take place is that a certain correlation function, usually 
 denoted by $u(q^2)$ and called ``Kugo function'', should satisfy $u(0)=-1$. Furthermore, 
in Landau gauge, the Kugo function is linked to the ghost dressing function such that
$F(0) (1+u(0))=1$ (This relation was first noted by T. Kugo~\cite{Kugo:1995km} and
very recentely, K-I. Kondo triggered an interesting 
discussion about this relation, in connection with the Gribov horizon condition and its implications 
on the Landau-gauge  Yang-Mills infrared solutions~\cite{Kondo:2009ug,Boucaud:2009sd}).
Therefore, the sufficient condition for the realization of Kugo-Ojima confinement 
scenario requires a divergent ghost dressing function. Then, for the two classes of solutions 
above discussed, only the scaling may satisfy this sufficient condition. A different mechanism should 
thus explain the confinement for decoupling solutions, 
 as for instance the one provided by 
the center vortices scenario~\cite{Cornwall:1979hz,Greensite:2003xf,Greensite:2004ur}. 
 
However, as stated in the brief review 
about ``Strong Coupling Continuum QCD'' recentely appeared in the proceedings of the 9th conference 
on ``Quark Confinement and the Hadron Spectrum''~\cite{Pennington:2010gy}, 
the physics of hadrons and the confinement do not depend very much on 
how gluon and ghost propagates over distances of atomic scales but mainly on those of the size of 
a nucleus, where the difference between scaling and decoupling solutions is not important. 
Thus, it is very reasonable to think that the real QCD confinement mechanism might not be of 
a great help to discriminate among the two types of low-momentum solutions.



\section{A first insight with analytical tools}
\label{sec:analytic}

\subsection{The low momentum solutions from the ghost propagator DSE}

\subsubsection{The ghost propagator DSE}

We will examine the Dyson-Schwinger equation for the ghost
propagator (GPDSE) which can be written diagrammatically as

\vspace{\baselineskip}
\begin{small}
\bea
\left(\ghostDr\right)^{-1}%
\left(\ghostBr\right)^{-1}%
- 
\ghostSD %
\nonumber
\eea\end{small}%
\noindent i.e., denoting by $F^{(2)}$ (resp. $G^{(2)}$) the full ghost (resp. gluon) propagator, 

\bea\label{SD}
(F^{(2)})^{-1}_{ab}(k) &=&-\delta_{ab} k^2  \\ 
&-& g_0^2 f_{acd} f_{ebf} 
\int \frac{d^4q}{(2\pi^4) }  F^{(2)}_{ce}(q)
(i q_{\nu'}) \widetilde{\Gamma}_{\nu'\nu}(-q,k;q-k) (i k_\mu) (G^{(2)})_{\mu\nu}^{fd}(q-k), \nonumber
\eea
where $\widetilde{\Gamma}$ stands for the bare ghost-gluon vertex (above defined in \eq{vertghost}),
\beq
\widetilde{\Gamma}_\nu^{abc}(-q,k;q-k) \ &=& \ i g_0 f^{abc} q_{\nu'}  
\widetilde{\Gamma}_{\nu'\nu}(-q,k;q-k) \nonumber \\
&=&
i g_0 f^{abc} \left( \ q_\nu H_1(q,k) + (q-k)_\nu H_2(q,k) \ \right) \ ,
\label{DefH12}
\eeq
while $q$ and $k$ are respectively the outgoing and incoming ghost momenta and $g_0$ is the bare coupling constant. 
Let us now consider eq.~(\ref{SD}) at small momenta $k$. After applying the decomposition for 
the ghost-gluon vertex in eq.~(\ref{DefH12}), omitting colour indices and dividing 
 both sides by $k^2$, it reads 
\begin{equation}
\label{SD1}
\begin{split}
\frac{1}{F(k^2)} & = 1 + g_0^2 N_c \int \frac{d^4 q}{(2\pi)^4} 
\left( \rule[0cm]{0cm}{0.8cm}
\frac{F(q^2)G((q-k)^2)}{q^2 (q-k)^4} 
\left[ \rule[0cm]{0cm}{0.6cm}
\frac{(k\cdot q)^2}{k^2} - q^2  
        \right]
\ H_1(q,k)
           \right) \ .
\end{split}
\end{equation} 
It should be noticed that, because of the transversality condition, $H_2$ defined in 
eq.~(\ref{DefH12}) does not contribute for the GPDSE in the Landau gauge.

\subsubsection{Renormalization of the Dyson-Schwinger equation}

The integral equation eq.~(\ref{SD1}) is written in terms of bare Green functions. It is actually 
meaningless unless one specifies some appropriate UV-cutoff,\footnote{We have written for simplicity the UV 
cutoff as a hard cut-off. It is preferable to use a gauge invariant regularization procedure in view of
the advantage of exploiting Ward-Slavnov-Taylor identities (see sec.~\ref{SlavTayl}). 
In practice we will derive our results from the subtracted GPDSE which incorporates gauge invariant 
UV regularisation.}
 $\Lambda$, and performs the replacements 
$F(k^2) \rightarrow F(k^2,\Lambda)$ \dots~ . 
 It can be cast into a renormalized form by dealing properly with UV divergences, {\it i.e.}
\beq
g_R^2(\mu^2) &=& Z_g^{-2}(\mu^2,\Lambda) g_0^2(\Lambda) \nonumber \\
G_R(k^2,\mu^2) &=& Z_3^{-1}(\mu^2,\Lambda) G(k^2,\Lambda) \nonumber \\
F_R(k^2,\mu^2) &=& \widetilde Z_3^{-1}(\mu^2,\Lambda) F(k^2,\Lambda) \ ,
\label{Ren}
\eeq
where $\mu^2$ is the renormalization momentum and $ Z_g, Z_3$ and $\widetilde Z_3$ the renormalization 
constants for the coupling constant, the gluon and the ghost respectively. $ Z_g$ is related to the 
ghost-gluon vertex renormalization constant (defined by
 $\widetilde{\Gamma}_R=\widetilde Z_1 \Gamma_B$) through $ Z_g= \widetilde{Z_1} (Z_3^{1/2}\,\widetilde Z_3)^{-1}$. 
Then Taylor's non-renormalization theorem, which states that $H_1(q,0)+H_2(q,0)=1$ in Landau gauge 
(see app.~\ref{app:Taylor}) and 
to any perturbative order, can be invoked to conclude that 
$\widetilde Z_1$ is finite. We recall that the renormalization point is arbitrary, except for the 
special value $\mu = 0$ which cannot be chosen without  a loss of generality (see, in this respect, 
the discussion in ref.~\cite{Lerche:2002ep}). Thus, 
\beq\label{SDRnS}
\frac 1 {F_R(k^ 2,\mu^2)} \ = \ \widetilde Z_3(\mu^2,\Lambda) 
+ N_C \widetilde Z_1 \ g_R^2(\mu^2) \ \Sigma_R(k^2,\mu^2;\Lambda)  
\eeq
 where
\beq
\Sigma_R(k^2,\mu^2;\Lambda) &=& \int^{q^2 < \Lambda^2} \frac{d^4 q}{(2\pi)^4} 
\nonumber \\ 
&\times&
\left( \rule[0cm]{0cm}{0.8cm}
\frac{F_R(q^2,\mu^2)G_R((q-k)^2,\mu^2)}{q^2 (q-k)^4} 
\left[ \rule[0cm]{0cm}{0.6cm}
\frac{(k\cdot q)^2}{k^2} - q^2  
        \right]
\ H_{1,R}(q,k;\mu^2) \right) \ . \nonumber \\
\label{sigma}
\eeq 
One should notice that the UV cut-off, $\Lambda$, is still required as an upper integration bound 
in eq.~(\ref{sigma}) since the integral is UV-divergent, behaving as 
$\int dq^2/q^2 (1+11 \alpha_S/(2\pi) \log{(q/\mu)}))^{-35/44}$. In fact, the 
cut-off dependence this induces in $\Sigma_R$ cancels \footnote{One can easily check that 
$\widetilde Z_3^{-1}(\mu^2,\Lambda) \Sigma_R(k^2,\mu^2;\Lambda)$ approaches  some finite limit 
as $\Lambda \to \infty$ since the ghost and gluon propagator anomalous dimensions and the
beta function verify the relation $2 \widetilde \gamma + \gamma + \beta = 0$  \cite{Chetyrkin:2004mf}.}  against the one 
of $\widetilde Z_3$ in the r.h.s. of eq.~(\ref{SDRnS}), in accordance with the fact that the l.h.s. does not 
depend on $\Lambda$.

Now, we will apply a MOM renormalization prescription. This means that all the Green functions 
take their tree-level value at the renormalization point and thus:
\beq
F_R(\mu^2,\mu^2) \ = \ G_R(\mu^2,\mu^2) \ = \ 1 \ .
\eeq
In the following, $H_1(q,k)$ will be approximated by a constant\footnote{ This  approximation is very usually used to solve GPDSE. Some lattice data are available for the 
ghost-gluon vertex,  although they are by far less numerous than the ones regarding the propagators (see section~\ref{ghglvertex}).The present   data do indicate that the zero gluon momentum  $H_1(q,q)$ is approximatively 
constant with respect to $q$~\cite{IlgenGrib}. Of course, more data for different kinematical configurations should  
be welcome to check that approximation.} with respect to both 
momenta and, provided that $H_1(q,0)=1$ at tree-level, our MOM prescription implies 
that $H_{1,R}(k,q;\mu^2)=1$ and $\widetilde Z_1$ is a constant in terms of $\mu$.

%

\subsubsection{A subtracted Dyson-Schwinger equation} 
\label{subtracted}

The renormalized GPDSE, eq.~(\ref{SDRnS}), should be carefully analysed. We aim to study the 
infrared behaviour of its solutions and therefore focus our analysis on the momentum region, 
$k \ll \Lambda_{\rm QCD}$, where the IR behaviour of the dressing functions (presumably in powers of 
the momentum) is supposed to hold. One cannot forget, though, that the UV cut-off dependences in both sides 
of eq.~(\ref{SDRnS})  match only in virtue of the previously mentioned relation between the ghost and gluon propagator anomalous 
dimension and the beta function. 

However, in order  not to have to deal with the UV cut-off, we prefer  to approach the study of the GPDSE
in the following manner: 
we consider eq.~(\ref{SDRnS}) for two different scales, $\lambda k$ and $\lambda \kappa k$ 
(with $\kappa<1$ some fixed number and $\lambda$ an extra parameter that we 
shall ultimately let go to 0) and subtract them 
\beq
\frac{1}{F_R(\lambda^2 k^2,\mu^2)} - \frac{1}{F_R(\lambda^2 \kappa^2 k^2,\mu^2)}  
\ = \  
N_C \ g_R^2(\mu^2) \ \widetilde Z_1 \ 
\left( \rule[0cm]{0cm}{0.5cm} \Sigma_R(\lambda^2 k^2,\mu^2;\infty) - 
\Sigma_R(\lambda^2 \kappa^2 k^2,\mu^2;\infty) \right) \ . \nonumber \\
\label{SDRS}
\eeq
Then the integral in the r.h.s. is UV-safe, thanks to the subtraction, and the limit 
$\Lambda \to \infty$ can be explicitely taken,
\beq
\label{LamInf}
 \Sigma_R(\lambda^2 k^2,\mu^2;\infty) - 
\Sigma_R(\lambda^2 \kappa^2 k^2,\mu^2;\infty) 
&=&  \int \frac{d^4 q}{(2\pi)^4} 
\left( \rule[0cm]{0cm}{0.8cm}
\frac{F(q^2,\mu^2)}{q^2} \left(\frac{(k\cdot q)^2}{k^2}-q^2\right) \right. 
\nonumber \\ 
 &\times& \left. 
\left[ \rule[0cm]{0cm}{0.6cm}
\frac{G((q-\lambda k)^2,\mu^2)}{\left(q-\lambda k\right)^4} -  
(\lambda \to \lambda \kappa)
\rule[0cm]{0cm}{0.6cm} \right]
\rule[0cm]{0cm}{0.8cm} \right) \ .
\eeq
This equation is evidently a necessary consequence of the original one~(\ref{SDRnS}). That,  conversely, it is  actually sufficient was shown in ~\cite{Boucaud:2008ji}.
For an accurate analysis of eq.~(\ref{SDRS}) it is convenient, in addition,  to split the 
integration domain of eq.~(\ref{LamInf}) into two pieces by introducing 
some new scale $q_0^2$ ($q_0$, typically of the order of $\Lambda_{\it QCD}$, 
is a momentum scale below which the deep IR power behaviour is a 
good approximation),
\beq\label{q0}
\Sigma_R(\lambda^2 k^2,\mu^2;\infty) - \Sigma_R(\lambda^2 \kappa^2 k^2,\mu^2;\infty) 
\ = \ 
I_{\rm IR}(\lambda) \ + \ I_{\rm UV}(\lambda)
\eeq
where $I_{\rm IR}$ represents the integral in eq.~(\ref{LamInf}) over $q^2 < q_{0}^2$ and 
$I_{\rm UV}$ over $q^2 > q_{0}^2$.
%
%
Only the dependence on $\lambda$ is written explicitly because we  shall let it go to zero 
with $k$, $\kappa$ and $\mu^2$ kept fixed. 
The relevance of the $q_0^2$  scale stems from the drastic difference  between the IR and UV behaviours of the integrand.  In particular, for $(\lambda k)^2 \ll q_{0}^2$, the 
following infrared power laws, 
\beq\label{dress}
F_{\rm IR}(q^2,\mu^2) &=& A(\mu^2) \left( q^2 \right)^{\alpha_F}
\nonumber \\
G_{\rm IR}((q-\lambda k)^2,\mu^2) &=& B(\mu^2) \left( (q-\lambda k)^2 \right)^{\alpha_G} \ ,
\eeq
will be applied for both dressing functions in $I_{\rm IR}$.

Now,  a straightforward power-counting argument shows that  $I_{\rm IR}$ is infrared convergent if :
\begin{eqnarray}
\label{cond_ward}
\alpha_F &>& -2 \qquad{\rm IR\; convergence \; at}\; q^2 = 0 \nonumber \\
\alpha_G &>& 0 \qquad{\rm IR\; convergence\; at}\; (q-k)^2 = 0 \; 
{\rm and} \;(q-\kappa k)^2 = 0
\end{eqnarray}
We shall suppose in the following that these conditions are verified. 
Let us first consider  $I_{\rm UV}$. Its  dependence on $\lambda$, which is explicit in the 
factor inside the square bracket of eq.~(\ref{LamInf}),  
should clearly be even in $\lambda$ : any odd power of  $\lambda$  would imply an odd 
power of $q \cdot k$ whose angular integral is zero.  
Since the integrand is identically zero at $\lambda = 0$ and the integral is ultraviolet 
convergent, it is  proportional to $\lambda^2$ (unless some accidental cancellation 
forces it to behave as an even higher  power of  $\lambda$). On the other hand, 
after performing the change of variable $q\to \lambda q$, the 
IR contribution of the integral in eq.~(\ref{LamInf})'s r.h.s can be rewritten
as:
\bea
I_{\rm IR}(\lambda) 
&\simeq& 
\left(\lambda^2 \right)^{(\alpha_F + \alpha_G )} A(\mu^2) B(\mu^2) 
\displaystyle \int^{q^2 < 
\frac{q_{0}^2}{\lambda^2}} \frac{d^4 q}{(2\pi)^4} \ \
(q^2)^{\alpha_F-1} \          
\left( 
\frac{(k\cdot q)^2}{k^2}-q^2
\right)  
\nonumber \\ 
& & \times 
\left[ 
  \left(  (q-k)^2 \right)^{\alpha_G  -2} -
  \left((q-\kappa k)^2\right)^{\alpha_G -2} 
\right] \ ,
\label{I1}
\eea
that, as it shall be seen in the next subsection, asymptotically 
behaves as
\bea
I_{\rm IR}(\lambda) \sim \left\{ 
\begin{array}{ll} 
\displaystyle 
\lambda^{2(\alpha_G+\alpha_F)}
 & \mbox{\rm if} \ \ \alpha_G+\alpha_F  < 1\\
\displaystyle 
\lambda^2 \ \ln{\lambda} &  \mbox{\rm if} \ \ \alpha_G+\alpha_F  = 1 \\
\displaystyle 
\lambda^2 & 
\mbox{\rm if}   \ \ \alpha_G+\alpha_F  > 1 \ .
\end{array} 
\right.
\label{cond_alpha_F_G}
\eea
Thus, in all the cases, 
the leading behaviour of $I_{\rm IR}+I_{\rm UV}$, as $\lambda$ vanishes, is given by $I_{\rm IR}$ 
in eq.~(\ref{cond_alpha_F_G}). The subtracted renormalised GPDSE reads 
for $\alpha_G+\alpha_F \leq 1$ as:
\beq
\frac{1}{F_R(\lambda^2 k^2,\mu^2)} - \frac{1}{F_R(\lambda^2 \kappa^2 k^2,\mu^2)}  
\ \simeq \  
N_C \ g_R^2(\mu^2) \ \widetilde Z_1 \ 
I_{\rm IR}(\lambda) \ ,
\label{SDRSF}
\eeq
for small $\lambda$. 
We have assumed that $H_1$ is constant when varying all the momenta but 
(\ref{cond_alpha_F_G},\ref{SDRSF}) remain true if one only assumes that 
$H_1$  behaves ``regularly'' for $q^2, k^2 \le q_{0}^2$
(i.e. is free of singularities or, at least, of any singularity worse than logarithmic).

\subsubsection{The integral for the ghost self-energy}

The present section is devoted to the quantitative 
analysis of the integral $I_{\rm IR}(\lambda)$, defined 
in \eq{I1}, which gives the contribution of the
 ghost loop to the renormalised GPDSE \eq{SDRSF}. If $\alpha_F+\alpha_G < 1$ it is possible to  
perform analytically the integral and to find a compact expression for it. In this case, one can write
\beq
I_{\rm IR}(\lambda) 
&\simeq& 
A(\mu^2) B(\mu^2) \ \left(\lambda^2 \right)^{(\alpha_F + \alpha_G )}
\left( \Phi(k;\alpha_F,\alpha_G) - \Phi(\kappa k;\alpha_F,\alpha_G) 
\rule[0cm]{0cm}{0.5cm} 
\right) 
\label{I2}
\eeq
where $A(\mu^2)$ and $B(\mu^2)$ were defined in \eq{dress} and
\beq
\Phi(k;\alpha_F,\alpha_G) = 
\int \frac{d^4 q}{(2\pi)^4} \ \
(q^2)^{\alpha_F-1} \ \left(  (q-k)^2 \right)^{\alpha_G  -2}         
\left( 
\frac{(k\cdot q)^2}{k^2}-q^2
\right)  \ ,
\label{BigPhi}
\eeq
provided that $\Phi(k;\alpha_F,\alpha_G)$ is not singular, so that the subtraction inside the bracket and 
the integral operator in \eq{I2} commute with each other.
Then, following~\cite{Bloch:2003yu}, we define
\beq
f(a,b) \ &=& \ \frac {16 \pi^2} {(k^2)^{2+a+b}} \ \int \frac{d^4 q}{(2\pi)^4} 
(q^2)^a
\left(  (q-k)^2 \right)^b \nonumber \\
&=&  
\frac{\Gamma(2+a) \Gamma(2+b) \Gamma(-a-b-2)}
{\Gamma(-a) \Gamma(-b) \Gamma(4+a+b)} \ ,\label{fgammas}
\eeq
and obtain
\beq\label{smallphi}
\Phi(k;\alpha_F,\alpha_G) \ = \ 
\frac{(k^2)^{\alpha_F+\alpha_G}}{16 \pi^2} \ \phi(\alpha_F,\alpha_G)
\eeq
where
\beq
\phi(\alpha_F,\alpha_G)  &=& 
-\frac 1 2 \left( 
  f(\alpha_F,\alpha_G-2) + f(\alpha_F,\alpha_G-1) 
  + f(\alpha_F -1,\alpha_G-1)
\right) \nonumber \\
&+& 
 \frac 1 4 \left(
  f(\alpha_F-1,\alpha_G-2) + f(\alpha_F-1,\alpha_G) 
  + f(\alpha_F+1,\alpha_G-2)
\right)  \ .\label{phifs}
\eeq
Thus, if $\alpha_F+\alpha_G < 1$,
\beq\label{IR<1}
I_{\rm IR}(\lambda) \ \simeq \ \frac {A(\mu^2) B(\mu^2)}{16 \pi^2} (\lambda^2 k^2)^{\alpha_F+\alpha_G}
(1-\kappa^{2(\alpha_F+\alpha_G)}) \ \phi(\alpha_F,\alpha_G) \ .
\eeq

\medskip

We will now compute the leading asymptotic behavior of $I_{\rm IR}$ as $\lambda \to 0$ when 
$\alpha_F+\alpha_G=1$. In that case, after performing in \eq{I1} the following expansion,
\bea\label{expa}
\left[ (k - q)^2 \right]^{\alpha_G -2} - 
\left[ (\kappa k - q)^2 \right]^{\alpha_G -2} &\simeq& (q^2)^{\alpha_G-2} \ (\alpha_G-2) (1-\kappa)
\\  
& \times & \left[ - 2 \ \frac{q \cdot k}{q^2} \ + \ (1+\kappa) \left( \frac {k^2}{q^2} 
+ 2 (\alpha_G-3) \frac{(q \cdot k)^2}{q^4} \right) \right] \ ,
\nonumber 
\eeq
and neglecting the term odd in $q_\mu \to -q_\mu$ 
one finds for the leading contribution 
\beq\label{IR=1}
I_{\rm IR}(\lambda) & \simeq &
- k^2 (1-\kappa^2) \frac {2 A(\mu^2) B(\mu^2)}{(2 \pi)^3} \ 
\lambda^2 \int^{q_0/\lambda} dq \ q^{2(\alpha_F+\alpha_G)-3}
\nonumber \\
&\times& \int_0^\pi d\theta \ {\rm sin}^4\theta 
\left( \alpha_G-2 + 2(\alpha_G-3)(\alpha_G-2) {\cos}^2\theta 
\rule[0cm]{0cm}{0.5cm}
\right)
\nonumber \\
& \simeq &
k^2 (1-\kappa^2) \frac {A(\mu^2) B(\mu^2)}{32 \pi^2} \ \alpha_G (\alpha_G-2) \lambda^2 \ln{\lambda} \ .
\eeq
We do not specify the lower bound of the integral over $q$ in \eq{IR=1} because it necessarily 
contributes as a subleading term, once the ghost-loop integral is required to be IR safe.
Then, as was indicated in anticipation  in \eq{cond_alpha_F_G}, $I_{\rm IR}$ 
diverges logarithmically as $\lambda$ goes to zero if $\alpha_F+\alpha_G=1$. 
In fact, since \eq{IR<1} is a reliable result for any $\alpha_F+\alpha_G < 1$ 
however close it may be to 1, such a divergence appears as a pole of a $\Gamma$ function of $\phi(\alpha_F,\alpha_G)$ 
in \eq{smallphi}.

\medskip

Finally, if $\alpha_F+\alpha_G > 1$, the leading contribution for $I_{\rm IR}(\lambda)$ 
as $\lambda$ vanishes can be computed 
after performing back the change of integration variable, $q \to q/\lambda$, in \eq{I1}. The first 
even term in \eq{expa} dominates again the expansion after integration, but now it does 
not diverge. Then, if we procceed as we did in \eq{IR=1}, we obtain
\beq\label{IR>1}
I_{\rm IR}(\lambda) \ \simeq \ - \frac{\alpha_G (\alpha_G-2)}{\alpha_F+\alpha_G-1} \
\frac{(q_0^2)^{\alpha_F+\alpha_G-1}}{64 \pi^2} \ A(\mu^2) B(\mu^2) \ k^2 \lambda^2 (1-\kappa^2) \ ,
\eeq
for small $\lambda$ and $\alpha_G+\alpha_F > 1$. It should be noticed that $I_{\rm IR}$ in \eq{IR>1} 
depends on the additional scale $q_0$ introduced in \eq{q0} to separate IR and UV integration domains.
In fact, if one takes $q_0 \to \infty$, $I_{\rm IR}$ diverges. This means that, when 
$\alpha_F+\alpha_G > 1$, the behaviour of the IR power laws  hampers their use for all momenta in the integral. The 
finiteness of the ghost-loop integral of the subtracted GPDSE can only be recovered  after taking into account the 
UV logarithmic behaviour for large-momenta 
dressing functions~\footnote{ The multiplicatively renormalisable (MR) truncation scheme corresponds to letting
$\Lambda_{\rm QCD} \to \infty$. Therefore, the scale $q_0$ being of the order of $\Lambda_{\rm QCD}$, power laws 
with $\alpha_F+\alpha_G > 1$ cannot be accepted as solutions of the GPDSE (see, for instance,~\cite{Bloch:2003yu}). The same argument holds also for 
$\alpha_F+\alpha_G=1$, because the ghost-loop integral in \eq{IR=1} diverges as $\lambda \to 0$ for any $q_0$ fixed 
as well as for $q_0 \to \infty$ for any fixed $\lambda$. }.  
  Furthermore, $I_{\rm UV}$,  behaving  too as $\lambda^2$, 
should also be added in the r.h.s. of \eq{SDRSF} in order to write the renormalised GPDSE.
Thus, the dependence on $\lambda$ but not the factor in 
front of it can be inferred 
from the GPDSE with only the information of the asymptotics for small-momentum dressing functions.

%

\subsubsection{The two classes of solutions}
\label{subsec:twoclass}

The starting point for the following 
infrared analysis will be the \eq{SDRSF} for small $\lambda$, where we will try to 
make the dependences on $k,\kappa$ and $\lambda$ of the two sides match each other. 

\paragraph{The case $\alpha_F \neq 0$ (scaling solution): } 

We will first study the case $\alpha_F \neq 0$. Then, the l.h.s. of \eq{SDRSF} can be 
expanded for small $\lambda$ as
\beq
\frac{1}{F_R(\lambda^2 k^2,\mu^2)} - \frac{1}{F_R(\lambda^2 \kappa^2 k^2,\mu^2)}  
\simeq 
\left( 1-\kappa^{-2 \alpha_F} \right) \
\frac{\left( \lambda^2 k^2 \right)^{-\alpha_F}}{A(\mu^2)} 
\label{lhsSDRSF}
\eeq
and one obtains from \eq{SDRSF}:
\beq
N_C \ g_R^2(\mu^2) \ \widetilde Z_1  A(\mu^2) \
\frac{I_{\rm IR}(\lambda)}
{\left( 1-\kappa^{-2 \alpha_F} \right) \left( \lambda^2 k^2 \right)^{-\alpha_F} }
\ \simeq \ 1 \ ,
\label{SDRS2}
\eeq
where the dependences on $k,\kappa$ and $\lambda$ of the  numerator and  the denominator
should cancel against each other. Using for  $I_{\rm IR}$ the form given  after \eq{cond_alpha_F_G}, 
we find three possible situations:

\begin{itemize}

\item If $\alpha_G+\alpha_F > 1$,  applying \eq{IR>1} in \eq{SDRS2}, we are led to the conclusion that 
only \fbox{$\alpha_F=-1$}~(and $\alpha_G  > 2$) satisfies this last 
equation and could be an IR solution for GPDSE. 
However, such  a solution appears to be in a clearcut contradiction with the current lattice simulations.

\item If $\alpha_G+\alpha_F=1$, there is no possible solution because the logarithmic 
behaviour of $I_{\rm IR}$ in \eq{IR=1} cannot be compensated by the powerlike one in the denominator 
of \eq{SDRS2}.  

\item If $\alpha_G+\alpha_F < 1$, \eq{IR<1} combined with \eq{SDRS2} implies the familiar relation   
\fbox{$2 \alpha_F+\alpha_G=0$} and we have then:
\beq
N_C \ g_R^2(\mu^2) \ \widetilde Z_1  \frac{(A(\mu^2))^2 B(\mu^2)}{16 \pi^2} \
\phi\left(-\frac{\alpha_G} 2,\alpha_G\right)
\ \simeq \ 1 \ ,
\label{SDRS2<1}
\eeq

\end{itemize}

An immediate consequence of this last condition is the freezing of the running coupling 
constant at small momentum. If the renormalization point, $\mu$, is arbitrarily chosen 
to be very small in order that the dressing functions observe the power laws 
at $k^2=\mu^2$, one obtains $A(\mu^2)=\mu^{-2 \alpha_F}$ and $B(\mu^2)=\mu^{-2 \alpha_G}$. 
Eq.~(\ref{SDRS2<1}) then reads
\beq\label{SDRS2<1deep}
N_C \ g_R^2(\mu^2) \ \widetilde Z_1 \  \phi\left(-\frac{\alpha_G} 2,\alpha_G\right) \ 
\simeq 16 \pi^2 \ ,
\eeq
and should be satisfied for any small value of 
$\mu$. Consequently, it should remain exact as $\mu \to 0$ and provides the small-momentum 
limit of the running coupling (which is independent of the infrared constants for 
ghost and gluon dressing functions). 

In particular, if $\alpha_G=1$, one has $\phi(-1/2,1)=8/5$ and thus
\beq\label{SDRS2<1aG=1}
N_C \ g_R^2(\mu^2) \ \widetilde Z_1 \  
\simeq 10 \pi^2 \ ,
\eeq

\paragraph{The case $\alpha_F=0$ (decoupling solution): }

The case $\alpha_F=0$ is particular in  that the leading contributions to the two occurrences of $F$ in the l.h.s of 
eq.~(\ref{SDRSF}) cancel against each other. 
We have then to go one step further, taking into account the subleading terms.   Defining $\widetilde F_{IR}$ by means of $ F_{IR}(q^2,\mu^2)=A(\mu^2) + \widetilde F_{IR}(q^2,\mu^2)$ we rewrite the  l.h.s of eq.~(\ref{SDRS}) as $ -(\widetilde F_{IR}(\lambda^2 k^2,\mu^2) -\widetilde  F_{IR}(\lambda^2 \kappa^2k^2,\mu^2))/A^2(\mu^2)$ and use the known IR behaviour of $I_{\rm IR}(\lambda)$ from eq.\ref{cond_alpha_F_G})) in the r.h.s. of eq.~(\ref{SDRSF}) to get 
\beq\label{FIR2}
F_{\rm IR}(q^2,\mu^2) \ = \ 
\left\{ 
\begin{array}{ll}
A(\mu^2) + A_2(\mu^2) q^2 \ln{q^2} & {\rm if~} \alpha_G=1 \\
A(\mu^2) + A_2(\mu^2) (q^2)^{\alpha_F^{(2)}} & {\rm otherwise} \ .
\end{array}
\right.
\eeq 
Furthermore, not only the subleading functional behaviour of the dressing function 
can be constrained but also the coefficient $A_2$ in \eq{FIR2}. In fact, if we
plug this equation into the l.h.s. of eq.~(\ref{SDRSF}) and expand we obtain :
\beq
- \frac{(A(\mu^2))^2}{A_2(\mu^2)} \ N_C \ g_R^2(\mu^2) \ \widetilde Z_1 \
I_{\rm IR}(\lambda) \ 
\simeq \ 
\left\{
\begin{array}{ll} 
k^2 (1-\kappa^2) \lambda^2 \ln{\lambda^2} & {\rm if~} \alpha_G=1 \\
(\lambda^2 k^2)^{\alpha_F^{(2)}} (1-\kappa^{2\alpha_F^{(2)}}) & {\rm otherwise} \ ,
\end{array}
\right.
\label{SDRS3}
\eeq
Let us consider now in more detail the three possible cases.
\begin{itemize}

\item If $\alpha_G < 1$, we obtain from eqs.~(\ref{IR<1},\ref{SDRS3}) that 
\fbox{$\alpha_F^{(2)}=\alpha_G$}~. Then, 
\beq
- \frac{(A(\mu^2))^3 B(\mu^2)}{A_2(\mu^2)} \ N_C \ g_R^2(\mu^2) \ \widetilde Z_1 \
\phi(0,\alpha_G) 
\simeq \ 
16 \pi^2 \ ,
\label{SDRS3<1}
\eeq
where, according to eqs.~(\ref{fgammas},\ref{phifs}) $\phi(0,\alpha_G)$ is given by 
\beq\label{solphi}
\phi(0,\alpha_G) \ = \ \frac 3 {2 \alpha_G (\alpha_G+1)(\alpha_G+2)(1-\alpha_G)}
\eeq

\item Similarly if $\alpha_G=1$, \eq{IR=1} applied to \eq{SDRS3} leads to
\beq
\frac{(A(\mu^2))^3 B(\mu^2)}{A_2(\mu^2)} \ N_C \ g_R^2(\mu^2) \ \widetilde Z_1 \
\simeq \ 
64 \pi^2 \ .
\label{SDRS3=1}
\eeq

\item At last, if $ \alpha_G > 1$, eqs. (\ref{IR>1}) and  (\ref{SDRS3}) imply:~\fbox{$\alpha_F^{(2)}=1$}~. 
{\it i.e.}, a ghost dressing function which behaves quadratically for small momenta,
In this case, however, as already said the ghost loop cannot be evaluated using the IR power laws  over the whole integration range and it is therefore not possible to solve the GPDSE  consistently, nor even to determine the small-momentum behaviour 
of the dressing functions, without matching appropriately those power laws to the UV perturbative 
formulas. Thus, we are not able to derive a constraint for the next-to-leading coefficient, $A_2(\mu^2)$.

\end{itemize}

In summary, the GPDSE admits IR solutions with $\alpha_F=0$ and any $\alpha_G > 0$, provided that 
\beq\label{solghost}
F_{\rm IR}(q^2,\mu^2) \ = \ 
\left\{
\begin{array}{ll}
A(\mu^2) \left( 1 -  \phi(0,\alpha_G) \displaystyle \frac{ \widetilde{g}^2(\mu^2)}{16 \pi^2} A^2(\mu^2) B(\mu^2)
(q^2)^{\alpha_G} \right) & \alpha_G < 1 \\
A(\mu^2) \left( 1 + \displaystyle \frac{\widetilde{g}^2(\mu^2)}{64 \pi^2} A^2(\mu^2) B(\mu^2) \ 
q^2 \ln{q^2} \right) & \alpha_G = 1 \\
\rule[0cm]{0cm}{0.5cm} A(\mu^2) + A_2(\mu^2) q^2 & \alpha_G > 1
\end{array}
\right.
\eeq
where $\widetilde{g}^2(\mu^2)=N_C \ g_R^2(\mu^2)  \widetilde{Z_1}$ and $\phi(0,\alpha_G)$ is given in \eq{solphi}. 
The gluon dressing function is supposed to behave as indicated in \eq{dress}. 
In particular for $\alpha_G=1$, the gluon propagator 
takes a finite (and non-zero) value at zero momentum, $B(\mu^2)$, 
after applying MOM renormalisation prescription at $q^2=\mu^2$.


\subsection{The low-momentum solutions and the gluon mass}
\label{subsec:next}

As will be seen in the next section, the current lattice data strongly supports a decoupling solution which does not obey 
$2\alpha_F+\alpha_G=0$ and in which $\alpha_G=1$. Furthermore, 
lattice data can also be very well accommodated within DS coupled 
equations in the PT-BFM scheme~\cite{Aguilar:2006gr,Aguilar:2008xm} and within 
the so-called refined Gribov-Zwanziger approach~\cite{Dudal:2007cw}, leading in both cases to 
decoupling solutions for gluon and ghost propagators. Then, as \eq{solghost} reads for, this 
implies a ghost dressing function proportional to $q^2 \log{q^2}$ while 
the gluon propagator takes a constant at vanishing momentum ($\alpha_G=1$) or, in other 
words, is ``{\it massive}''. 

Indeed, it is well known that the Schwinger mechanism of mass generation~\cite{Schwinger1962} can be incorporated 
ino the gluon propagator DSE
through the fully-dressed non-perturbative three-gluon vertex and gives rise to the generation of a dynamical 
gluon mass such that~\cite{Aguilar:2006gr,Cornwall:1981zr}, 
\beq\label{schwinger}
D^{-1}(q^2) \sim q^2 + M^2(q^2) \ .
\eeq
In particular, it has been shown that a power-law running mass, 
\beq\label{runningmass}
M^2(q^2) \ = \ \frac{m_0^4}{m_0^2+q^2} 
\left( \frac{\ln{\left(\frac{q^2+2m_0^2}{\Lambda_{\rm QCD}}\right)}} 
{\ln{\frac{2m_0^2}{\Lambda_{\rm QCD}}}} \right)^3 \ ,
\eeq
appears as a solution for the coupled ghost and gluon propagator DSE in the PT-BFM truncation 
scheme~\cite{Aguilar:2007ie} and, also, as a consequence of the dimension-4 gluon condensate 
in the OPE expansion of the gluon self-energy in the Pinching Technique framework~\cite{Lavelle:1991ve}.

Having this in mind, the authors of ref.~\cite{Boucaud:2010gr} applied the following simple 
model~\footnote{This is a renormalized 
massive gluon propagator, as given by \eq{schwinger}, where the gluon running mass appears to be 
approximated by $M(q^2) \simeq M(0) = m_0 \equiv M$. This is, for instance, a very good 
low-momentum approximation for the running mass given by \eq{runningmass} with $m_0 \sim 0.5$ GeV and 
$\Lambda_{\rm QCD} \sim 0.3$ GeV (see fig. 11 of ref.~\cite{Aguilar:2009nf}).}:  
\beq\label{gluonprop}
D_{\rm IR}(q^2,\mu^2) \ = \ \frac{G_{\rm IR}(q^2,\mu^2)}{q^2} &\simeq& \frac{B(\mu^2)}{q^2 + M^2} 
\ = \ \frac{B(\mu^2)}{M^2} \left( 1 \ - \ \frac{q^2}{M^2} + \mathcal{O}\left(\frac{q^4}{M^4}\right) \right) \ ,
\eeq
for a massive gluon propagator, in order to compute the ${\cal O}(q^2)$-correction for the 
low-momentum ghost dressing function and then prove that this low-momentum behaviour 
is controlled by that gluon mass and by the zero-momentum value of the 
effective charge defined from the Taylor-scheme ghost-gluon vertex in 
ref.~\cite{Aguilar:2009nf}.

The work of ref.~\cite{Boucaud:2010gr} can be easily overviewed if we re-write \eq{SDRS} 
as follows
\beq\label{eq:JP1}
\frac 1 {F_R(k)} - \frac 1 {F_R(p)} & = & \frac{g_R^2 N_C}{(2\,\pi)^4}  
\int_{q<q0} d^4q \frac{F_R(q)}{q^2}\left\{ \frac{1}{k^2} D_R(q-k) \frac{(k\cdot q)^2-k^ 2 q^2}{(q-k)^2} -k\to p \right\} 
\nonumber \\
&=&
N_C g^2_R \ \left( I_{\rm IR}(k^2) -   I_{\rm IR}(p^2) \right)
\eeq
for the two momenta $k$ and $p$, which will be taken to be small (as compared with $M$ and $q_0$, the cut-off separating 
IR and UV modes in the ghost-self-energy integral), and where the ghost-gluon transverse 
form factor, $H_1$, has been explicitly replaced by 1 and the dependence on the renormalization momentum, $\mu^2$, omitted 
by simplicity. 
Then, if one inserts the form  \eq{gluonprop} of the gluon propagator into the integrand 
and focuses on the low-momentum behaviour of the ghost dressing function for the 
decoupling case ($\alpha_F=0)$, the ghost dressing function will be replaced by 
the constant leading term, $A(\mu^2)$ from \eq{dress}, and one will obtain:
\beq\label{eq:JP2}
I_{\rm IR}(k^2) &=&  - \frac{1}{4\,\pi^3}\frac{A(\mu^2) \ B(\mu^2)}{ M^2} \int_0^{q_0} q^3 F(q)\,dq\int_0^\pi d\theta \sin^4(\theta)
\nonumber \\
&\times& \left( \frac{1}{k^2+q^2 -2 k q \cos(\theta)} -   \frac{1}{M^2+k^2+q^2 -2 k q \cos(\theta)}\right) 
\nonumber \\
&=& - \frac{1}{32\,\pi^2}\frac{A(\mu^2) \ B(\mu^2)}{ M^2}  \int_0^{q_0} q^3 F(q)\,dq\left( \theta(k-q) \frac{3 k^2 - q^2}{k^4} +  \theta(q-k) \frac{3 q^2 - k^2}{q^4}\right.
\nonumber \\
&& \left.  -\frac{6 k^2 q^2 (M^2+k^2+q^2) -(M^2+k^2+q^2)^3 + ((M^2+k^2+q^2)^2-4 k^2 q^2)^{3/2}}{2 k^4 q^4}\right)
\ .
\eeq\label{eq:JP3}
The integral given by \eq{eq:JP2} can be first analytically obtained and  then expanded up to the first order in $k^2/M^2$ 
to give:
\beq
I_{\rm IR}(k^2) &=& -\frac{1}{32\,\pi^2}\frac{A(\mu^2) \ B(\mu^2)}{ M^2} 
\nonumber \\
&\times& \left( \frac{3}{2}M^2\log(\frac{M^2+q_0^2}{M^2})+ \frac{k^2}{2} \log(\frac{k^2}{M^2})- \frac{11}{12} k^2 + \frac{k^2}{2} \log(1+\frac{M^2}{q_0^2}) - \frac{k^2}{2} \frac{q_0^2 M^2}{(q_0^2 +M^2)^2} \right) 
\eeq
from where one needs only to keep
\beq\label{eq:JP4}
I_{\rm IR}(k^2) &=& -\frac{1}{64\,\pi^2}\frac{A(\mu^2) \ B(\mu^2)}{ M^2} \
\left( k^2 \ \log(\frac{k^2}{M^2})- \frac{11}{6} k^2 \ + \cdots \right) \ ,
\eeq
because the terms in ${\cal O}(M^2/q_0^2)$ are neglected, while the constant terms and the ones 
logarithmically divergent ($\sim \log{(q_0})$) happen to be cancelled when applying the subtraction for 
two different momenta in the r.h.s. of \eq{eq:JP1}.

Thus, if one replaces $I_{\rm IR}$ by the result of \eq{eq:JP4} in the r.h.s. of \eq{eq:JP1}, the subleading 
term for the ghost dressing function should be:
%
\beq\label{solFIRJo}
F_{\rm IR}(q^2,\mu^2) = F_{\rm IR}(0,\mu^2) \left( 1 \ + \ 
\frac{N_C H_1}{16 \pi} \ \overline{\alpha}_T(0) \ 
\frac{q^2}{M^2} \left[ \ln{\frac{q^2}{M^2}} - \frac {11} 6 \right]
\ + \ {\cal O}\left(\frac{q^4}{M^4}, \frac{M^2}{q_0^2} \right) \right)
\eeq
%
where:
\beq\label{coefC2}
\overline{\alpha}_T(0) = \lim_{q \to 0} \left(q^2 + M^2 \right) \frac{\alpha_T(q^2)}{q^2} 
&=& 
M^2  \ \frac{g^2_R(\mu^2)}{4 \pi} F_{\rm IR}^2(0,\mu^2) D_{\rm IR}(0,\mu^2)  
\nonumber \\
& =& 
\frac{g^2_R(\mu^2)}{4 \pi} A^2(\mu^2) B(\mu^2) \ .
\eeq
Here $\alpha_T=g_T^2/(4\pi)$ is the perturbative strong coupling 
defined in the Taylor scheme~\cite{Boucaud:2008gn}, while 
$\overline{\alpha}_T$ is the non-perturbative Taylor effective charge defined as an extension 
of the Taylor ghost-gluon coupling in ref.~\cite{Aguilar:2009nf}, 
similarly to the way the ``pinching technique''~\footnote{The ``pinching technique''~\cite{Cornwall:1981zr} implies a resummation of the 
diagrams for the perturbative gluon propagator expansion leading to a redefinition of the 
propagator such that it observes a QED-like Ward identity, thus providing us with a way to 
construct an IR effective charge as happens in QED.} (PT) effective charge was  from the gluon 
propagator in ref.~\cite{Aguilar:2008fh}.

In the forthcoming  sections \ref{sec:lattice} and \ref{sec:num}, \eq{solFIRJo} will be confronted to lattice and numerical DSE estimates for 
the ghost dressing function in the low-momentum domain.


\subsection{Some constraints from Slavnov-Taylor identities}
\label{SlavTayl}

In the previous section, we have analysed the infrared behaviour of the  GPDSE solutions and found that the 
ghost dressing function  can either diverge at vanishing momentum 
($\alpha_F=-\alpha_G/2$ with $\alpha_G > 0$) or take a finite value 
($\alpha_F=0$ with any $\alpha_G > 0$). As appendix \ref{App} shows, the GPDSE themselves can be 
derived from the general Ward-Slavnov-Taylor equation~\cite{Taylor}.
The Ward-Slavnov-Taylor identities (WSTI) can be derived formally from 
the gauge invariance of the path integral, eq.~(\ref{action}) of appendix \ref{App}, 
as shown in \cite{IZ}. This is  the case in lattice simulations. If
the path integral is limited to a domain of the configuration space such as in 
the Gribov-Zwanziger approach, used in~\cite{Dudal:2007cw}, the STI may not be satisfied. We assume 
a gauge invariant path integral and will now invoke the WSTI 
for general covariant gauges relating the 3-gluon, $\Gamma_{\lambda\mu\nu}(p,q,r)$, 
and ghost-gluon vertices, 
\begin{equation}
\label{STid}
\begin{split}
p_\lambda\Gamma_{\lambda \mu \nu} (p, q, r) & \frac{F(p^2)}{G(r^2)} (\delta_{\rho\nu} r^2 - r_\rho r_\nu) \widetilde{\Gamma}_{\rho\mu}(r,p;q) 
\\ & -
\frac{F(p^2)}{G(q^2)} (\delta_{\rho\mu} q^2 - q_\rho q_\mu) \widetilde{\Gamma}_{\rho\nu}(q,p;r) \ .
\end{split}
\end{equation}
to shed some light on that matter~\cite{ours-ST,Davydychev:1996pb}. 
Using for the ghost-gluon 
vertex the  general decomposition\footnote{We work, of course, on the energy-momentum shell, 
so that the relation $p + q + r \equiv 0$ holds}~\cite{Ball:1980ax}
\beq\label{Def2}
 \widetilde{\Gamma}_{\nu\mu}(p,q;r) &=& 
\delta_{\nu\mu} \ a(p,q;r) \ - \ r_\nu q_\mu \ b(p,q;r) 
\ + \ p_\nu r_\mu \ c(p,q;r)\nonumber
\\ 
&+& r_\nu p_\mu \ d(p,q;r) \ + \  p_\nu p_\mu \ e(p,q;r) \ ,
\eeq
and multiplying by $r_\nu$  both sides of \eq{STid}, one obtains:
\beq\label{STid2}
r_\nu p_\lambda \Gamma_{\lambda\mu\nu}(p,q,r) = 
\frac{F(p^2)}{G(q^2)} X(q,p;r) \ 
\left[ (q \cdot r) q_\mu - q^2 r_\mu  \right]  \ ;
\eeq
where 
\beq\label{combi}
X(q,p;r) \ = \ a(q,p;r)- (r \cdot p) \ b(q,p;r) + (r \cdot q) \ d(q,p;r) \ .
\eeq
Since the vertex function, $\Gamma$, in the l.h.s. of \eq{STid2} is antisymmetric under $p \leftrightarrow r$
and $\lambda \leftrightarrow \nu$, one can  conclude that\cite{Chetyrkin:2000dq,ours-ST}:
\begin{equation}\label{Ghexp}
F(p^2) X(q,p;r)= F(r^2) X(q,r;p) \ .
\end{equation} 
This last result is a compatibility condition required for the WSTI to be satisfied that 
does not involve the 3-gluon vertex and implies a strong correlation between the infrared 
behaviours of the ghost-gluon vertex and the ghost propagator. 
Now, under the only additional hypothesis that those scalars of the ghost-gluon vertex 
decomposition in \eq{Def2} which contribute to the scalar function $X$ defined 
in \eq{combi} are regular\footnote{Note also that, for our purposes, it will actually be enough to 
restrict, but not forbid, the possible presence of singularities in the scalar coefficient 
functions provided that they could be compensated by kinematical zeroes 
stemming from the tensors.} 
when one of their arguments goes to zero while the others 
are kept non-vanishing, one can consider the small $p$ limit in \eq{Ghexp} and obtain:
\beq
F(p^2) X(q,0;-q)= F(q^2) X(q,-q;0) +{\cal O}(p^2)\ 
\eeq
This has to be true for {\bf any} value of $q$, which implies that $F(p^2)$  {\bf goes to some finite and non-zero value when $p$ goes to zero}, since neither $X(q,0;-q)$ nor $X(q,-q;0)$ are presumably zero for all values of $q$. Rephrased in terms of infrared exponents, the latter argument  implies that  $\alpha_F=0$.

To reach the above conclusions we did not appeal to the properties of the  3-gluon vertex, apart from the symmetry under the exchange of gluon legs. If one assumes in addition that the longitudinal 
part of the 3-gluon vertex also behaves regularly when anyone of its arguments goes to $0$, the 
others being kept non-vanishing, a divergent gluon propagator at vanishing momentum will be 
obtained from \eq{STid2}~\cite{Boucaud:2005ce,Boucaud:2007va,ours-ST}:
\beq
\lim_{q \to 0} \ p_\lambda r_\nu \Gamma_{\lambda \mu \nu}(p,q,r) &=& - \ 
r_\lambda r_\nu \Gamma_{\lambda \mu \nu}(-r,0,r) 
\nonumber \\ 
&=&
\lim_{q \to 0} \left[ F(p^2) \frac{q^2}{G(q^2)} \ 
\left(r_\mu - \frac{(q \cdot r)}{q^2} q_\mu \right) X(q,p;r)
\right] \ = \ 0 \ .
\eeq
implying that
\beq
\lim_{q \to 0} \frac{q^2}{G(q^2)} \ = \ 0 \ .
\eeq
Of course, as far as it involves a vertex with longitudinal gluons which have not been very extensively studied, 
this last conclusion is not as clean as the previous one about the ghost dressing (according to authors of
ref.~\cite{Alkofer:2008jy} a soft kinematical singularity appears for the landau-gauge 3-gluon vertex,
however it does not concern our proof relying on the regularity of the longitudinal-longitudinal-transverse 3-gluon vertex).

In ref.~\cite{ours-ST}, we showed that only a very mild divergence, for example of logarithmic type, could be compatible 

(although very unlikely) 
with current LQCD results for the gluon propagator. 
The IR analysis of the previous section can be straightforwardly extended  to this case 
by generalizing
\beq
G_{\rm IR}(q^2,\mu^2) \ = \ B(\mu^2) \left(q^2\right)^{\alpha_G} \log^{\nu}\left({\frac 1 {q^2}}\right) \ ,
\eeq
the effect of which is to modify \eq{solghost} with
 
\beq
\label{solghostlog}
F_{\rm IR}(q^2,\mu^2) \ = \ 
\left\{
\begin{array}{ll}
A(\mu^2) \left( 1 -  \phi(0,\alpha_G) \displaystyle \frac{ \widetilde{g}^2(\mu^2)}{16 \pi^2} A^2(\mu^2) B(\mu^2)
(q^2)^{\alpha_G} \log^{\nu}{(q^{-2})} \right) & \alpha_G < 1 \\
A(\mu^2) \left( 1 - \displaystyle \frac{\widetilde{g}^2(\mu^2)}{(\nu+1) 64 \pi^2} \ A^2(\mu^2) B(\mu^2) \ 
q^2  \log^{(\nu + 1)}{(q^{-2})}\right) & \alpha_G = 1 \\
\rule[0cm]{0cm}{0.5cm} A(\mu^2) + A_2(\mu^2) q^2 \log^{\nu}{(q^{-2})} & \alpha_G > 1
\end{array}
\right.
\eeq
where only the power of the logarithm is then modified.

Sticking now to the case where $\alpha_F$ is zero (for the reasons explained above) 
and $\alpha_G$ is 1 (as suggested by the lattice results) we are left with 
%
%
\beq\label{FIR-lat}
F_{\rm IR}(q^2,\mu^2) \ = \ F_{\rm IR}(0,\mu^2) \left( 1 - \frac{\widetilde{g}^2(\mu^2)}{(\nu + 1)64 \pi^2} \ 
F_{\rm IR}(0,\mu^2)^2 B(\mu^2) \ q^2 \log^{(\nu + 1)} {\left(\frac{M^2}{q^2}\right)}  \right)
\ ,
\eeq
according to whether there are logarithmic corrections to the gluon propagator ($\nu \neq 0$) or 
not ($\nu=0$). Here, $M$ is some scale which is out of the scope of the IR analysis we performed in the 
previous section and, if $\nu=0$,  $B(\mu^2)=G_{\rm IR}^{(2)}(0,\mu^2)$ is the gluon propagator 
at zero momentum.



\section{Low-momentum Green functions lattice results}
\label{sec:lattice}

As was recalled in the previous section, the mechanisms usually invoked to explain the confinement imply specific behaviours in the infrared for the Green's functions of the theory  :
\begin{itemize}
 \item A sufficient condition for the Kugo-Ojima criterion\footnote{ The Kugo-Ojima scheme also implies that there be no massless pole in the transverse gluon propagator (cf~\cite{Kugo:1995km}), a condition which is weaker than the vanishing advocated in  the Gribov-Zwanziger scenario. } to be satisfied would  be the divergence of the ghost {\sl  dressing function}, $F^{-1}(q^2)\,\sim 0$ as $q^2\to 0 $ (see ref.~\cite{Kugo:1979oj,Kugo:1995km})
 \item The Gribov-Zwanziger scenario implies the vanishing of the gluon {\sl propagator}. i.e.  $G(q^2)/q^2\, \sim 0$ as $q^2\to 0 $ (ref~\cite{Zwanziger:1991gz})
\end{itemize}

In addition, the dressing functions should obey Dyson-Schwinger equations, from which it was inferred  that the ghost and gluon infrared exponents (cf. equation~(\ref{param}) above) should satisfy the relation  $\alpha_G  + 2 \alpha_F\, = 0 $ (referred to in the following as Rel$\alpha$) 
Accordingly Zwanziger ~\cite{Zwanziger:2001kw} and Lerche and von Smekal~\cite{Lerche:2002ep} predicted  a value 
$ \alpha_G  = -2 \alpha_F\,\simeq 1.19$

Since those  results  rely on theoretical conjectures and since using the  Dyson-Schwinger equations demands an unescapable and not fully controlled  truncation of their infinite tower, the technique of lattice simulations has become an alternative and totally independent means to get reliable model-independent information on the Green's functions. The first attempts to measure propagators took place at the end of the eighties for the gluon \cite{ Mandula:1987rh,Gupta:1987zc,Bernard:1992hy} and in the second half of the nineties for the ghost~\cite{Suman:1995zg,Cucchieri:1997dx,Nakajima:1999dq}. Since then, thanks to the huge increase in the performance level of the computers, it has become possible to reach larger volumes while keeping a small enough lattice spacing. This circumstance is essential to get an ever more detailed insight in the infrared behaviour of the propagators and vertices. 

The simulations have been performed using a variety of setups :
\begin{itemize}
\item choice of the gauge group ($SU(2)$ or $SU(3)$)  Theoretical arguments lead to the conclusion that the qualitative features of the IR Green's functions should be independent of this choice~\cite{ Zwanziger:2001kw, Lerche:2002ep}.
\item dimensionality $d$ of the space, ranging from 2 to 4. The relation~(Rel$\alpha$) is actually the restriction to the $d=4$ case of a more general $d$-dependent one.
\item choice of the gauge action : either standard Wilson or improved versions, quenched  or unquenched.
\item lattice geometry (isotropic or not), spacing, lattice size.
\end{itemize}

 We present in this section an overview of the results of these simulations, with emphasis on  the case of an $SU(3)$ gauge group and of the pure Yang-Mills theory (quenched QCD). The case of  the $SU(2)$ gauge group has been considered very thoroughly by Cucchieri and Mendes (see  for instance \cite{Cucchieri:2010xr} and the references therein).  A comparison of the dressing functions for $SU(2)$  and $SU(3)$ has been performed in ref.~\cite{Sternbeck:2007ug} and shows that the dressing functions are remarkably close for both the gluon  and the ghost, in accordance with the theoretical expectations.

\subsection{Ghost and gluon propagator results in Landau gauge}

We summarize in table \ref{chronokappa}  the  results concerning the IR properties of the propagators  which have been reported in the literature. They are usually, but not always,  given in terms of values of the exponents $\alpha_F$ and $\alpha_G$. The reader should keep in mind that in many of the latter  cases, the authors have attempted to describe both the gluon and the ghost propagator with only one parameter, automatically  taking into account the relation~(Rel$\alpha$), but simultaneously  introducing some kind of bias in the fit, all the more as, as we shall see, this relation  is not satisfied by the data. Actually such  a one-parameter fit revealed impossible in several cases \cite{Sternbeck:2005qj}.


\begin{table}[!t]\centering
\begin{tabular}{|c|c|c|c|c|c|}
\hline
Ref&year& $\beta$&Lat. Size&gluon&ghost\\\hline
\cite{Bonnet:2000kw,Bonnet:2001uh}&2001&5.7&$16^3 \times 32 $&&\\
-id-&-id&4.38$^{(+)} $&$16^3 \times 32 $&&\\
-id-&-id&3.92$^{(+)} $&$ 10^3 \times 20$&finite&\\
-id-&-id&3.75$^{(+)} $&$8^3  \times16$&non zero&n.m.\\
-id-&-id&3.92$^{(+)} $&$ 16^3 \times 32$&$D(0)$&\\
-id-&-id&4.1$^{(+)} $&$12^3  \times24$&&\\
-id-&-id&6.0&$ 32^3 \times 64$&&\\\hline
\cite{Sternbeck:2005tk}&2005&5.8&$16^4-32^4$&$D(q^2)$&\\
-id-&-id&6.0&$16^4, 24^4,$& decreases&$ \alpha_F \simeq .25 $\\
-id-&-id&6.2&$16^4, 24^4$&with $q^2$&\\\hline
\cite{Boucaud:2005ce, Boucaud:2005gg}&2005&5.75&$32^4$&$ \alpha_G = .864(16)$&$ \alpha_F = -.153(22)$\\\hline
\cite{Silva:2005hb,Silva:2006av}&2006&6.0&$(8^3 -,18^3)\times 256$&$ \alpha_G = .996 -- 1.05$&n.m.\\\hline
\cite{Sternbeck:2006cg}&2006&5.8&$24^4, 32^4$&finite&$ \alpha_F \simeq .2 $\\
-id-&-id&6.0&$16^4- 48^4$&non-zero.&or\\
-id-&-id&6.2&$16^4, 24^4$&D(0)&log-like\\\hline
\cite{Bogolubsky:2007ud}&2007&5.7&$56^4-80^4$& -id-&not power-like\\\hline
\cite{Oliveira:2007dy}&2007&6.0&$(8^3-18^3)\times 256, $&&\\
-id-&-id&-id&$16^3\times 128, 48^4, 64^4 $&$ \alpha_G =1.07 (28) $&not power-like\\
-id-&-id&6.2&$64^4$& &\\\hline
\cite{Oliveira:2008uf}&2008&6.0&$16^4-64^4,$&not&\\ 
-id-&-id&-id&$(8^3-18^3)\times 256$&conclusive&n.m.\\\hline
\cite{Oliveira:2009eh,Oliveira:2009nn}&2009&6.0&$16^4-80^4$&not&n.m.\\
-id-&-id&5.7&$8^4- 44^4$&conclusive&\\\hline
\cite{Iritani:2009mp,Suganuma:2009zs,Suganuma:2010mm}&2009&6.0&$32^4  $&&\\
-id-&2009&5.8&$20^3 \times 32$&non zero&n.m.\\
-id&2009&6.0&$16^3 \times 32 $&D(0)&\\\hline
\cite{Bogolubsky:2009dc}&2009&5.7&$ 64^4-96^4$& $\alpha_G = 1.$& $\alpha_F = 0.$\\\hline
\end{tabular}
\caption[]{Summary of the infrared  behaviour  of  the gluon and ghost propagators from lattice simulations,  restricted to the  $SU(3)$ case; the infrared exponents, $\alpha_G$ and $\alpha_F$, are given when available; ``n.m.'' stands for ``not measured''. The ``$^{(+)}"$ mark refers to the use of the L\"uscher-Weisz improved action (cf. infra).}
\label{chronokappa}
\end{table}

While the  first measurements on the lattice  resulted in a value of $ \alpha_G$ compatible with 2, i.e.  with a gluon dressing function behaving as $k^4$ at small $k$'s  \cite{ Bernard:1992hy}, the table shows  that there is nowadays a general agreement (see also ref.~\cite{Cucchieri:2010xr}) in favour of a solution incorporating :
\begin{enumerate}
 \item a gluon dressing function going to 0 like $q^2$, leading to a propagator remaining finite and non-zero (``massive gluon'')
\item  a ghost dressing function going to a non-zero constant, i.e a ghost propagator behaving as $1/q^2$ (``free ghost'')
\end{enumerate}
Recent  numerical data (regarding the quenched case ) on very large lattices are to be found in \cite{Bogolubsky:2009dc}. They are visualised in figure~\ref{propagators} below, in which  the left panel is borrowed from \cite{Bogolubsky:2009dc}. Similar simulations are now underway for the unquenched cas. The preliminary results appear to be qualitatively compatible with the general picture we have described. We defer the discussion of the different artefacts which can affect  the results  to a special subsection but, meanwhile we can only make ours the statement 
of ref.~\cite{Cucchieri:2010xr} : `` The current paradigm is that of a massive gluon and a free ghost". 

\vspace{5mm}

\begin{figure}[h!]
\begin{center}
 \begin{tabular}{cc}
 \includegraphics[width=7.5cm]{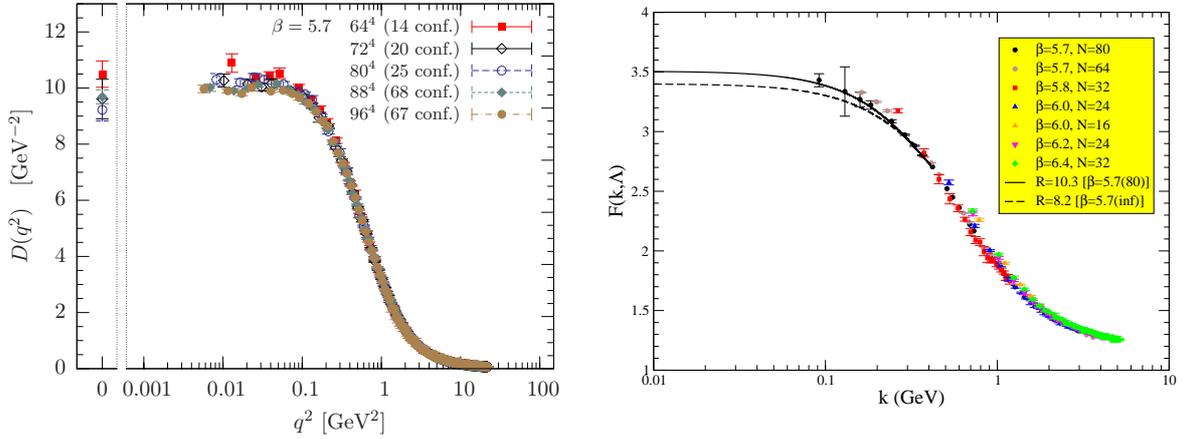}
 & \includegraphics[width=7.5cm]{all-gh.eps}\\
 \end{tabular}

\end{center}
\caption{Gluon propagator (left, borrowed from  \cite{Bogolubsky:2009dc}) and ghost dressing function (right, from  \cite{Boucaud:2009sd})}
 \label{propagators}
\end{figure}


\subsubsection{Lattices vs Dyson-Schwinger equations}

The fact that the most recent  results of lattice simulations, $\alpha_G = 1$ and  $\quad \alpha_G = 0 $, were actually compatible with the Dyson-Schwinger equations was first demonstrated in reference~\cite{Boucaud:2008ji} by a numerical analysis 
of the GPDSE where the gluon propagator lattice data were plugged into the kernel and the transverse form factor was supposed 
to be constant for all momenta. We will come back to this result in sec.~\ref{sec:num}, where the agreement of lattice data 
and the numerical DSE results for the decoupling case will be manifest from fig.~\ref{fantomefig}. 

\begin{figure}[hbt]
\begin{center}
\includegraphics[width=10cm]{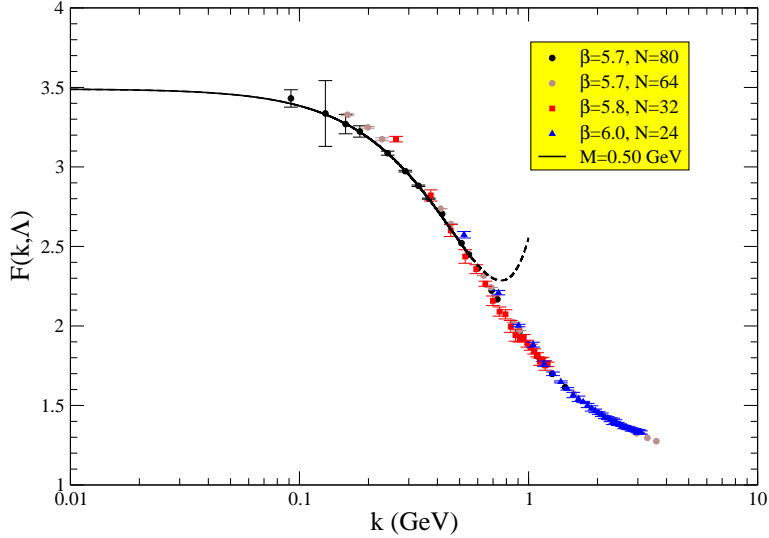} 
\end{center}
\caption{\small The ghost dressing function from lattice data~\cite{Boucaud:2009sd,Bogolubsky:2007ud} is pretty well described by 
the low-momentum formula of \eq{solFIRJo} with $M=0.50(2)$ GeV, as shown in ref.~\cite{Boucaud:2010gr} from  where we borrow the 
figure. The curve for the low-momentum formula is drawn as solid black line inside the fitting region and 
as dashed one outside.}
\label{fig:low-ghost}
\end{figure}

Furthermore, as we discussed in the previous section, in the subsection~\ref{subsec:next}, the authors 
of ref.~\cite{Boucaud:2010gr} provided us with a low-momentum analytic expression for the ghost 
dressing function, \eq{solFIRJo}, derived from the asymptotic analysis of the GPDSE. This result 
was also succesfully confronted to the lattice data for the ghost propagator dressing, in particular with those 
obtained from very large lattice~\footnote{These lattice data were also obtained with 
 the SA gauge-fixing algorithm so as to  deal as well as possible with the Gribov ambiguity but, for instance in 
 determining the gluon mass by fitting \eq{solFIRJo} to the ghost dressing data, some systematic uncertainty should be admitted 
 to come from the gauge-fixing procedure.} simulations in ref.~\cite{Bogolubsky:2007ud}; this comparison was performed  in~\cite{Boucaud:2010gr}. 
from which we borrow  Fig.~\ref{fig:low-ghost} where the ghost propagator lattice data 
are shown to behave pretty well as \eq{solFIRJo} asks for with a 
gluon mass, $M=0.50(2)$ GeV, in the right ballpark (roughly from 400 MeV to 700 MeV) 
defined by phenomenological tests~\cite{Halzen:1992vd} or direct lattice measurements from 
the gluon propagator~\cite{Bonnet:2000kw,Bonnet:2001uh,Iritani:2009mp,Oliveira:2010xc}.

\subsection{The ghost-gluon vertex}\label{ghglvertex}

Among the numerous  possibilities to define the QCD renormalised coupling constant $g_R$ a particularly attractive one,  for   use in lattice simulations,  is  based on the ghost-ghost-gluon 3-point function together with a specific $MOM$-type renormalisation scheme in which the incoming  ghost has zero momentum.  We have intoduced in eq.~(\ref{DefH12}) the 2 form-factors $H_1$ and $H_2$ of the vertex; in terms of those factors the generic definition of $g_R$ would   be 

\beq\label{alpha_gene}
g^2_R(\mu^2)=  \ \lim_{\Lambda \to \infty} g_0^2(\Lambda^2) Z_3(\mu^2,\Lambda^2)\widetilde{Z}_3^{2}(\mu^2,\Lambda^2) (H_1(q,k;\Lambda) + H_2(q,k;\Lambda))\vert_{q^2=\mu^2}
\eeq
 
In the specific kinematical situation we have just mentioned,  Taylor's non renormalisation theorem states that  $H_1(q,0;\Lambda) + H_2(q,0;\Lambda) = \widetilde{Z}_1(\mu^2)=1 $ from which follows that 

\beq\label{alpha} 
\alpha_T(\mu^2) \equiv \frac{g^2_T(\mu^2)}{4 \pi}=  \ \lim_{\Lambda \to \infty} \frac{g_0^2(\Lambda^2)}{4 \pi} G(\mu^2,\Lambda^2)\, F^{2}(\mu^2,\Lambda^2).
\eeq

The remarkable feature of  \eq{alpha} is that  it involves only $G$ and $F$ so that no measure of the ghost-gluon vertex is needed for the determination of the coupling constant.

\begin{figure}[h!]
\begin{center}
 \begin{tabular}{c}
\includegraphics[width=10cm]{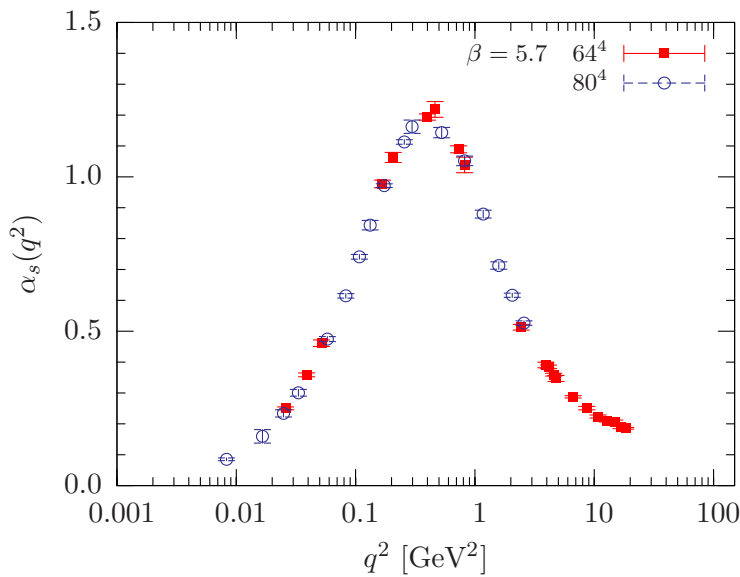}\\
 \end{tabular}
\end{center}
\caption{Recent and very accurate data for $\alpha_T(q^2)$ borrowed from ref.~\cite{Bogolubsky:2009dc}  (figure 5), which shows a 
low-momentum behaviour clearly compatible only with the decoupling solution.}
\label{fig:decoupling}
\end{figure}

 It is thus particularly suitable for use on the lattice, since measuring it does not demand any delicate 3-point function computation ; therefore it has been extensively advocated and studied  (see for instance  reference~\cite{vonSmekal:1997is}).  Were the relation~(Rel$\alpha$) fulfilled, this quantity should go to a finite non-zero limit in the infrared. 
  What is observed on the lattice is quite different from this expectation : the data invariably show a fast decrease to 0, as displayed in figure~\ref{fig:decoupling} (see also Fig.~\ref{fig:F2GSD} in the next section \ref{sec:num}).
 The mechanism which leads to the apparition of those 2 types of solutions has been analytically discussed at length in the second section. Here we just want to stress the fact that the lattice simulations undoubtedly favor the ``decoupling'' case. Of course this argument does not say anything on  $\alpha_F$ and $\alpha_G$ separately but it shows, at least, that one should use  2 {\sl independent} exponents to describe the infrared behaviour of the propagators.

To our knowledge, the direct measurement of the ghost-gluon vertex that has been performed with the $SU(3)$ gauge group appears to be very noisy and helps to conclude nothing beyond the fact that the transverse form factor, $H_1$, in \eq{alpha_gene} is pretty close to 1, as refs.~\cite{Sternbeck:2005qj,Sternbeck:2005re} clearly showed for the case of a ghost-gluon vertex with vanishing incoming gluon momentum. There exist also simulations in the $SU(2)$ case~\cite{Cucchieri:2004sq,Cucchieri:2008qm}, which constitute a very clear direct check of  Taylor's theorem. 
On the other hand, it is not clear whether  the data presented in this reference for $ \alpha_s$ favour the scaling or the decoupling solution.

Furthermore, as will be discussed at the end of app.~\ref{app:Taylor}, \eq{alpha} missed some non-perturbative corrections 
which, although playing no role for the running in the UV domain, need to be taken into account when integrating numerically the GPDSE.

\subsection{The 3-gluon vertex}

It is of course also possible to define the strong coupling constant directly from the three gluon  vertex. 
 Let us recall shortly this  definition  of
 $\alpha_s(p^2)$~\cite{Alles:1996ka, Boucaud:1998bq}. 
 We consider the three-gluon Green function 
 ${G^{(3)}}_{\mu_1\mu_2\mu_3}^{a_1 a_2 a_3}(p_1,p_2,p_3)$
 at the symmetric point,
 $p_1^2=p_2^2=p_3^2\equiv \mu^2$.

The tree-diagram three-gluon vertex is given by $g_s\,T^{tree}$ 
with $T^{tree}$ defined by
\beq
 T^{tree}_{\mu_1\mu_2\mu_3}=\big[\delta_{\mu_1'\mu_2'}
 (p_{1}-p_{2})_{\mu_3'}
 + \hbox{cycl. perm.}\big]
 \prod_{i=1,3} \left(\delta_{\mu_i'\mu_i}-\frac {p_{i\,\mu_i'}p_{i\,\mu_i}}
 {p_i^2}\right) \label{tree}
 \eeq 
 The three-gluon Green function may be expanded on a basis
 of tensors. We are interested in the scalar function $G^{(3)}(\mu^2,\mu^2,\mu^2)$
 which multiplies $T^{tree}$. It is obtained by the following contraction
 \[
 G^{(3)}(\mu^2,\mu^2,\mu^2)=\frac {-i} {18 \mu^2}\,\frac{f^{a_1 a_2 a_3}}{24}\,
  {G^{(3)}}_{\mu_1\mu_2\mu_3}^{a_1 a_2 a_3}(p_1,p_2,p_3)\]
 \beq \left[ T^{tree}_{\mu_1\mu_2\mu_3} + \frac{(p_1-p_2)_{\mu_3}
(p_2-p_3)_{\mu_1}(p_3-p_1)_{\mu_2}}{2 \mu^2}\right]\label{projsym}
 \eeq 
 
 The Euclidean two point Green function in momentum space writes in the
 Landau Gauge: 
\beq
        {G^{(2)}}_{\mu_1\mu_2}^{a_1 a_2}(p,-p)=G^{(2)}(p^2) 
        \delta_{a_1 a_2} \left(\delta_{\mu_1\mu_2}-
        \frac{p_{\mu_1}p_{\mu_2}}{p^2}\right)\label{prop}
\eeq
where $a_1, a_2$ are the color indices ranging from 1 to 8. 

Then the renormalised coupling constant is 
 given by \cite{Alles:1996ka} 
 \beq
 g_R(\mu^2)= \frac{G^{(3)}(p_1^2,p_2^2,p_3^2) Z_3^{3/2}(\mu^2)}
 {G^{(2)}(p_1^2)G^{(2)}(p_2^2)G^{(2)}(p_3^2)}\label{gr}
 \eeq 
where 
\beq
        Z_3(\mu^2)= \mu^2 G^{(2)}(\mu^2).\label{Z3}
\eeq

\begin{figure}[ht]
\begin{center}
\begin{tabular}{c}
\includegraphics[width=10cm]{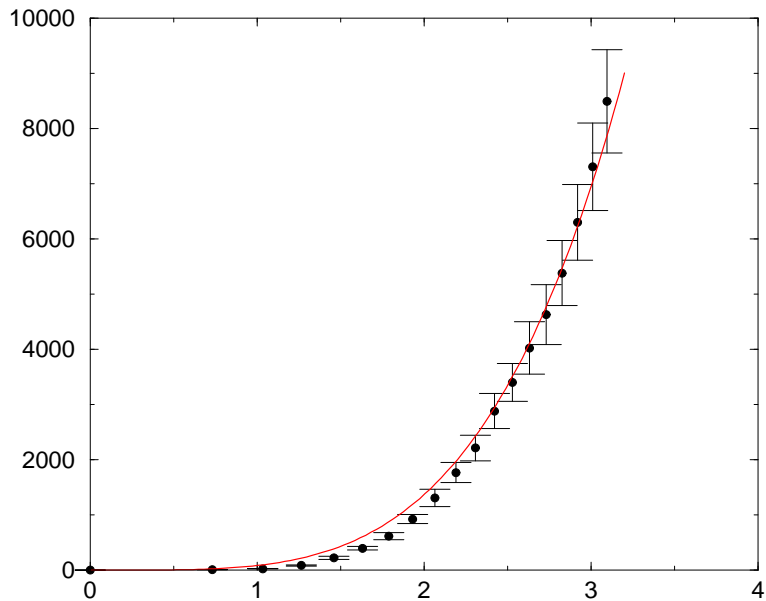} 
\end{tabular}
\caption{\small Coupling constant in a cooled lattice ($L=24$,$\beta=6.0$) 
after 200 cooling sweeps. The solid line corresponds to the fit discussed in 
the text. The horizontal axis is given in GeV, assuming for simplicity
the lattice spacing of the thermalised configurations, 
$a^{-1}=1.97$ GeV.
  }
\label{alpha_sym_cool}
\end{center}
\end{figure}

This method has been used in the infrared regime in ref.~\cite{Boucaud:2002fx}. It leads to a coupling constant which can be nicely fitted by a $p^4$ law compatible with an interpretation in terms of an instanton gas,  as is shown in figure \ref{alpha_sym_cool}.

\subsection{Reflexion positivity violation}

Although the link between the confinement property and the infrared behaviour of the propagators is not fully understood, it is generally admitted that the spectral functions of the latter should not  be positive definite, since, in the coloured sector, no physical positive norm state can contribute.  That this is actually the case has been verified by several groups~\cite{Sternbeck:2006cg, Bowman:2007du,Suganuma:2009zs,Suganuma:2010mm,Iritani:2009mp}. To perform this check one considers the quantity :

$$ C(t)= \sum_{\mu,\vec{x}} \langle A_\mu(\vec{x},t) A_\mu(\vec{x},0)\rangle  $$

The 0-momentum gluon propagator can be obtained from $C$ by integrating it over $t$. Therefore $C(t)$ must change sign in order that $D(0)$ vanish, but this necessary condition is of course not sufficient.
Note that $C(t)$ is similar to what one considers in hadronic physics simulations to ``measure'' the masses (typically the pion mass). It had already been noticed by Bernard  et al. (\cite{Bernard:1992hy}) that the effective mass $ m(t)= Log(C(t+1)/C(t))$ {\sl increases} with $t$, in contradiction with what happens in the ``physical'' situation where only the lightest state survives as time increases. Sternbeck an his collaborators  \cite{Sternbeck:2006cg, Bowman:2007du} have shown by direct inspection that $C(t)$ becomes negative for large enough $t$.\footnote{They have furthermore checked that this remains true in the unquenched case.}  As for the japanese group~(\cite{Iritani:2009mp,Suganuma:2009zs,Suganuma:2010mm}), performing a fit to the lattice data and computing the spectral density of the fitting function they show that this density is almost everywhere negative.  Cucchieri and Mendes~\cite{Cucchieri:2009zt} consider the gluon propagator in $x$-space as a function of the 4-dimensional distance and also observe that it becomes negative as the distance increases. Thus the four methods reach the same conclusion, namely that the  gluon propagator as measured  in lattice simulations does violate the reflexion positivity.


\subsection{The artefacts}
\label{artefacts}
\subsubsection{Finite spacing effects and rotational invariance violation}\label{h4artefacts}

On the lattice, in dimension 4, the usual $O(4)$ symmetry which prevails in continuum euclidean field theory is broken down to the hypercubic $H(4)$ invariance. As a consequence the number of independent scalar invariants is increased to 4 : $ p^{[n]}=\sum_\mu p_\mu^n, n=2,4,6,8$. Evidently, dimensional arguments show that the higher invariants should actually appear as $ a^2 p^{[4]}, a^4 p^{[6]} \cdots$ so that they are also higher order corrections in terms of lattice spacing.
 This manifests itself through the ``fanning'' or ``fishbone'' effect~: when displaying a scalar function versus $p$ one observes a series of characteristic fringes due to the dependence  on the higher invariants. It makes the signal noisy at large momenta and one has to cure this effect in order to identify the ``physical'' value. The simplest way  to realize this goal consists in keeping only the momenta which minimize  $ p^{[4]}$ as compared with $ p^{[2]}$; in practice  this amounts to keeping only those momenta the components of which differ at most by a few units from the diagonal case (cf~ \cite{Bonnet:2001uh}). A more efficient technics has been devised later on~\cite{Becirevic:1999uc,Becirevic:1999hj,deSoto:2007ht}. The idea is to perform an expansion of the scalar functions into a series ot the invariants :

\begin{equation}
      f(p^{[2]},p^{[4]},p^{[6]},p^{[8]})\,=\, f(p^{[2]},0,0,0)\,+\,p^{[4]} \, \frac{\partial{f}}{\partial{p^{[4]}}} (p^{[2]},0,0,0)\,+\cdots
 \end{equation} 

The first term of the series is the desired continuum scalar function, up to rotationally invariant finite lattice spacing effects.  The latter are expected not to be important  in the deep infrared domain since in this region $pa \ll 1$.  At moderate momenta the situation regarding those discretisation effects depends on the quantity under scrutiny ; for a given physical momentum a comparison between data at $\beta = 5.8$ and $\beta = 6.0$ shows a perceptible but moderate increase of  the gluon dressing function  $G$ while the ghost dressing function seems unsensitive to the effect~\cite{Sternbeck:2005tk}.

\subsubsection{Finite volume effects}
The size of the lattice is, for several reasons  a very important parameter. 

First of all  it determines  how deep in the infrared it is possible to go. The 0-momentum value of the gluon propagator and of its dressing function can be obtained directly on the lattice. Of course the zero-momentum value cannot be used directly to fit $\alpha_G$, since including it in the data would force it to be equal to 1. But it is useful in making it possible to compare the data to possible analytic forms, or to look for possible discontinuities, provided sufficiently small momenta are available. The situation is different for the ghost: in this case the zero momentum is not accessible. Then the determination of the ghost infrared exponent relies crucially on the possibility of getting as close as possible to zero. 

Second, the DSE equations on a torus in the continuum  have been studied in reference~\cite{Fischer:2007pf}. The authors compare numerical solutions for various volumes to the infinite volume one and show that  a minimum volume of  $(10-15 fm)^4 $ is necessary in order to observe the onset of the infrared behaviour of the Green's function and that  only much larger ones could allow a reliable estimate of the infrared exponents. 

A third reason why volume effects might be important has to do with gauge fixing and  the problem of Gribov copies. It is believed  that, as the  volume increases,   the integration measure over field configurations should concentrate  on the boundary of the Gribov region. This entails that the ghost propagator must increase with the volume, because the smallest eigenvalue of the Faddeev-Popov operator is located on the boundary and goes to 0 in the infinite volume limit.

A number of strategies have been elaborated in order to determine whether or not the measured lattice quantities approach the physical values. The most direct one consists evidently in  measuring $D(0),\, \alpha_G\,  \mbox{and}\, \alpha_F$  on a set  of lattices of increasing volumes, to study their volume dependence and, if possible, to determine their infinite volume limit, either first determining  for each finite volume the infrared exponent, then extrapolate to get $\alpha_\infty$ or  extrapolating the raw data for the propagators and then determine  directly $\alpha_\infty$. Others use a set of V-dependent bounds to constrain the IR behaviour of the propagators~\cite{Cucchieri:2007rg}. Finally, in ref.~\cite{Oliveira:2007dy}, the authors have developed a method which minimizes the volume effects while checking the power-like dependence.

 The situation is not completely settled yet, but : 
\begin{enumerate}
 \item  The sizes mentioned above as minimal requirements to observe the switching to the infinite volume regime have now  been passed. Volumes of  $(16 fm)^4$ in~\cite{Bogolubsky:2009dc} and up to $(27 fm)^4$  for $SU(2)$ in~\cite{Cucchieri:2010xr}) have bee reached without  any sign of a change  in the curves showing up.   
\item While some authors~\cite{Bogolubsky:2009dc, Bonnet:2001uh} present  results according to which the small momentum gluon propagator is only slightly dependent on the volume and conclude to   a finite infinite-volume limit, according to others (see for instance~\cite{Oliveira:2009nn}) the present data are compatible with both a finite and a vanishing popagator.
\item A similar situation prevails for the ghost propagator : Bogolubsky et al. \cite{Bogolubsky:2009dc} conclude to   finite $F(0)$,  but others~\cite{Oliveira:2009nn} consider that $F$ cannot be described by a simple power law. 
\end{enumerate}

Altogether  the curves for $G$ show a remarkable consistency between the small momentum plateau and the direct 0-momentum measurement. As volumes are increased, less and less room is left for a possible bending down.
Then, taking also into account the very  neat results of the $SU(2)$ study of Cucchieri and Mendes on very large lattices we are led to decide in favour of the decoupling solution, although  more work is still needed to be absolutely sure
 that neither $G(0) \neq 0$ nor a finite  $F(0)$  are finite volume artefacts.

\subsubsection{Lattice anisotropy effects}

A means to reach small values of the momentum without having to deal with prohibitively large lattices consists in using  anisotropic lattices with a very large number of nodes in one of the directions (usually the temporal one).  The question whether  this asymmetry might induce   artefacts has been raised in the $SU(2)$ case by Cucchieri and Mendes (\cite{Cucchieri:2006za}) who concluded that, although the IR behaviours of the Green's functions looked qualitatively similar, they were quantitatively dependent on the geometry~;  later on it has been extended to $SU(3)$ and considered in detail by  Oliveira and Silva   (\cite{Oliveira:2007px}), who compare, in the anisotropic case, the  purely  temporal momenta to spatial ones, as well as  the isotropic and anisotropic situations for various types of cuts. Neither of these comparisons show any effect in the infrared region.  However the same authors, but using a different approach (Cucchieri-Mendes bounds) conclude in~\cite{Oliveira:2008uf} that the data from  symmetric and asymmetric lattices are incompatible, acknowledging at the same time that there is no theoretical  explanation for the phenomenon. The situation still demands clarification.

\subsubsection{Influence of the action}
To conclude this section we mention that the possibility that the choice of the action might introduce artefacts has been considered by Bonnet et al.\cite{Bonnet:2001uh}.  They compare the results obtained for the  gluon propagator from the standard Wilson action with the outcome of  the L\"uscher-Weisz $ {\cal O}(a^2)$-improved action in nearly similar conditions of lattice spacing and volume. The raw data in the second case exhibit considerably reduced finite spacing artefacts, as was to be expected. Once the data have been treated for those artefacts (see above in subsection~\ref{h4artefacts}) the agreement is excellent.  This study has been extended in ~\cite{Bowman:2006zk} to the unquenched case. Again, the gluon propagators obtained under the 2 hypotheses present the same qualitative infrared features.

\subsection{The Gribov problem}
\label{AmbGribov}
As was discussed in sec.~\ref{subsec:GZ}, 
Gribov~\cite{Gribov:1978} first realised that in a non-Abelian 
gauge theory there remains a gauge ambiguity even if imposing,
in the case of Landau gauge, the constraint 
$\partial_\mu A^a_\mu = 0$.  This is also true on a lattice~\cite{Giusti:2001xf}
 where it is referred to as the 
 ``lattice Gribov ambituity".
On the lattice Landau gauge is fixed by minimising the functional
\bea
F_U[g] = -\sum_{x, \mu} \Re \left [\tr \left(U_\mu(x)\right)\right]
= \int d^4_x\, g_0^2\, a^2 \,A^2(x)  + {\cal O}(a^4).\label{eq:functional}
\eea
This verifies $\partial_\mu A^a_\mu = 0$ since $\partial_\mu A^a_\mu$ is the
derivative of the functional in~\eq{eq:functional} as 
a function of the infinitesimal gauge transformations.
Picking a minimum implies that we choose a gauge such that the Faddeev-Popov
operator is positive (which is Gribov's prescription) since the Faddeev-Popov
operator is the second derivative of the functional in~\eq{eq:functional} as 
a function of the infinitesimal gauge transformations. But even this
restriction, which eliminates local maxima and saddle points, is not sufficient
because there are many local minima. It was suggested to choose the absolute 
minimum on the gauge orbit~\cite{Dell'Antonio:1991xt}, however this is
numerically out of reach. A reachable method was used~\cite{IlgenGrib,Cucchieri:1997dx} 
consisting in taking a number of random copies, selecting the ``best copy''
which provides the minimum of the functional in this sample and compare it
to the ``first copy'' appearing in the random sample. This allows to 
estimate the effect of decreasing the functional and the amount of scattering of
the considered quantity. It has been applied to the gluon and ghost propagators.
The general conclusion is that there is no dependence on the Gribov copy except
for the smallest momenta (infrared). In that region the gluon propagator depends
moderately on the copy, while the ghost does depend significantly~: the ghost
propagator is smaller for the best copy that for the first copy.

Moreover, a further SU(2) investigation~\cite{Bogolubsky:2009dc} 
based on the application of the so-called ``simulated annealing algorithm'' (SA) 
gauge-fixing algorithm leads to the same conclusion: lattice results for ghost and 
gluon propagator behave as expected for a decoupling solution. 
This SA is a ``stochastic optimization method'' allowing quasi-equilibrium tunneling through functional 
barriers, with a statistical weight which is $\propto \exp{-F_U[g]/T}$ where $T$ 
is a ``temperature'' that should be taken to decrease~\cite{Bogolubsky:2007bw}. 
Then, in principle, with the appropriate T decrease and number of cooling steps, the SA 
requires the gauge fixing functional to take the extrema as arbitrarily close 
to the global extremum as wished; in other words, SA permits to select the ``best copy'' by 
fixing the Landau gauge in the ``fundamental modular region''. 
Although it has been recently guessed that finding the  
Gribov copies as close as possible to the global extremum may maximally enhance the infrared 
asymptotics of the ghost dressing function, as it would correspond with the scaling 
solution~\cite{Maas:2009se}, the comparison of SA results with the ones obtained by applying 
the standard gauge-fixing method based on the ``over-relaxation'' algorithm seems not 
to support such a conjecture~\cite{Bogolubsky:2009qb,Bornyakov:2009ug}. 
In particular, the authors of ref.~\cite{Bornyakov:2009ug} found the Gribov copy effect not to have any impact 
for momenta above a given $p_{\rm min}$ depending on the physical lattice size ($p_{\rm min}$ decreases 
when lattice size, $a L$, increases) and concluded that   the SU(2) gluon propagator was compatible 
with a decoupling solution. These authors compared the gluon propagator results at 
any fixed momentum obtained by fixing the Landau gauge by the random choice of a first Gribov copy (fc) with 
the ones obtained by the choice of the best copy (bc) as the extrema of the gauge fixing functional. They concluded 
that the discrepancy clearly tends to disappear as the lattice volume in physical units increases (see the 
figs. 2, 3 and 4 of ~\cite{Bornyakov:2009ug}). Thus, although some gauge-fixing ambiguity for the very low-momentum 
gluon propagator data is reported, it is claimed either to disappear in the large-volume limit or to be avoided in practice 
by applying the SA algorithm to fix the Landau  gauge and the best Gribov copy.

On the other hand, the continuum limit of one Gribov copy is impossible to perform, but 
some statistical quantities related to these copies can be studied.
This has been performed on the $SU(2)$ non-Abelian gauge 
theory~\cite{Lokhov:2005ra}. It was found that the probability to find a second
copy strongly depends on the size of the lattice in physical units.  There is a
fast transition between high probability and small probability at a lattice
length of about $2.75/\sqrt{\sigma}$ where $\sigma$ is the asymptotic string
tension and is estimated in QCD to be $\sqrt{\sigma}\sim 0.5$ GeV. This implies
a critical length above which Gribov copies become frequent of $\sim 1$ fm. It
is not surprising that in the small volume limit, close to the perturbative
regime, Gribov copies tend to disappear. They are typically a non-perturbative
phenomenon. That their existence depends on the physical volume supports the
idea that they are related with low modes. And in fact it was shown 
in~\cite{IlgenGrib} that the major difference between the best copy and the
first copy lies in the low modes of the Faddeev-Popov operator which increase in
number for a decreasing minimum of the gauge functional.

\subsection{Coulomb gauge results}

The propagators have also been measured on the lattice in the Coulomb gauge. The reason why it is specially interesting
 to consider this gauge is that relations between the behaviour of the propagators  and the confining properties are 
particularly transparent. Let us first  describe the continuum formalism, which was established by Christ and Lee~\cite {Christ:1980ku} .  The Yang-Mills  theory can be written in  terms of  the Hamiltonian :
\beq
H & = & \frac{1}{2}\,\int d^3x  \left( G_{\perp i}^a(x)   G_{\perp i}^a(x)  +  \frac{1}{2}\,\int d^3x  G_{ij}^a(x)   G_{ij}^a(x) \right) 
\nonumber \\ 
&+& \int   d^3x  d^3y \rho^a(x) K^{ab}(x,y)  \rho^b(y) 
\eeq
 In this formula $G_{\perp}$ is the transverse part of the chromoelectric field (conjugate to $A_{\perp}$), the kernel $K$ can be expressed as the  convolution of the Faddeev-Popov operator with itself  and  the colour density  $\rho$ can itself be written in terms of $G_{\perp}$ and $A_{\perp}$ (in the quenched case which we are considering).
It was further suggested by Gribov~\cite{Gribov:1978} that one should add to this Hamiltonian an extra term proportional to $m^4 \int d^3x A_{\perp i}  (\nabla^{-2}) A_{\perp i}$ in order to restrict the field  configurations to what is nowadays known as  the Gribov region.
Doing so results in an equal-time transverse propagator $D_\perp(\vec{k},0) \propto\vert\vec{k}\vert /(m^4 +(\vec{k}^2)^2.$.  At the same time, the time component $D_{44}$ of the propagator assumes the form 
\beq
 D_{44} = V(\vec{x}) \delta(t) + \mbox{non instantaneous term}
 \eeq
 with $V$ the vacuum expectation value of the kernel $K$ above.  $V(x)$ is different from the static quark potential, but, as has been shown by Zwanziger~\cite{Zwanziger:2002sh}, it constitutes an upper bound for the latter. Thus a divergence of $D_{44}$ at large $\vec{x}$ is a necessary condition for the static quark potential to  be confining.

 The Coulomb and Landau gauges can actually be considered as special cases of the so-called $\lambda-$gauge specified by the condition  $ \lambda \partial_i A_i + (1-\lambda)  \partial_4 A_4 =0$. 
 
 Some authors~\cite{Cucchieri:1998ta, Maas:2006fk,Iritani:2011zg,Maas:2011ej} have studied  this general gauge and shown how the propagators pass continuously from the Landau to the Coulomb schemes. 
 
 Since, however,  those studies are still preliminary this section will be mainly dedicated to the case $\lambda= 1$ (Coulomb gauge).

The relationship between the Landau and Coulomb gauge propagators has been studied in a more empirical way by Burgio et al \cite{Burgio:2009xp}.  Defining a Coulomb gauge scalar function $D_C(\vert\vec{p}\vert) $ as the $ p_0$-integrated and renormalized scalar coefficient of the spatial tensor and comparing it to the Landau gauge function $D(p^2)=G(p^2)/p^2$  they propose that the relation assumes  the form 
$$ D(p^2) = \frac{1}{\vert\vec{p}_C(p)\vert} D_C(\vert\vec{p}_{C}(p)\vert); $$ the UV asymptotic form of the rescaling function $p \to p_{C}(p)$ is known perturbatively.  The important point, in what concerns the infrared region, is that  $ p_{C}(p)/p$ goes to a constant as $p$ goes to zero. This implies that a linearly vanishing Coulomb propagator, such as proposed by Gribov (see above), corresponds to an infrared finite and non-vanishing Landau gauge gluon propagator.

Let us finally recall that the Coulomb gauge is not complete : even after the Coulomb condition $\partial_i A_i = 0$ has been imposed, it remains possible to perform  a space-independent but time-dependent gauge transformation.  Therefore a further step in gauge fixing is necessary; this residual gauge symmetry is a {\sl continuous} one ; taking care of it does not mean that one does not have any longer to consider the {\sl discrete} Gribov problem.   It has been shown (\cite{Quandt:2008zj}) that this second step in gauge fixing is important in order to reduce scaling violations for the transverse propagator although it does not affect the ghost and  temporal ones.

\subsubsection{Lattice results}
In table~\ref{chronokappaCoulomb} we present the results of the several groups that have actually attempted to check numerically the expectations we have  presented in the beginning of this section in a form similar to what we have done for the Landau gauge.  The small $\vec{p}$ behaviour is parametrised as $(\vec{p}^2)^\alpha$, meaning, in particular, that,  for the transverse propagator,  Gribov's suggestion corresponds to $ \alpha = 0.5$.  Because of the very small number of simulations performed with $SU(3)$ and since the general arguments about the IR behaviour appear to be independent of the gauge group we have included in this case  the data obtained for $SU(2)$. 
 ref.~\cite{Langfeld:2004qs} measures directly the Coulomb potential rather  than the temporal propagator. Their result, an IR behaviour slightly more singular than $p^{-4}$ may also have been seen in~\cite{Quandt:2008zj} but  has not been reproduced in subsequent work on $D44$. Greensite and his collaborators also  measure the potentials (both coulombic and static interquark) in $x$-space. Their results agree with a linear long-distance behaviour and with the dominance of the Coulomb potential over the static one.

\begin{table}[!t]\centering
\begin{tabular}{|c|c|c|c|c|c|c|c|}
\hline
Ref&N&year&$\beta$&Lat. Size&D$_\perp$&D$_{44}$&ghost\\\hline
\cite{Cucchieri:2000gu}&2&2000&2.2&$14^4, 16^4,.,, 30^4$&.49-.51&$ -1.9$&n.m.\\\hline
\cite{Greensite:2003xf}&2&2003&2.2, 2.3, 2.4,2.5&$16^4-24^4$&n.m.&-2.&n.m.\\\hline
\cite{Langfeld:2004qs}&2&2004&2.2-2.8&$26^4, 32^4, 42^4$&0.&-2.05&-.245\\\hline
\cite{Quandt:2007qd}&2&2007&2.15-2.6&$36^4 $&.12&n.m.&n.m.\\\hline
\cite{Quandt:2008zj, Burgio:2008yg, Burgio:2010wv,PhysRevLett.102.032002}&2&2008&2.15-2.6&($24^4, 32^4$)&.5&-2.&-1.22\\\hline
\cite{Nakagawa:2008zza,PhysRevD.77.034015}&3&2008&5.9&$18^4,24^4,32^4$&n.m.&-2&n.m.\\\hline
Osaka~\cite{Nakagawa:2009zf}&3&2009&5.8-6.2&$18^4,24^4,32^4$&n.m.&-1.351&-1.22.\\
Berlin~\cite{Nakagawa:2009zf}&3&2009&5.8-6.2&$18^4,24^4,32^4$&n.m.&-1.13&-1.22.\\\hline
\cite{Greensite:2009cj,Greensite:2009iv,Greensite:2009mi}&2&2009&2.2, 2.3, 2.4&$12^4-22^4$&n.m.&-2.&n.m.\\\hline
\cite{Nakagawa:2009is,Nakagawa:2010eh,Nakagawa:2011ar}&3&2009,&5.8-6.2&$18^4,24^4,32^4$&.15&-1.61&n.m.\\
&3&2011&5.8-6.2&Anisotropic&.08&-1.92&n.m.\\\hline
\end{tabular}
\caption[]{Summary of the infrared  exponents  of  the gluon and ghost propagators from lattice simulations in  Coulomb gauge; the notations are similar to the ones used in  Landau gauge. N corresponds to the choice of gauge group. In refs.\cite{Nakagawa:2008zza,PhysRevD.77.034015} the results are given in $x$ space rather than in $p$. The value ``-2'' which we quote corresponds to the linear potential which they report. See in text for what regards refs.\cite{Langfeld:2004qs,Greensite:2003xf,Greensite:2009cj,Greensite:2009iv,Greensite:2009mi}}\label{chronokappaCoulomb}
\end{table}

A few words about the artefacts are also in order here. 

The scaling violations  in the  propagator, which are very important  as has been noticed for long and can be seen for example in figures 2 and 3 of ref.~\cite{Nakagawa:2011ar},  resist the usual cutoffing techniques devised to reduce the discretization effects. Their origin has been traced back to the definition of the instantaneous propagators, which induces a spurious dependence over $\vert\vec{p}\vert$. A solution to overcome this difficulty consists in the use of  anisotropic lattices in which the temporal spacing is much smaller than the spatial one.  This is compensated for by a larger number of points in the time direction.  The efficiency of his procedure has been checked in ~\cite{Nakagawa:2011ar}.

To summarize, the situation is not completely settled yet. However, in contrast with the situation regarding the  {\sl Landau gauge},  all {\sl Coulomb gauge} simulations seem to be in qualitative agreement with Gribov's and Zwanziger's  statements :

\begin{itemize}
\item  the transverse gluon propagator vanishes  in the infrared.
\item the equal-time temporal propagator $D_{44}$ diverges with $R$.
\end{itemize}

Still, the precise infrared exponents are  not exactly known yet. For $D_{44}$ a linear divergence with $R$
($1/p^4$ in Fourier space) appears to be compatible with all data.  Although there is a debate about the value of the coefficient, it is admitted that it is larger than the known string tension.  This is in agreement with the statement that  $D_{44}$ is an upper bound for  the static quark-antiquark potential. The crucial role of the lowest Faddev-Popov eigenvalues in building this confining potential  is explicited very clearly in ref.~\cite{Nakagawa:2010eh}. For the transverse propagator, there is a problem with data giving an IR exponent smaller than $0.5$. According to the argument of ref.~\cite{Burgio:2009xp} this would correspond to a diverging Landau gauge gluon propagator, which is excluded. On the opposite no Coulomb gauge simulation results in a transverse gluon infrared exponent greater than $.5$, which, according to the same reasoning, means that all of them are in contradiction with any Landau gauge  result predicting a vanishing gluon propagator. All these points  still need to be clarified. 

On the other hand, as shall be discussed in sec.~\ref{subsec:coulomb}, the authors of ref.~\cite{Watson:2010cn} have very recently demonstrated that Gribov's formula for the equal-time spatial gluon propagator and the corresponding 
ghost dressing might be seen to admit both scaling and decoupling behaviour for the resulting GPDSE in Coulomb gauge.
Thus, as shown in ref.~\cite{RodriguezQuintero:2011vw}, the picture for both Coulomb and Landau gauge DSE solutions would be 
pretty the same, although the current lattice data, for the available momentum range, 
appear to be compatible with both classes of solutions.

\subsection{The dimension-two gluon condensate from the lattice}
\label{sec:A2lat}

As seen in sections 3.1 and 3.2 and refs.~\cite{Boucaud:2000nd,Boucaud:2001st,DeSoto:2001qx,Boucaud:2005xn,Boucaud:2008gn}, 
the ghost and gluon propagators as well as the resulting strong coupling constant
\eq{alpha}, do not run as perturbation theory 
requires in the energy range 2.5 - 7. GeV. This is surprising since it is widely
believed that the perturbative regime is good above 2 GeV or at least 3 GeV.

Having to deal with a non-perturbative correction to perturbative QCD we 
resort to the Operator Product Expansion (OPE) approach~\cite{SVZ}. 
In Landau gauge  (which is the only one we shall consider in this section)   there is only one dimension-two operator which has the vacuum
quantum numbers:  $A^\mu_a \,A_\mu^a \equiv A^2$. The fact 
that it is a dimension-two
operator implies that it behaves as $1/p^2$ up to logarithms. This explains why
the non-perturbative effects are stronger when computed in Landau gauge
 (and more generally in any fixed gauge) than on gauge invariant quantities
 since the dominant gauge invariant gluonic operator is 
 $G^{\mu\nu}_a\,G_{\mu\nu}^a$ which has dimension-four and thus is $\propto
 1/(p^2)^2$ up to logs.

\subsubsection{Several comments about OPE using $A^2$}
The use of OPE with $A^2$ in Landau gauge has been criticised. We would like to
present our justification before going further.
\begin{itemize}
\item 
$A^2$ is not a gauge invariant quantity although, in Landau gauge, it is 
invariant for infinitesimal gauge transformations as well as BRST
transformations. As was already pointed by the authors of ref.~\cite{Lavelle:1992yh},
it is legitimate to apply OPE with a gauge dependent quantity
in a gauge theory. 

Indeed, to our knowledge, all the arguments used to prove OPE for a Lagrangian field theory
can be applied to the theory defined by adding the gauge-fixing term to 
the gauge invariant QCD Lagrangian, including then the non-physical ghost fields to 
restore unitarity, i.e. to QCD in a fixed gauge.


\vspace*{0.35cm}
\item 
$\VA$  is ultraviolet divergent like the cut-off squared. As stressed 
in~\cite{SVZ}, when speaking of a condensate we think of the infrared modes 
of $\VA$. How can we discriminate in a theoretically sound manner the infrared
modes from the ultraviolet ones ? A cut off in the loop momenta would be much
too crude. The best is precisely to use the OPE expansion: given a quantity 
$Q(p^2)$ we perform the following expansion
\bea
Q(p^2)=   Q_{\rm pert}(p^2,\mu^2)  + C^Q_{\rm wilson}(p^2,\mu^2) 
\langle A^2(\mu^2)\rangle  + ....
\label{eq:OPE}
\eea 
A well defined renormalisation procedure and renormalisation scale is mandatory 
to be allowed to compare $\VA$ computed from one quantity $Q(p^2)$ and an other
one $Q'(p^2)$. Two different quantities will have different Wilson coefficients:
$Q_{\rm pert}(p^2,\mu^2)\ne Q'_{\rm pert}(p^2,\mu^2)$   and $C^Q_{\rm
wilson}(p^2,\mu^2)\ne C^{Q'}_{\rm wilson}(p^2,\mu^2)$, but the same  $\VA$.
Indeed $\VA$ is a property of the vacuum. {\it It is not a lattice artefacts, it
is defined in the continuum limit, at vanishing lattice spacing. It depends on
the gauge, and on the vacuum properties: the number and masses of the dynamical
 quarks}. 

\vspace*{0.35cm}
\item  
The method just advocated consists in separating in a quantity $Q(p^2)$ the
perturbative contribution from the dominant non-perturbative one. However it is
known that the perturbative series, i.e. the contribution of the identity
operator in the OPE, is only asymptotically convergent. This means that higher
order terms sum up in what is called ``renormalons" which precisely behave like
$1/p^2$ up to logs and are apparently non distinguishable form the effect of 
$A^2$. It is not even clear that the distinction has a well defined theoretical
meaning. This difficulty applies as well to gauge invariant operators and thus
to all the activity around what is called ``QCD sum rules". This has been
discussed  in~\cite{Martinelli:1996pk}. To make it short, the authors
concentrate on the issue: can we use a condensate estimated from one quantity,
say  $Q(p^2)$ for the expansion of another quantity, say $Q'(p^2)$. They show
that in $Q(p^2)-c\,Q'(p^2)$ ($c$ being a relevant prefactor) 
 the renormalon ambiguity cancels. In other
words, the properly defined difference between the two expansions contains a
convergent perturbative series. If we stop at order $n$ in the perturbative
expansion, and assume that the sum of the expansion from $n+1$ to $\infty$ 
is bounded by the term of order $n$, we get the following 
criterium of validity. We can indeed use the condensate estimated from $Q(p^2)$
in the expansion of $Q'(p^2)$ if the non-perturbative contribution
($\propto 1/p^2$) is significantly larger than the last (highest order)
perturbative contribution. This, of course depends on the energy. When the
energy is very large, the last perturbative contribution dominates any $1/p^2$
term and one can be satisfied with the perturbative series.  In intermediate
energies the $1/p^2$-term dominates over the last perturbative one provided one has gone
far enough in the perturbative expansion. At even lower energies all higher
dimension operators contribute and OPE is no more  applicable.   
\end{itemize}

\subsubsection{Computing the Wilson coefficient}
A momentum dependent quantity $Q(p^2)$ with vacuum quantum numbers 
can be inverse power expanded as in~\eq{eq:OPE}.
 $Q$ can be the strong
coupling constant, the quark field renormalisation constant, other
renormalisation constants, etc~\cite{Pene:2011kg}. The coefficients 
$C^Q_{\rm wilson}$ are often called Wilson coefficients.
All Wilson coefficients $C^Q_{\rm wilson}(p^2,\mu^2)$  are of the type 
\bea\label{eq:CW}
C^Q_{\rm wilson}(p^2,\mu^2)= d^Q_{\rm tree}\;
g^2(\mu^2)\;\frac{1+\cal{O}(\alpha)}{p^2} 
\eea
where
\bea
d^Q_{\rm tree} = \left\{ 
\begin{array}{lc}
\displaystyle
\frac 1 {12} & \rule[0cm]{0.5cm}{0cm} {\rm for} \ Zq \\
\displaystyle
\rule[0cm]{0cm}{0.75cm}
\frac 9 {32} &  \rule[0cm]{0.5cm}{0cm}{\rm for} \ \alpha_T 
\end{array}
\right.
\eea

From \eq{eq:OPE} and \eq{eq:CW} we see that there is always the same factor
$\langle g^2(\mu^2) A^2(\mu^2)\rangle_{\rm \overline MS}$,  (throughout this
section we choose the $\rm \overline {MS}$ scheme and the renormalisation
scale $\mu=10$ GeV).  We will therefore give the fitted values of the condensate
as  $\langle g^2(\mu^2) A^2(\mu^2)\rangle_{\rm \overline MS}$.

In practice one can show that the best and most general fitting formula is:
\bea\label{eq:fact}
Q(p^2)=  Q_{\rm pert}(p^2,\mu^2) \left (1 + 
\frac{C^Q_{\rm wilson}(p^2,\mu^2)}{Q_{\rm pert}(p^2,\mu^2)}\;\;
\langle A^2(\mu^2)\rangle_{\rm \overline MS} \right )
\eea
At leading logarithm for the non-perturbative correction ~\cite{Boucaud:2001st}, 
\bea \label{eq:LL}
\frac {C^Q_{\rm wilson}(p^2,\mu^2)}{Q_{\rm pert}(p^2,\mu^2)}\;\; 
\langle A^2(\mu^2)\rangle_{\rm \overline MS} \ =  \ \frac{d^Q{\rm tree}}{p^2}\;\;
\langle g^2(\mu^2) A^2(\mu^2)\rangle_{\rm \overline MS} \left(\frac
{\alpha(p^2)}{\alpha(\mu^2)}
\right)^e
\eeq
where
\bea
e \ = \ \frac{27}{132 - 8 N_f} \ = \ \left\{
\begin{array}{lc}
\displaystyle
\frac{27}{116} & \rule[0cm]{0.5cm}{0cm} {\rm for} \   N_f=2\\
\displaystyle
\rule[0cm]{0cm}{0.75cm}
\frac 9 {44} &  \rule[0cm]{0.5cm}{0cm}{\rm for} \  N_f=0 \ .
\end{array}
\right.
\eea
Notice that $e$ is small. Therefore the corrective factor in \eq{eq:LL} is
almost scale invariant. It is noticeable that the exponent is the same for all
quantities $Q(p^2)$~\footnote{For the coupling constant computed from the three
gluon coupling with one vanishing momentum, the formula~\eq{eq:LL} does not
apply~\cite{DeSoto:2001qx}.}. 
In \eq{eq:LL} the only term which depends on the measured
quantity is  $d^Q{\rm tree}$, which is given in table~\ref{tab:dQtree}
for different quantities. 
\begin{table}[h]
\centering
\begin{tabular}{||c|c|c|c|c||}
\hline
$Q(p^2)$ &
gluon prop & ghost prop &$\alpha_T$ & $\alpha_{3g\,sym}$\\
\hline
&3/32&3/32&9/32 &9/32 \\ \hline
\end{tabular}
\caption{$d^Q{\rm tree}$ as defined in \eq{eq:LL}. $\alpha_{3g\,sym}$
is the coupling constant extracted from the three gluon coupling with 
all momenta equal to $p^2$~\cite{Boucaud:2005ce}.}
\label{tab:dQtree}
\end{table}

Beyond the leading logarithm, the Wilson coefficient have been computed at three
loops by Chetyrkin and Maier~\cite{Chetyrkin:2009kh} for the propagators, and
hence also for $\alpha_T$.

\subsubsection{Numerical results} 

Once a quantity $Q(p^2)$ has been computed on the lattice we need to 
treat the lattice artefatcs of order $a^2$. There are two types of such
artefacts, the ones which originate in the hypercubic geometry of the lattice,
related to the $H_4$ group, and the ones which have the continuum geometry, 
$O(4)$.  

The first type of artefacts are corrected via a non-perturbative 
method~\cite{Becirevic:1999uc,deSoto:2007ht,Boucaud:2003dx} 
which fits from the data themselves for different orbits of the discrete group,
the dependence of the considered quantity on the invariants of the discrete
group which $H_4$ are not invariants of $O(4)$. An extrapolation to the value
zero of the latter invariants produces a result free of the hypercubic
artefacts. Let us call the latter $Q_{\rm smooth}(p^2)$.    

Having got rid of hypercubic artefacts we still have to take into account the
$O(4)$ invariant ones. The dominant one, at order $a^2$ is 
$\propto a^2\,p^2$. We just add such a term to the fitting
equation~\eq{eq:fact}:
 
\bea\label{eq:factsmooth}
Q_{\rm smooth}(p^2)=  Q_{\rm pert}(p^2,\mu^2) \left (1 + 
\frac{C^Q_{\rm wilson}(p^2,\mu^2)}{Q_{\rm pert}(p^2,\mu^2)}\;\;
\langle A^2(\mu^2)\rangle_{\rm \overline MS} \right ) + c_{a2p2}\, a^2\,p^2.
\eea
The function $Q_{\rm pert}(p^2,\mu^2)$ is known from perturbation theory up 
to a multiplicative factor $Q_{\rm pert}(\mu^2,\mu^2)$.  The ratio 
$C^Q_{\rm wilson}/Q_{\rm pert}$ is also known, but $\VA$ has to be fitted.
Altogether, taking into account the coefficient $c_{a2p2}$, we have three quantities to be fitted 
from the function $Q_{\rm smooth}(p^2)$ which provides one  data for each 
$p^2$ in our fitting interval. This is sufficient to perform a fit but  there is a
source of instability  due to the fact that $1/p^2$ decreases with
$p^2$, leading to a possible partial cancellation of the contribution of the
artefact and the condensate. However the lattice spacing dependence of the 
term $\propto a^2\,p^2$ and the condensate $\propto 1/p^2$ up to logs are very
different, which allows to check the quality of the fit.  
 
\begin{figure}
\begin{center}
\begin{tabular}{cc}
\includegraphics[width=7.5cm,height=6cm]{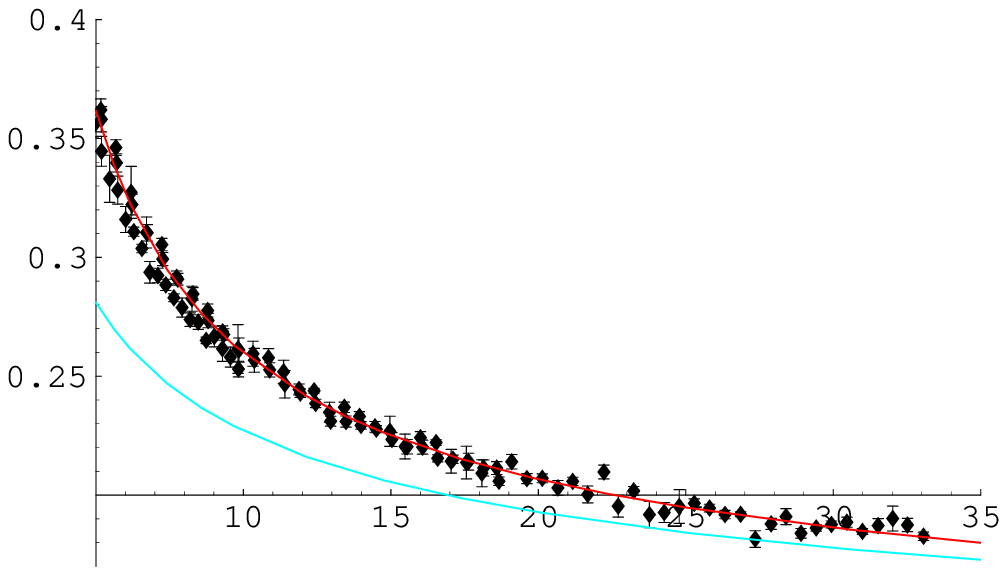} &
\includegraphics[width=7.5cm,height=6cm]{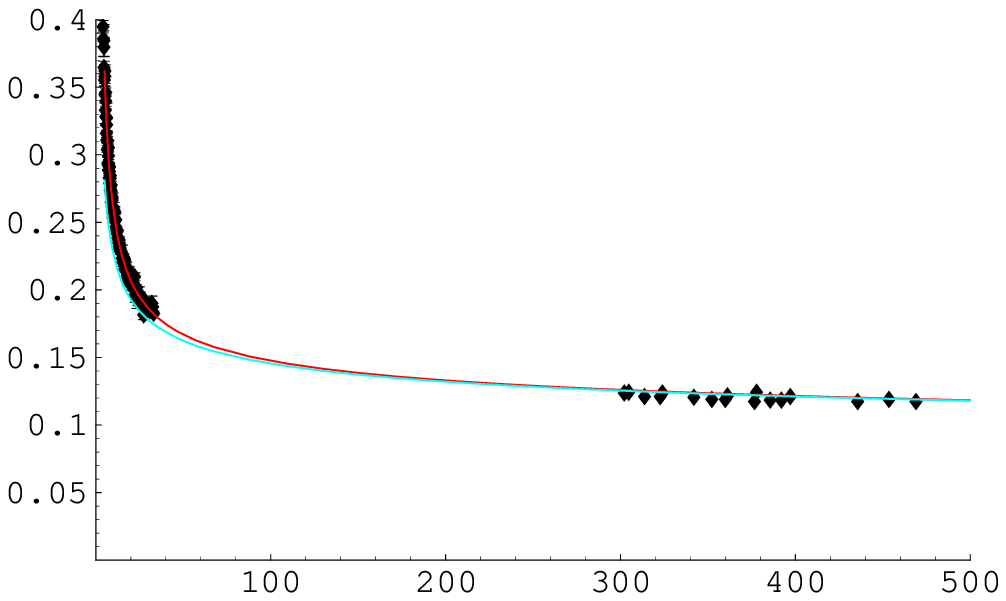} 
\end{tabular}
\end{center}
\caption{\small (Left) Comparison, borrowed from ref.~\cite{Boucaud:2008gn}, of lattice estimates 
for the running coupling in Taylor scheme and the non-perturbative prediction including the 
dimension-two gluon condensate, $\VA$; $\alpha_T$ in vertical axis is plotted in terms of the square of 
the momentum in horizontal axis. (Right) The same at very high momenta to reach the 
perturbative regime, where the gluon condensate effects can properly be neglected,  
and check the consistency of the results, in particular the estimate of $\Lambda_{\overline{\rm MS}}$.}
\label{fig:alphaT-OPE}
\end{figure}

We will now consider several $Q(p^2)$ and provide the estimated 
values of the condensate. This is shown in Tab.~\ref{tab:global_comp}. 
For the particular case of $\alpha_T$, which we paid special interest to 
in this work, we also borrow from ref.~\cite{Boucaud:2008gn} the plots shown 
in Fig.~\ref{fig:alphaT-OPE}, where the very good agreement between 
the prediction from Eqs.~(\ref{eq:fact},\ref{eq:LL}) and the lattice data for 
the Taylor running coupling is manifest.

\begin{table}[h]
\centering
\begin{tabular}{|c|}
\hline
 fitted $g_R^2 \VA$ \\
\begin{tabular}{|c|c|c|c|c|c|}
\hline
 order  $g^2 \VA$ &  gluon propagator & ghost propagator & $\alpha_T$ & 
 3-gluon $\alpha$ asym & 3-gluon $\alpha$ sym\\
\hline
Tree & 2.7(4) & 2.7(2)  & & & \\
\hline
 LL & & & 5.2(1.1) & 10(3) &6.8(1.5)\\
\hline
  $O(\alpha^4)$ & & & 3.7(8) & &\\
\hline 
\end{tabular} 
\\
\hline 
\end{tabular}
\caption{\small Comparison of estimates of $g^2 \VA$ from different quantities.
All are taken at the scale $\mu= 10$\, GeV. Tree means  tree level for the
Wilson coefficient. The data in this line come from~\cite{Boucaud:2005xn}. LL
means leading logarithm for the Wilson coefficient. The data in this line are
taken from ref.~\cite{Boucaud:2008gn}. $O(\alpha^4)$ refers to Chetyrkin and Maier
computation. }
\label{tab:global_comp}
\end{table}

As will be also mentioned in App.~\ref{app:Taylor}, this dimension-two gluon condensate is seen to have an impact 
on the ghost-gluon vertex and has to be taken into 
account when solving the GPDSE and comparing with lattice data for the ghost dressing function~\cite{Boucaud:2011eh}.



\section{DSE numerical solutions}
\label{sec:num}

\subsection{Solving the ghost propagator DSE}
\label{subsec:num1}

Let us analyze first in this section the general picture for the low-momentum Landau-gauge DSE solutions by applying 
the approach proposed for the first time in ref.~\cite{Boucaud:2008ji}:{\it  to combine lattice gluon propagator 
results with a ghost propagator DSE truncated by a well supported hypothesis for the 
ghost-gluon vertex}. The same procedure has been very recently followed for the analysis of the Coulomb-gauge 
DSE solutions in ref.~\cite{Watson:2010cn} leading, as we shall discuss below, to a very analogous picture. 
It should be noted too that the authors of ref.~\cite{Aguilar:2010cn}  also successfully applied the same strategy 
of combining lattice data and DSE, in particular by invoking the lattice data for the truncation of the gap equation and 
studying the chiral symmetry breaking. 

\subsubsection{Landau gauge solutions}
\label{subsec:landau}

We shall present now the results of ref.~\cite{Boucaud:2008ji} for the ghost solutions in Landau  gauge. 
The goal was to see whether the two types of solutions ($\alpha_F=0$ and $2\alpha_F+\alpha_G=0$) suggested by the previous analytical 
discussion in section \ref{sec:analytic} actually exist for the same gluon propagator input. A positive answer came out by solving 
numerically the ghost SD equation for  given gluon propagator and vertex, as we shall briefly describe in the following. 
To this goal, we invoke again the renormalized DSE given by (\ref{SD1}) and cast it into the appropriate subtracted form:

\bea\label{SDRcst2}
\frac{1}{F_R(k^2)}&=& 1-\widetilde {g}^2 \int\frac{d^4q}{(2\pi)^4} 
\left(1-\frac{(k.q)^2}{k^2 q^2} \right)  \nonumber \\
&\times& \left[ \left.\frac{G_R((q-k)^2)}{((q-k)^2)^2}-\frac{G_R((q-k^\prime)^2)}{((q-k^\prime)^2)^2}\right]  F_R(q^2) \right\arrowvert_{{k^\prime}^2=\mu^2} \ 
\eea

\noindent
where we work in the $MOM$ scheme, and set ${k^\prime}^2$ appearing in eq.(\ref{SD1}) as the squared renormalisation scale $\mu^2$ ($\mu$ 
has been chosen at an optimum $1.5$~GeV,  not too high to allow the lattice data to be safe,  and not too small in order that  the differences 
between solutions at small momenta can be clearly displayed). An IR finite gluon propagator ($\alpha_G=1$) extracted from lattice data in pure Yang-Mills theory,  
with Wilson gauge action,  $\beta = 5.8$ and a lattice volume equal to $32^4$, is used for momenta lower than $1.5$~GeV ; this choice is 
justified to have moderate UV artefacts. This is then extended to  larger momenta using a one loop asymptotic expansion (with $\Lambda_{MOM}=1$~GeV 
corresponding to the standard $\Lambda_{\overline MS}=0.240$~GeV of lattice quenched QCD).
As for  the ghost-gluon transverse form factor in \eq{DefH12}, $H_1(q, k)$,  it is taken to be constant with respect to both 
momenta \footnote{This cannot be an exact statement,  as already shown in perturbation by the calculations of 
ref. \cite{Chetyrkin:2000dq,Davydychev:1996pb} : although finite, the vertex invariants do depend on the momenta through the running $\alpha_s$. }.   
As we said above,  this is suggested by the lattice data for $q=k$ (i.e. for zero gluon momentum),  but we extend it to all values of $q$ and $k$. The authors of ref.~\cite{Sternbeck:2005re} find a bare vertex very close to $1$ in this zero momentum gluon configuration for a large range of $\sqrt{q^2}$. Note that we have re-defined the 
coupling as $\widetilde {g}^2=N_C \widetilde{Z}_1 g_R^2(\mu^2)$ in \eq{SDRcst2}, which, as far as the constancy for the 
ghost-gluon transverse form factor is assumed, depends only on the renormalisation point chosen 
for the propagators; it is furthermore independent of the particular way used to define the renormalisation 
of the vertex. On the other hand, the redefined coupling $\widetilde {g}$ can be also 
written in terms of bare quantities or related to the well-known running coupling in the Taylor scheme:

\beq
\widetilde {g}^2 \equiv N_c g_{R}^2 \widetilde{Z}_1 \ H_{1R} 
&=& N_C g_B^2 Z_3 \widetilde Z_3^2/\widetilde{Z}_1 \ H_{1R} \nonumber \\
&=& N_C g_B^2 Z_3 \widetilde Z_3^2 \ H_{1B} \nonumber \\
&=& N_C g_T^2(\mu^2) \ H_{1B} 
\label{g2eff}
\eeq
where it should be remembered that $H_{1R}=1$, in MOM scheme.
 \eq{SDRcst2} can be still transformed to a new form which makes the numerical calculation and the presentation of the various solutions easier; for this,  we subtract the equation at $k=0$,  to let the value of $F_R(k)$ at the origin appear and to eliminate the reference to the particular renormalisation point $\mu$,  and we redefine also the unknown function to be calculated as $\widetilde {F}(k)=\widetilde {g}F_R(k)$. Then the reference to the value of $\widetilde {g}$ also disappears; we end with :

\bea\label{SDRcst3}
\frac{1}{\widetilde {F}(k^2)}&=& \frac{1}{\widetilde {F}(0)}- \int\frac{d^4q}{(2\pi)^4}
\left(1- \frac{(k.q)^2}{k^2 q^2}\right)  \nonumber \\
&\times& \left[ \frac{G_R((q-k)^2)}{((q-k)^2)^2}-\frac{G_R((q)^2)}{((q)^2)^2}\right]  \widetilde {F}(q^2)
\eea

Equation \eq{SDRcst3} can be solved for $\widetilde {F}(k^2)$, for a set of values of  $\widetilde {F}(0)$. It is easy to see that, from this,  the desired solution of eq.~(\ref{SDRcst2}) can be straightforwardly reconstructed for any renormalisation point and any value of $\widetilde {g}$. Indeed, as the MOM renormalization condition imposes $\widetilde {g}(\mu)=\widetilde {F}(\mu^2)$,  for any given  $\mu$ and $\widetilde {g}$,  we have just to identify the value of $\widetilde {F}(0)$ such that $\widetilde {F}(\mu^2)= \widetilde {g}$ and reconstruct  then ${F_R}(k^2)$ through 
\beq\label{eq:conect}
{F_R}(k^2)= \frac{\widetilde {F}(k^2)} {\widetilde {g}(\mu)} \ .
\eeq
By construction,  all the solutions found in this way are finite at the origin and they correspond to the ``decoupling'' family of 
solutions that were described above in section~\ref{sec:analytic}. On the other hand, the solution which diverges at vanishing momentum 
appears as an end-point for the solutions of this ``decoupling'' family that can be approached by making $\widetilde {F}(0)$ 
larger and larger so as to get the limiting case: $\frac{1}{\widetilde {F}(0)} \to 0$. Then, the critical ``scaling'' solution will 
be found by setting $\frac{1}{\widetilde {F}(0)}=0$ in eq.~(\ref{SDRcst3}). As we will discuss below, this strategy  for solving the DSE, after it had been
followed in \cite{Boucaud:2008ji}, was further applied by the authors of ref.~\cite{Fischer:2008uz} 
to the analysis of the coupled ghost and gluon DSE in Landau gauge, where they also concluded that the 
choice of $\widetilde {F}(0)$ amounted to fix a boundary condition for the DSE system and, consequently, to determine the class to which the actual solution belongs 
: either a ``decoupling'' one for any finite value of $\widetilde {F}(0)$ or the {\it unique} ``scaling'' 
one for $\widetilde {F}(0) \to \infty$. Unfortunately, the authors of  ref.~\cite{Fischer:2008uz} missed the 
connection of $\widetilde {F}(0)$ and $\widetilde{g}$, and hence with the coupling at the renormalization point, 
$g_R(\mu^2)$, given by \eq{eq:conect}. 

 In ref.~\cite{Boucaud:2008ji}, the solutions of eq.(\ref{SDRcst3}) with the integral cut in the UV at $q=30$~GeV were 
obtained after discretization in $k$ and $q$ and solving by iteration\footnote{It should be noted that minus the integral in the r.h.s. is positive,  allowing an easy convergence. We linearize it at each step,   following the Newton method,  to accelerate the convergence of the iteration procedure,  as suggested by Bloch~\cite{Bloch:2003yu}.}. The results are the following:
\vspace*{0.5cm}

1) \textbf{Critical case,  scaling solution}: One finds a solution with {{\penalty 10000} $\frac{1}{\widetilde {F}(0)}=0$},  i.e. $\widetilde {F}(0)=\infty$, the corresponding "critical" constant being
\bea 
\widetilde {g_c}^2=\widetilde {F}(1.5~{\mathrm{ GeV}})= 33.198.... 
\eea
The relation of eq.(\ref{SDRS2<1}) for $\alpha_G=1$, obtained in sec.~\ref{sec:analytic}, should be 
verified by the numerical solutions and it happens to be very well satisfied:
\bea
\widetilde {g_{c}}^2  A^2(\mu^2) G^{(2)}(0) \frac{1}{10\pi^2}  \thickapprox 0.994....
\eea

\noindent
The integration near $k=0$ can be improved by taking explicitly into account the analytical behavior of the kernel,  
and assuming that the solution behaves as $1/k$ at small $k$. 
This imposes eq.~(\ref{SDRS2<1}),  and one indeed can check that 
\beq
\lim_{k \to 0} \ \frac{1}{10\pi^2} \ \widetilde {g}^2_c  k^2 F(k^2)^2 G^{(2)}(k^2) \  \to 1 \ ,
\eeq
although very slowly.

\vspace*{0.3cm}

\begin{figure}[hbt]
\begin{center}
\includegraphics[height=8cm]{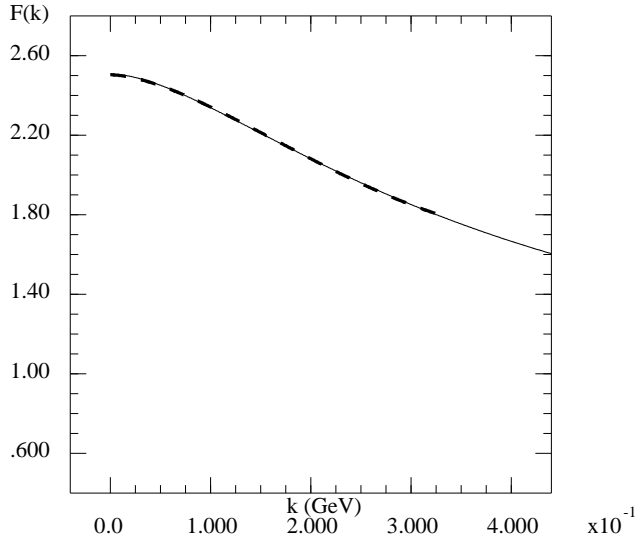}
\vskip -0.5 cm
\caption{\small The  $a+b k^2 \log(k^2)$ fit  at small momentum (dashed line) to our continuum SD prediction for the ghost dressing function, renormalised at $\mu=1.5$~GeV for $\widetilde g^2=29.$(solid line)~; the slope of the $k^2 \log(k^2)$ term is $4.06$~; the agreement with the expected coefficient of $k^2 \log(k^2)$,  $4.11$ from the eq.~(\ref{solghost}),  is striking.}
\label{pente}
\end{center}
\end{figure}

2) \textbf{Regular case, decoupling solution:} One finds a solution and only one for any $\widetilde {F}(0)>0$ with the method of solution  described above.  As can be seen below, a numerical solution at $\widetilde g^2 \simeq 29$ corresponds to the best description of lattice data (see Fig. \ref{fantomefig}, borrowed from ref.~\cite{Boucaud:2008ji}). Furthermore, 
the asymptotic low-momentum behaviour for the ghost dressing function, giving a next-to-leading $k^2 \log(k^2)$ term, 
(cf.   \eq{solghost} )  is checked in ref.~\cite{Boucaud:2008ji} (this is shown in Fig.\ref{pente}, also borrowed from 
this work) and the slope appears to agree pretty well with the prediction from \eq{solghost}: $4.06$ against $4.11$.

\vskip 0.5 cm

The critical value of the coupling constant,  as well as the corresponding curve of $\widetilde {F}(k)$,  can be very well approximated by the regular decoupling solutions at very large $\widetilde {F}(0)$. When $\widetilde {F}(0)$ is larger and larger,  \eq{solghost}  remains valid only in a smaller and smaller region near $k=0$ while, in an intermediate region, one observes the expected $1/k$ behaviour. 
In this way it is possible to show that, as long as the coupling constant does not exceed a critical value $\widetilde{g}_{\rm crit}$, where the 
scaling solution emerges  (see also sec.~\ref{subsec:limit} ), the solution goes to a constant in the infrared : one is in the decoupling case. As soon as the coupling constant reaches that critical value~\footnote{The existence and the value of this critical coupling  had already been noticed, see for instance ref.~\cite{Bloch:2003yu}.} the solution converts to the infrared-infinite one (``scaling situation"). Actually, the  DSE solution accounted for the  scaling solution when $\widetilde{g}_{crit}\simeq 33.2$ at the renormalization point $\mu=1.5$ GeV, as it is shown in Fig.~\ref{fantomefig}. 
Admittedly, a real resolution of the DSE equations would require solving reciprocally the DSE for the gluon and verifying  the compatibility of  the solution with the  gluon dressing function which was used as an input to build the kernel for the same g-value. We will deal with this below.

In conclusion,  in the case $\alpha_G=1$, a continuum set of IR finite decoupling solutions for arbitrary $F(0)$ emerges,  
and a unique singular scaling solution for $\widetilde {g}^2=\widetilde {g}^2_c$,  with $\alpha_F=-\frac{1}{2}$, which appears 
to be the end-point for the previous ones.

\subsubsection{Comparison with ghost lattice data} 
\label{subsec:comp-latt}

After the analytic study of the general solutions of the GPDSE in sec.~\ref{sec:analytic}  
and the previous numerical analysis exhibiting both types of solutions for the ghost dressing functions, 
either regular (decoupling) or singular (scaling), the question also 
addressed in ref.~\cite{Boucaud:2008ji} is: which one is effectively realised on the lattice, 
and therefore in true QCD? 

\begin{figure}[hbt]
\begin{center}
\vspace*{0.5cm}
\includegraphics[height=7.5cm]{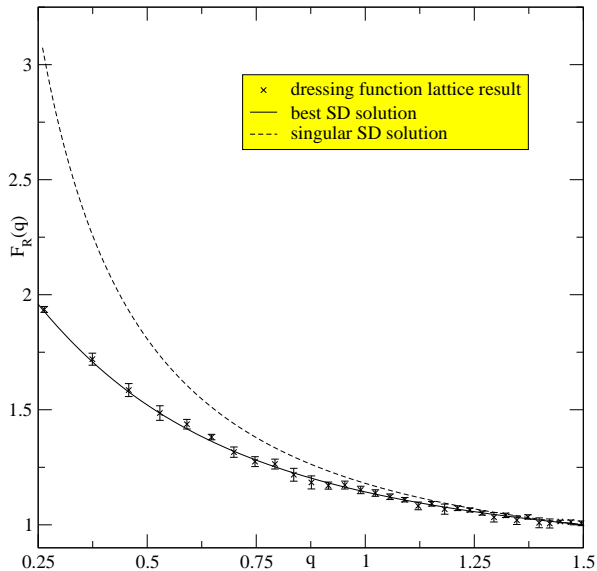} 
\caption{\small Comparison, borrowed from ref.~\cite{Boucaud:2008ji}, between the lattice SU(3) data at $\beta=5.8$ and with  a volume $32^4$ for the ghost dressing function and our continuum SD prediction renormalised at $\mu=1.5$~GeV for $\widetilde g^2=29.$ (solid line)~; the agreement is striking; also shown is the singular solution which exists only  at $\widetilde g^2=33.198....$ (broken line),and  which is obviously excluded.}
\label{fantomefig}
\end{center}
\end{figure}

A better means to provide us with an answer is offered by the numerical calculation in previous subsection 
that, predicting the behavior of the respective solutions for the ghost over the \textbf{whole range} of momentum, 
can be confronted to the lattice estimate for the ghost dressing function and that can lead us to identify 
which one offers the best agreement with the data~\footnote{At this stage,  it is useful to stress the advantage of working with the renormalised form of the SD equations; indeed the continuum and lattice versions are more directly comparable than the bare ones. As the authors of ref.~\cite{Boucaud:2005ce} have shown,  the bare lattice equation for the ghost is affected by an important artefact which vanishes only very slowly with the cutoff,  being of order ${\cal O}(g^2)$. In the renormalised version,  this effect is included in the renormalisation constant $\widetilde Z_3$,  and we are left only with the much smaller cutoff effects of the type ${\cal O}(a^n)$.}. This is done in ref.~\cite{Boucaud:2008ji} and displayed here in 
fig.~\ref{fantomefig}, where one can see that the singular scaling solution appears to be clearly discarded by the 
lattice data around and below $k=0.5$~GeV. On the contrary, a very good description of the lattice data in the 
range $\widetilde g^2=28.3-29.8$ (clearly below the critical value) is found. This striking agreement, although for 
a narrow momentum window, is illustrated by Fig. \ref{fantomefig} (As an indication, we quote the IR limit $F_R(0)=2.51$ for the same $\mu=1.5$ and $\widetilde g^2=29$). 

An additional consistency test is obtained from using \eq{g2eff} to connect the 
continuum $\widetilde {g}^2$ to the lattice bare quantities,
\beq\label{g2eff2} 
\widetilde {g}^2 = N_c g_R^2 \widetilde z_1 = N_c \frac 6 {\beta} F_B^2(\mu^2) \ G_B(\mu^2) \ H_{1B} \ ,
\eeq
and then checking whether our range $\widetilde g^2=28.3-29.8$ is reasonably consistent 
with the r.h.s. of eq.(\ref{g2eff2}) as given by lattice data. In spite of the crude approximation 
made~\footnote{First, it is valid up to finite cutoff effects, as well as volume effects; second, we have 
replaced the lattice vertex invariant $H_{1B}(q, k)$ by the constant $H_{1B}$ which, as we discussed above, 
is a rough approximation over the momenta which are actually implied in our calculation. In 
app.~\ref{app:Taylor}, some non-perturbative corrections for this invariant, that appear to 
be pretty well in agreement with some SU(2) lattice estimates~\cite{Boucaud:2011eh}, will be discussed.} 
to be left with \eq{g2eff2}, the result of the check is very encouraging:
indeed, from the  above value of $\widetilde g^2$ found in the continuum on the one hand and the lattice data  $\beta=5.8$, $G_B(\mu^2)\simeq 2.89$ and $F_B(\mu^2)\simeq 1.64$ ($\mu$ is here chosen 
as $1.5$~GeV) on the other, one finds $H_{1B}\backsimeq 1.2$ to satisfy equation (\ref{g2eff2}), 
which represents some kind of average on momenta. This number should be compared  to the SU(3) lattice measurements 
which are performed at $q=k$ and have been submitted to a renormalization such that the result  takes the value 1 at $q=3$ GeV, 
and which give about $1.1$~\cite{Sternbeck:2005re}, with large errors. 

\begin{figure}[hbt]
\begin{center}
\includegraphics[height=11cm,angle=-90]{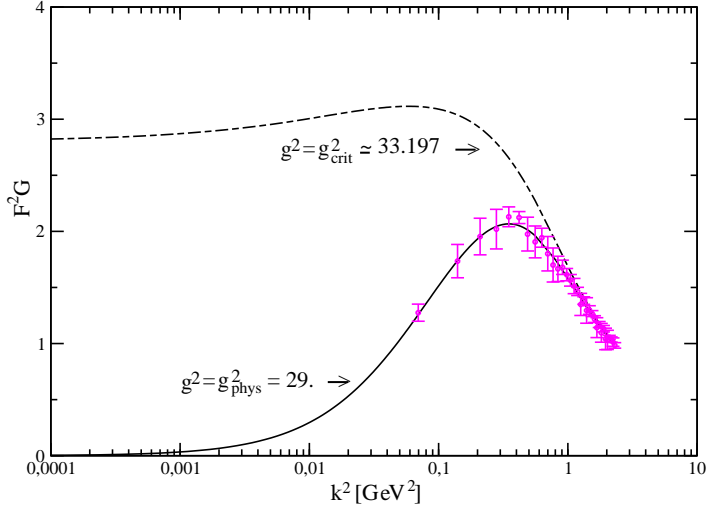} 
\end{center}
\caption{\small The same than in Fig.~\ref{fantomefig} but for 
$G_R(k) F_R(k)^2$;  up to a factor $g^2/4 \pi$, the scaling (dotted) curve corresponds to the $\alpha(k^2)$ presented 
in Fig.~8 by Fischer~\cite{Fischer:2006ub}; the shape is very similar.}
\label{fig:F2GSD}
\end{figure}

Another striking way of presenting the difference between the regular decoupling solution and the singular scaling 
one is in terms of the familiar product, $G_R(k) F_R(k)^2$, which is proportional, at least for the UV domain, to the 
running coupling in Taylor scheme. This is shown in Fig.~\ref{fig:F2GSD}.

In ref.~ \cite{Bogolubsky:2009dc} 
new ghost propagator data are provided, coming from larger volume lattice simulations and 
covering a wider momentum range, with smaller momentum data, to compare with. 
The confrontation of those data with the results from the integration of the GPDSE truncated 
with the help of a lattice-based gluon propagator as done in sec.~\ref{subsec:landau} requires, 
to account succesfully for the very low-momentum data, to go beyond the hypothesis 
of constancy of the ghost-gluon transverse form factor, $H_1$. 
A possibility, studied in ref.~\cite{Boucaud:2011eh} consists in applying an ansatz for the transverse 
form factor, $H_1$, which is inspired by the OPE description for the non-perturbative corrections 
of the ghost-gluon vertex, obtained by applying the same procedure outlined in sec.~\ref{sec:A2lat} 
and by taking the gluon condensate value of table~\ref{tab:global_comp}.  

%
%

\subsubsection{Coulomb gauge solutions}
\label{subsec:coulomb}

As we previously mentioned, the authors of ref.~\cite{Watson:2010cn} recently performed a  
 study (very analogous to the one in ref.~\cite{Boucaud:2008ji} for Landau gauge) 
of the GPDSE in Coulomb gauge, obtained within the (second order) functional formalism, 
in order to investigate the low-momentum ghost dressing 
solutions. They took  Gribov's equal-time spatial gluon propagator 
dressing function~\footnote{In very good agreement with the Euclidean SU(2) lattice results 
obtained for small lattice couplings in ref.~\cite{Burgio:2008jr}.}, 
\beq\label{Gribov}
G^{T}(\vec{k}^2) = \int_{-\infty}^{\infty} \frac{dk_4}{2\pi} \ 
\frac{G\left(k_4^2,\vec{k}^2\right)}{k_4^2+\vec{k}^2} \ = \ 
\frac 1 2 \frac{\sqrt{\vec{k}^2}}{\sqrt{\vec{k}^4+m^4}} \ ,
\eeq 
as the input required to build a kernel and solve the GPDSE, again with the approximation 
of replacing the fully dressed spatial ghost-gluon vertex by the bare one (this is, also in 
Coulomb gauge, an exact result in the limit of a vanishing incoming ghost 
up to all perturbative orders~\cite{Watson:2006yq}). 
Thus, the GPDSE is rewritten as follows:
\beq
\frac 1 {F\left(\vec{k}^2,\mu^2\right)} 
\ = \ 
\frac 1 {F\left(\vec{p}^2,\mu^2\right)}   
\ - \ 
N_C \frac{g^2(\mu)}{(4 \pi)^2} 
\int_{0}^{\infty} \frac{d\vec{q}^2}{\vec{q}^2} F(\vec{q}^2, \mu^2)  
\ 
\left( I\left(\vec{k}^2,\vec{q}^2; m \right) 
- I\left(\vec{p}^2,\vec{q}^2; m \right) \right) \ ,
\label{SDRSCou}
\eeq
where $I$ represents the angular integration, 
\beq
I\left(\vec{k}^2,\vec{q}^2; m \right) 
\ = \ 
\int_{-1}^{1} dz \left(1-z^2\right)  
\left(1+\frac {\vec{k}^2}{\vec{p}^2} 
- 2 z \sqrt{\frac {\vec{k}^2}{\vec{p}^2} } \right)^{-1/2}
\left[
\left(1+\frac {\vec{k}^2}{\vec{p}^2} 
- 2 z \sqrt{\frac {\vec{k}^2}{\vec{p}^2} } \right)^2 + \frac{m^4}{\vec{p}^4}
\right]^{-1/2}
\ .
\eeq
It should be emphasized that the ghost propagator dressing function in Coulomb 
gauge is strictly independent of the energy, $k_4^2$, as a non-perturbative consequence of the 
Slavnov-Taylor identities~\cite{Watson:2007vc}.

Assuming a pure powerlaw behaviour, $F(\vec{k}^2) \sim (\vec{k}^2)^{\alpha_F}$, for the ghost dressing 
function and analyzing asymptotically  Eq.~(\ref{SDRSCou}), one is left in Coulomb gauge 
again with the two same cases we have encountered in Landau gauge: (i) $\alpha_F=0$ (decoupling) 
and (ii) $\alpha_F \neq 0$ (scaling). As well in Landau as in Coulomb gauge, a massive gluon propagator  generated
via the Schwinger mechanism or  Gribov's fomula for the equal-time spatial dressing 
leads to $\alpha_G=1$ and thus $\alpha_F=-1/2$. In particular, 
from the next-to-leading analysis in sec.~\ref{subsec:next} of \eq{SDRS}, 
one obtains~\cite{Boucaud:2008ky} :

\beq \label{solsFs}
F(q^2,\mu^2) \simeq 
\left\{
\begin{array}{lr}
\displaystyle
\left(
\frac {10 \pi^2}{N_C H_1 g_R(\mu^2) B(\mu^2)} 
\right)^{1/2}
\ \left(\frac {M^2} {q^2} \right)^{1/2} 
&
\mbox{\rm if } \alpha_F \neq 0 \ ,
\\
\rule[0cm]{0cm}{0.8cm}
\displaystyle
F(0,\mu^2) \left( 1 +   
\frac{N_C H_1}{16 \pi} \ \overline{\alpha}_T(0) \ 
\frac{q^2}{M^2} \left[ \ln{\frac{q^2}{M^2}} - \frac {11} 6 \right] 
\ + \ {\cal O}\left(\frac{q^4}{M^4} \right) \right)
&
\mbox{\rm if }  \alpha_F = 0 \ .
\end{array}
\right.
\eeq
If $\alpha_F \neq 0$, the perturbative strong coupling defined 
in the Taylor scheme~\cite{Boucaud:2008gn}, $\alpha_T=g_T^2/(4\pi)$, goes to a non-zero
constant at zero-momentum,
\beq
\lim_{q^2 \to 0} \alpha_T(q^2) \ = \
\lim_{q^2\to 0} \left(
\frac{g^2(\mu^2)}{4 \pi}  q^2 \Delta(q^2,\mu^2) F^2(q^2,\mu^2) 
\right)
\ = \ 
\frac{5 \pi}{2 N_C H_1} \ ,
\label{alphaT0}
\eeq
as can be obtained from Eqs.~(\ref{gluonprop},\ref{solsFs}).
In the case $\alpha_F = 0$, the subleading correction to the non-zero finite value for 
the zero-momentum ghost dressing function, given by \eq{solsFs},  
is controlled by the well-defined zero-momentum limit of $\overline{\alpha}_T(q^2)=(M^2/q^2) \alpha_T(q^2)$, 
which is the extension to the Taylor ghost-gluon coupling case~\cite{Aguilar:2009nf} 
of the non-perturbative effective charge defined from the gluon  
propagator in ref.~\cite{Aguilar:2008fh}. 

\begin{figure}
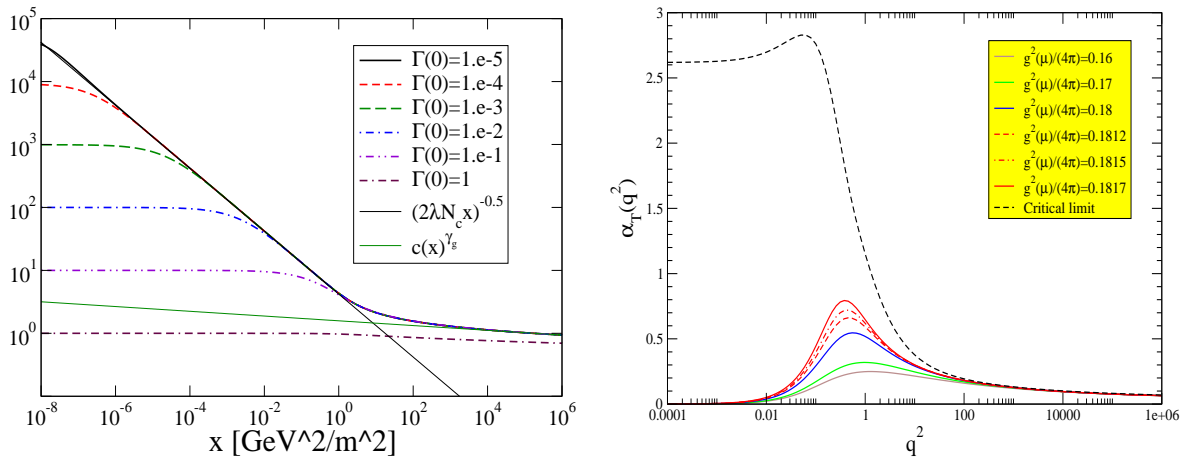

\begin{center}
\begin{tabular}{cc}
\includegraphics[width=7.5cm,height=6cm]{ghostd00.eps} &
\includegraphics[width=7.5cm,height=6cm]{F2Gcrit.eps} 
\end{tabular}
\end{center}
\caption{\small (Left) Ghost dressing in Coulomb gauge plotted for different values of the inverse of 
boundary condition fixed for the dressing at zero-momentum, $\Gamma(0)$ in terms of $x=\vec{k}^2/m^2$, 
where $m$ is the Gribov mass in \eq{Gribov}. This plot is borrowed from ref.~\cite{Watson:2010cn}. 
(Right) The Taylor coupling defined by \eq{alphaT0} and computed from the gluon and ghost propagator 
results of ref.~\cite{RodriguezQuintero:2010wy} for $g^2(\mu)/(4\pi)=0.16,0.17,0.18,0.1812,0.1815,0.1817$, 
with the subtraction point $\mu=10$ GeV. The curve for the critical limit is obtained by applying the 
results of ref.~ \cite{Boucaud:2008ji}, as explained in the text.}
\label{fig:alphaT}
\end{figure}

The same two cases result from the analysis of \eq{SDRSCou} for the Coulomb gauge 
in ref.~\cite{Watson:2010cn}, where a ghost propagator dressing function 
behaving asymptotically as either a constant or $F(\vec{k}^2) \sim (\vec{k}^2)^{-1/2}$ is 
analytically found and confirmed by a numerical study. This can be seen in the left plot of 
Fig.~\ref{fig:alphaT} which we borrowed from ref.~\cite{Watson:2010cn}. 
In Coulomb gauge,  the lattice results for the 
ghost propagator, within the available momentum window, may 
agree with decoupling and scaling solutions, and cannot help to discriminate.

 It should be noted that the lagrangian approach to the Coulomb gauge  
(the continuum functional formalism is based on the QCD Lagrange density) is not the 
most widely used. However, and despite the technical difficulties mostly related to the 
inherent non-covariance of the Coulomb gauge, some recent progresses have been made in 
order to derive explicitely the DSE~\cite{Watson:2006yq,Watson:2007vc,Popovici:2008ty} 
(allowing the previous analysis) or studying the 
Bethe-Salpeter equation for heavy quarks~\cite{Popovici:2010mb}. 
The method which is most widely applied in the continuum 
is the canonical formalism based on the QCD Hamiltonian density 
operator~\cite{Epple:2007ut,Szczepaniak:2001rg,Reinhardt:2004mm,Schleifenbaum:2006bq,Epple:2006hv}. 
Dyson-Schwinger-like equations (in a space with one less dimension) for the equal-time 
correlators are obtained in the canonical 
formalism and again the two types of solutions, critical and subcritical, are found for the propagator 
dressing function~\cite{Epple:2007ut}. 
It seems to happen  that,  in the Coulomb-gauge canonical formalism, two different values for the infrared exponents emerge
in the critical case  and that the favoured one is the one 
which produces the most singular ghost dressing, which diverges as $1/|\vec{k}|$ in the infrared, 
{\it i.e.} similarly to the critical solution  discussed above (with $\alpha_F=-1/2$).

\vspace*{3mm}

\subsubsection{The critical limit of decoupling solutions from the GPDSE analysis}
\label{subsec:limit}

In summary, the GPDSE in \eq{SDRS} with the input of a gluon propagator borrowed from 
lattice QCD calculations can be numerically solved 
and two kinds of solutions result and appear to be controlled 
by the size of the coupling at the renormalization point~\footnote{In QCD, one needs to 
provide a physical scale and a standard manner to proceed is by fixing the size of the coupling at 
a given momentum scale. This can be seen as a boundary condition to solve the DSEs.}, $g(\mu)$. 
For any coupling below some critical value, $g_{\rm crit}$, an infinite number of 
regular or decoupling solutions for the ghost dressing, behaving as \eq{solsFs} 
indicates, are found; for $g(\mu)=g_{\rm crit}$, a unique 
critical or scaling solution behaving as \eq{solsFs} is found, and no other 
type of solutions  appears  to exist. 
In ref.~\cite{Boucaud:2008ji}, for a subtraction 
point $\mu=1.5$ GeV, a critical coupling $g_{\rm crit} \simeq 3.33$ and a very 
good description of ghost propagator lattice data with a regular solution of \eq{SDRS} for 
$g(\mu) \simeq 3.11$ were obtained. 

As we shall discuss in the next section, these results were also 
recently confirmed~\cite{RodriguezQuintero:2010wy} by studying the coupled system of ghost and gluon 
propagator DSE in the PT-BFM scheme~\cite{Binosi:2009qm}. This last work paid attention to 
the critical solution limit by studying how $F(0,\mu^2)$ diverges 
as $g(\mu) \to g_{\rm crit} \simeq 1.51$, with a subtraction point $\mu=10$ GeV. 
In addition, the author of ref.~\cite{RodriguezQuintero:2011vw} applied the 
perturbative definition of the Taylor strong coupling in \eq{alphaT0} to compute 
this coupling with the gluon and ghost solutions of \cite{RodriguezQuintero:2010wy} 
in order to see how the critical limit is approached. 
This is shown by the right plot of Fig.~\ref{fig:alphaT}, borrowed 
from ref.~\cite{RodriguezQuintero:2011vw}, 
where it can be also seen that all the curves for $\alpha_T$ obtained for different values of $g(\mu)$ 
tend to superpose over each other as $q^2/\mu^2$ increases (right). As a striking check of consistency, 
the curve for the critical limit in this right plot of Fig.~\ref{fig:alphaT} is obtained by rescaling, 
up to giving $\alpha_T(0)$ from \eq{alphaT0} with $H_1=1$ at zero-momentum, 
the results at the critical limit for $q^2 \Delta(q^2) F^2(q^2)$ numerically 
obtained in ref.~\cite{Boucaud:2008ji} and plotted here in Fig.~\ref{fig:F2GSD} of the previous 
subsection. Indeed, the critical value for the coupling at $\mu=10$ GeV 
can be read from the critical curve in Fig.~\ref{fig:alphaT} and one gets $g(\mu) \simeq 1.56$, 
in fairly good agreement with the value of ref.~\cite{RodriguezQuintero:2010wy}.

On the other hand, the numerical analysis of \eq{SDRSCou} for the Coulomb gauge 
in ref.~\cite{Watson:2010cn} also shows both the regular and the critical solutions to exist, 
but  being  controlled by $F(0,\mu)$ (or $\Gamma(0,\mu)=1/F(0,\mu)$) as a boundary condition with 
the size of the coupling fixed to be $g^2(\mu)=\overline{g}^2=4 \pi \times 0.1187$ 
for $N_C=3$ (see the right plot of Fig.~\ref{fig:alphaT}). 
Regular or decoupling solutions appear for finite values of $F(0,\mu^2)$ and 
critical or scaling solution for $F(0,\mu^2) \to \infty$. 
Nevertheless, the authors of ref.~\cite{Watson:2010cn} concluded that, 
as far as all the solutions join each other in the perturbative domain 
 (see the right plot of Fig.~\ref{fig:alphaT}) and 
can be found for a fixed coupling, the boundary condition 
is not connected to the renormalization and claimed for a contradiction 
with the Landau-gauge results of ref.~\cite{Boucaud:2008ji}.

Finally, the two pictures for the low-momentum solutions from GPDSE, 
either a family of Landau-gauge DSE solutions labelled by the size of the coupling 
at the renormalization point or a family of Coulomb-gauge ones labelled by the zero-momentum 
ghost dressing value as a boundary condition independent of the renormalization, were 
reconciled by the work of ref.~\cite{RodriguezQuintero:2011vw}. 
The key point stems from the different renormalization prescriptions applied 
to the ghost propagator in both analyses. 
MOM scheme in the Landau-gauge analysis of 
refs.~\cite{Boucaud:2008ji,Boucaud:2008ky,RodriguezQuintero:2010wy}, 
and the prescription applied to the ghost propagator in eq.~(3.20) of ref.~\cite{Watson:2010cn} 
for the renormalization constant $Z_c(\Lambda,[\overline{g},\Gamma(0)])$, where $\Gamma(0)=1/F(0,\mu^2)$.
In particular, this last renormalization constant depends on the boundary condition, $\Gamma(0)$, 
in such a manner that the value for this boundary condition is rescaling the ghost 
dressing function (and, as can be clearly seen in Fig.~2 of \cite{Watson:2010cn}, it does not 
take the tree-level value, 1, as happens in MOM prescription for the subtraction point). 
Thus, according to ref.~\cite{RodriguezQuintero:2011vw}, the non-trivial connection between solutions 
in both schemes, that relies on the relation previously shown by \eq{eq:conect}, comes out from 
the following property of Eqs.~(\ref{SDRS},\ref{SDRSCou}): 
let $F(q^2,\mu^2)$ be a MOM solution of \eq{SDRS} for 
arbitrary coupling, $g(\mu)$; if we then 
apply the following transformation:
\beq\label{MOM2New}
g(\mu) \to s \ g(\mu) \ , \quad F(q^2,\mu^2) \to \frac 1 s F(q^2,\mu^2) \ .
\eeq
 for any c-number $s$, the transformed dressing function verifies the DSE equation with the 
transformed coupling (of course, MOM prescription implies $s=1$). 
Then, one only needs to choose $s=\overline{g}/g(\mu)$ and to apply the transformation 
to every solution of the MOM family and one will be left with a one-to-one correspondence 
between these solutions and the new ones
\beq
\overline{F}(q^2,\mu^2)\ \equiv \ \frac {g(\mu)}{\overline{g}} F(q^2,\mu^2) \ ,
\eeq
for the fixed coupling $\overline{g}$, which can be identified by 
the zero-momentum value, $\overline{F}(0,\mu^2)$. This new family of transformed 
solutions obeys the same pattern as the Coulomb gauge family in ref.~\cite{Watson:2010cn} 
and corresponds, up to the fixed number $\overline{g}$ (which does not even depend on the renormalization point), 
to the family of solutions $\widetilde{F}(q^2)$ obtained by the analysis for Landau gauge in previous 
sec.~\ref{subsec:landau}. It is interesting to note that the strong coupling 
defined in the Taylor scheme can be also obtained from the transformed solutions as
\beq\label{TralphaT}
\alpha_T(q^2) \ \equiv \frac{\overline{g}^2}{4 \pi} \ q^2 D(q^2,\mu^2) \overline{F}(q^2,\mu^2) 
\ \equiv \ \frac{g^2(\mu)}{4 \pi} \ q^2 D(q^2,\mu^2) F(q^2,\mu^2) \ ,
\eeq
although it is obvious that neither $\overline{F}$ nor the coupling are in MOM scheme. 

Thus, the same picture for the low-momentum Green function solutions emerges in both Landau and Coulomb 
gauge from the analysis of the GPDSE: a family of MOM-renormalized regular decoupling solutions, characterised 
by the value of the coupling at the renormalization point; and a singular scaling solution as an 
end-point for the family of regular ones. An interesting final remark is that 
the input parameter for the solutions in ref.~\cite{Watson:2010cn}, the zero-momentum ghost dressing, 
can be put in connection with the Gribov problem~\cite{Watson:2010cn}; while, for Landau gauge and MOM scheme, 
$g(\mu)$ is related to the strong coupling in Taylor scheme, as was shown in \eq{g2eff}. This is not the case for 
the size of the fixed coupling, $\overline{g}$, after applying \eq{MOM2New} which is physically meaningless.

\subsection{The ghost and gluon propagator coupled DSEs}

Satisfying the GPDSE, as was required in sec.~\ref{sec:analytic}, is a necessary but not sufficient condition for a 
DSE solution to exist. Of course, the existence of a solution can only be confirmed by treating the infinite tower 
of DSEs, but this is an impossible task. In the previous section, sec.~\ref{subsec:num1}, this infinite 
tower of DSEs was truncated by plugging into the one among them to be solved, the GPDSE in that case, the available 
lattice data, or a model compatible with them, for the gluon propagator and the ghost-gluon vertex. 
On the other hand, the usual approach consists in applying a truncation scheme based on hypotheses and 
approximations that preserve the main properties of the theory and 
that leave us with a closed system of equations to deal with. 
The former approach can be seen to provide with a consistency analysis of the lattice and DSE picture 
for the solutions and benefits of not  ``polluting''  the conclusions with the possible implications 
of any particular truncation scheme. However, the DSE picture should be completed by also 
applying the latter usual approach. 

As we shall discuss below, both scaling and decoupling solutions 
have also been proven  to emerge when the DSEs are truncated so as to give a coupled system for 
the ghost and gluon propagators. 

\subsubsection{Scaling solutions}

 As we have repeated insistantly in this paper, it has been recognized for a long time that the set of solutions of the DSEs for the ghost and gluon propagators consists in a continuum of so-called ``decoupling" solutions augmented with a unique ``scaling" one (cf. ref.~\cite{Lerche:2002ep}); which one is actually encountered depends on the value of the coupling constant. Nevertheless, for quite a time, attention has mainly be paid to the scaling one which,  
 in practice, was obtained by replacing the fully dressed 
vertices by ans\"atze which take into account as much information as possible (see 
ref.~\cite{Alkofer:2000wg} for a first review on the subject).
The loops in DSEs had been proven to be dominated by the infrared contributions for the 
scaling solution~\cite{vonSmekal:1997is,vonSmekal:1997vx} which had been fully worked-out in 
ref.~\cite{Lerche:2002ep} (see also ref.~\cite{Zwanziger:2001kw}). The infrared exponents for the power 
behaviour on the momentum for both ghost and gluon dressing functions being related by 
$2 \alpha_F + \alpha_G=0$, the value for one of them, usually the one for the ghost, $\alpha_F$, completely 
characterizes the low-momentum solution. Under the assumption of a constant ghost-gluon transverse form factor, 
the only solution to emerge in  the interval $[-1,-1/2]$ ( \cite{Lerche:2002ep}) is 
$\alpha_F \simeq 0.595$, as was numerically put forward by the authors of ref.~\cite{Fischer:2002eq}, 
and independently confirmed by the analysis performed in ref.~\cite{Pawlowski:2003hq} with the help of 
renormalization group methods (RGE). Then, the uniqueness of the the above-mentioned low-momentum solution 
defined by $\alpha_F\simeq 0.595$ was discussed in two papers~\cite{Fischer:2006vf,Fischer:2009tn}, 
first (wrongly) claimed to be true in general for the coupled DSE system~\cite{Fischer:2006vf} 
and later on  concluded to happen only for scaling-type solutions~\cite{Fischer:2009tn}, {\it i.e.} 
provided that the relation $2 \alpha_F + \alpha_G=0$ is ``{\it a priori}'' assumed (we will pay attention 
to this in a next subsection). 

Very recently, the authors of ref.~\cite{Fischer:2008uz} 
re-analysed the problem of the low-momentum properties of the 
Yang-Mills Green functions by following both DSEs and RGE approaches and found both scaling and decoupling 
solutions to exist. They paid special attention to the truncation schemes and also claimed that only 
the (unique) scaling solution satisfies BRST invariance, while the decoupling ones would be 
at odds with it.  This was presented as an incitation to prefer the scaling solution as the ``real" QCD one but, as 
was previously mentioned when discussing the Gribov-copies problem, Gribov or Gribov-Zwanziger 
(either refined or not) 
approaches to avoid the copies already imply a BRST breaking and this only prevents the 
Kugo-Ojima confinement scenario from working.  Other confinement scenarios are of course possible and 
nothing indeed prevents the decoupling solution from being, as lattice appears to indicate, 
the real QCD one.

The properties and implications of the scaling-type  low-momentum solutions have been extensively 
discussed in the literature. We address the interested reader to reviews like the ones in 
ref.~\cite{Alkofer:2000wg,Bloch:2003yu,Fischer:2006ub} as well as to the numerous  works qoted above or to others 
like ref.~\cite{Watson:2001yv} focusing on the Kugo-Ojima criterium implications, 
refs.~\cite{Schleifenbaum:2006bq,Alkofer:2004it,Schleifenbaum:2004id} 
on the infrared behaviour of vertices, ref.~\cite{Alkofer:2008jy} on the analysis of infrared 
sigularities or ref.~\cite{Huber:2009wh} about the study of scaling solutions in 
the maximally abelian gauge. 

We will now end this section by adding a few words about the elusiveness of decoupling solutions, 
after the scaling one has been proposed.

\subsubsection{Why have the decoupling solutions been so elusive?}

The decoupling or regular solutions have been missed for almost ten years. Why? 
they could not have been previously obtained by the proponents of the relation~(Rel$\alpha$)  
because,  as it seems to us,  they discarded them from the very beginning, and thereby chose the critical 
value of the coupling constant,  by making an implicit assumption when solving  the so-called ``infrared equation" 
for the ghost SD equation. This can be seen, for instance, in ref.~ \cite{vonSmekal:1997vx}, eqs. (43) and (44),  
or in the detailed discussion of Bloch~\cite{Bloch:2003yu}),  eqs. (55) to (58).

Let us explain this briefly. They consider the above unsubtracted equation (note that this requires then an 
UV cutoff,  which we avoid in our previous analysis by considering the subtracted form); 
we write again the unsubtracted form:
\beq 
\frac{1}{F_R(k^2)}=\widetilde Z_3-N_c g_{R}^2 \widetilde z_1 \int\frac{d^4 q}{(2\pi)^4}
\left(1- \frac{(k.q)^2}{k^2 q^2} \right)  \nonumber \\
\left[ \frac{G_R((q-k)^2)H_{1R}(q, k)}{((q-k)^2)^2}\right] F_R(q^2) \label{Rsubtracted}
\eeq
One must try to match the small $k^2$ behaviour of the two sides of \eq{Rsubtracted}. 
This is done for example in eq.~(58) of~\cite{Bloch:2003yu}. A condition is then written 
which consists in equating the coefficient of $(k^2)^{-\alpha_F}$ with the corresponding one in the r.h.s.. 
However,  one notices that on the r.h.s.,  there is a constant contribution  $\varpropto(k^2)^0$. 
Therefore  unless the constant term  $\tilde Z_3$ is cancelled by the integral contribution for $k \to 0$,  
we have necessarily $\alpha_F=0$. To have $\alpha_F<0$ as the author finds,  one needs this cancellation. 
This is what is \textbf{implicitly assumed},  but not stated explicitly. The condition of cancellation is :
\beq
\widetilde Z_3 \ = \  N_c g_{R}^2 \widetilde z_1 \int\frac{d^4 q}{(2\pi)^4}
\left.\left(1- \frac{(k.q)^2}{k^2\,q^2} \right)F_R(q^2)\right\arrowvert_{k=0}
\eeq
However,  this additional equation {\bf does not derive} from the starting SD ghost equation,  and indeed it is not 
satisfied in general by the solutions of this basic equation,  as we show by displaying actually IR finite solutions. 
In fact,  it  can be valid only for a \textbf{particular value of the coupling constant,  the critical one} which is
solution to the equation of Bloch,  his  eq.~(58),  and which we derive rigorously through the subtracted equation. A similar conclusion is obtained in the analysis of ref.~\cite{Fischer:2008uz}, although its  authors 
missed the connection between their boundary condition, the zero-momentum value of the renormalized 
ghost dressing function, and the coupling size at the renormalization momentum.

\subsubsection{Decoupling or massive solutions}

A decoupling behaviour has also been proven to result as a solution of a coupled system of gluon and ghost 
propagators DSEs. First, the authors of ref.~\cite{Aguilar:2004sw} implemented  
some ans\"atze based on Slavnov-Taylor identities for the involved full vertices, applied a particular 
angular approximation when integrating the ghost self-energy and thus obtained a ``massive'' gluon propagator 
($\alpha_G=1$), although they claimed this to be compatible with an enhanced ghost propagator.  Then, 
as  mentioned above, the Schwinger mechanism of mass generation~\cite{Schwinger1962} was proven to be consistently 
incorporated into the gluon propagator DSE through the fully-dressed non-perturbative three-gluon 
vertex and to give rise to the generation of a dynamical gluon mass~\cite{Cornwall:1981zr,Aguilar:2006gr}. 
Then, a massive solution for the coupled ghost and gluon propagator DSE was shown to apppear ~\cite{Aguilar:2007ie} in 
the PT-BFM truncation scheme~(see also~\cite{Binosi:2002ft}).  The lattice data result to be 
furthermore very well accommodated within coupled DSEs in the PT-BFM scheme~\cite{Aguilar:2008xm}. As a matter 
of fact, as will be seen below, the PT-BFM DSEs solutions have been shown to asymptotically behave 
as Eqs.(\ref{gluonprop}, \ref{solFIRJo}) predict for a decoupling 
solution~\cite{RodriguezQuintero:2010wy}. The authors of ref.~\cite{Fischer:2008uz} also confirmed 
the decoupling solutions to be present by the analysis of the coupled DSEs and the functional 
Renormalization group equations (FRGs). They also obtained an infinite family of decoupling solutions, 
as was discussed in sec.~\ref{subsec:limit} but they used the zero-momentum 
ghost propagator as the boundary condition for the DSEs integration and missed its connection with the value 
of the coupling at the renormalization momentum ({\it i.e.} the particular value of $\Lambda_{\rm QCD}$ one 
applies to build the solutions) or the critical coupling 
the scaling behaviour requires to emerge. This connection is an important ingredient because it provides us with a manner, 
through a comparison with the physical strong coupling, to discuss whether the scaling critical DSE solution could 
be allowed by the data.

\begin{figure}[hbt!]
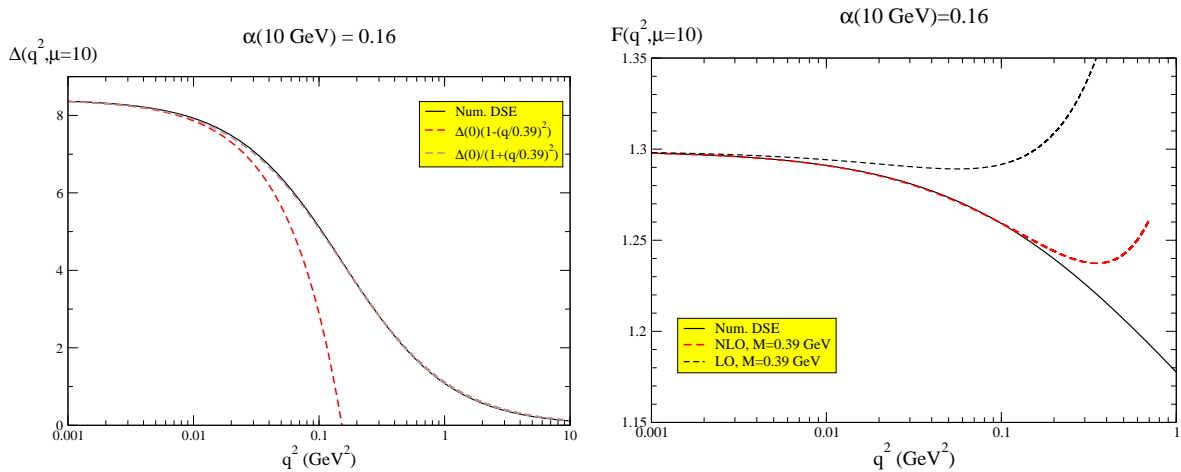

\begin{center}
\begin{tabular}{cc}
\includegraphics[width=7.5cm]{gluon016.eps} &
\includegraphics[width=7.5cm]{ghost016-newF.eps} 
\end{tabular}
\end{center}
\caption{\small Gluon propagators (left) and ghost dressing functions (right) after the numerical integration of  
the coupled DSE system for $\alpha(\mu=10 \mbox{\rm GeV})=0.16$, 
confronted to \eq{gluonprop} and \eq{solFIRJo}, taken from~\cite{RodriguezQuintero:2010wy}. The black dotted 
line corresponds to the asymptotical expression including the leading correction and the red dotted to the 
one including the next-to-leading.}
\label{fig:ghgl}
\end{figure}

We will now present, in the following, the comparison performed in 
ref.~\cite{RodriguezQuintero:2010wy} of the decoupling analytical 
low-momentum expressions, given here by eqs.~(\ref{gluonprop},\ref{solFIRJo}), 
and the PT-BFM solutions shown to provide a quantitative description of 
lattice data~\cite{Aguilar:2008xm,Aguilar:2010gm}.
The main feature in the PT-BFM scheme is that it guarantees the transversality of the gluon self-energy 
order-by-order in the dressed-loop expansion, thus leading 
to a gauge-invariant truncation of the gluon DSE~\cite{Binosi:2002ft}.
In this PT-BFM scheme for the coupled DSE system, the ghost propagator DSE 
is the same as the one given by eqs.~(\ref{SDRS}), where the bare ghost-gluon 
vertex is approximated by $H_1=1$. The gluon DSE is given by 
\beq\label{coupledDSE}
\frac{(1+R(q^2))^2}{D(q^2)} \left( g_{\mu\nu} - \frac{q_\mu q_\nu}{q^2} \right) =   
q^2 g_{\mu\nu} - q_\mu q_\nu + i \sum_{i=1}^4 \left( a_i \right)_{\mu\nu}
\eeq
where 
\beq\label{gluondiags}
a_1 = \gluonSDi ,  & a_2 = \gluonSDii 
\nonumber \\
a_3 = \rule[0cm]{0cm}{1.5cm} \gluonSDiii , & a_4= \gluonSDiv .
\eeq
In the diagrams of (\ref{gluondiags}) for the gluon DSE, \eq{coupledDSE}, the external gluons 
are treated, from the point of view of Feynman rules, as background fields 
(these diagrams should be also properly regularized, as explained in \cite{Binosi:2009qm}). 
This is what justifies the four field coupling of two background gluons and two ghosts leading to 
the contribution $a_4$. The function $1+R(q^2)$ is defined in ref.~\cite{Grassi:1999tp} through a 
so-called background quantum identity~\cite{Binosi:2009qm} and can be, 
in virtue of the ghost propagator DSE, connected to the ghost propagator~\cite{Aguilar:2009nf,Aguilar:2010gm}. 
The coupled system is to be solved, by numerical integration, 
with the two following boundary conditions as the only required inputs:
the zero-momentum value of the gluon propagator and that of the coupling at 
a given perturbative momentum, $\mu=10$ GeV in this particular case, that will be used as 
the renormalization point. This is done by varying the boundary conditions (In particular, 
as explained in ref.~\cite{RodriguezQuintero:2010wy}, by keeping the zero-momentum value of 
the gluon propagator fixed while $\alpha(\mu^2=100\mbox{\rm~GeV}^2)$ is ranging 
from 0.15 to 0.1817) and leaves us with a family of massive or decoupling solutions 
that eqs.~(\ref{gluonprop},\ref{solFIRJo}) must account for, in the low-momentum domain. 
The successful confrontation can be seen in Fig.~1 of~\cite{RodriguezQuintero:2010wy}, 
from which we extract, as an example, the plot in Fig.~\ref{fig:ghgl}. 

The ghost dressing function at vanishing momentum, $F(0,\mu^2)$, is 
also shown to diverge as  $\alpha=\alpha(\mu^2) \to \alpha_{\rm crit}$, 
by obeying a power behaviour, s
\beq
F(0) \ \sim \ (\alpha_{\rm crit} - \alpha(\mu^2))^{- \kappa(\mu^2)} \ ,
\eeq
where the coefficient $\kappa(\mu^2)$ is a positive critical exponent
 (depending presumably on the renormalization point, $\mu^2$) which governs
 the transition from decoupling ($\alpha < \alpha_{\rm crit}$) to the 
 scaling ($\alpha = \alpha_{\rm crit}$) solutions. This is seen in 
 Fig.~2 of ref.~\cite{RodriguezQuintero:2010wy} and confirms the 
 results from the GPDSE analysis in sec.~\ref{subsec:limit}.

Had one let $\alpha_{\rm crit}$ be a free parameter to be fitted 
by requiring the best linear correlation for $\log[F(0)]$ in terms of 
$\log[\alpha_{\rm crit}-\alpha]$, one would have  obtained a best correlation 
coefficient of 0.9997 for $\kappa(\mu^2) = 0.0854(6)$  and $\alpha_{\rm crit}=0.1822$ 
(which is pretty close to the critical value of the coupling above which the coupled DSE 
system does not converge any more). 
This last critical value at $\mu=10$ GeV for the coupling can be pretty well translated 
to that of $\Lambda_{\rm QCD}$ in $\overline{\rm MS}$ (see for instance eqs.(22,23) 
of ref.~\cite{Boucaud:2008gn}) and then compared to $\Lambda_{\overline{\rm MS}}$ 
in pure Yang-Mills from the lattice~\footnote{It should 
be noted that the procedures for the lattice determination of $\Lambda_{\overline{\rm MS}}$ mainly work in 
the UV domain, where IR sources of uncertainties as the Gribov ambiguity or volume effects are indeed negligible. 
In fact, there are unquenched lattice determinations with $N_f=5$ staggered fermions for the strong 
coupling~\cite{Davies:2008sw} which are pretty consistent with the PDG value.}. The latter  
is estimated to be 238(19) MeV~\cite{Luscher:1993gh}, clearly below the former, 434 MeV, for 
the critical limit for the PT-BFM DSE in pure Yang-Mills. 

In summary, one can clearly conclude that the analysis of the coupled DSEs also agrees with 
the existence of both decoupling and scaling classes of solutions and with the pattern for 
them described in sec.~\ref{subsec:limit}. In particular, the analysis of the solutions in 
the PT-BFM scheme proved the scaling one to appear as an end-point for the decoupling 
family in Landau gauge, when the coupling at the renormalization point approaches a critical value. This 
critical value, at $\mu=10$ GeV, is well above the lattice estimates for the coupling, in 
total consistence with the conclusion of lattice QCD favouring the decoupling solution 
presented in sec.~\ref{sec:lattice}. 

Of course, gluon and ghost propagators and the vertices involving them, altogether with 
quark propagators and the quark-gluon vertex, are basic building blocks to study the 
QCD bound states. It might be that the exact very low-momentum behaviour of gluon 
and ghost propagators, that we paid attention to, is not very relevant for much of 
the hadron physics. In particular, the quark functions can be studied by modelling 
the product of the dressed gluon propagator and the quark-gluon vertex and, regardless of 
whether the gluon is suppressed or massive, the key region for physics is the momentum 
region of $p \simeq \Lambda_{\rm MOM}$~\cite{Maris:1999nt,Maris:1997hd}. Nevertheless, 
the dressed quark propagator, using massive gluons with non-singular interaction in the quark-gluon 
vertex, do become like expected by the heavy quark effective theory~\cite{Pennington:2010gy}.  
Furthermore, many phenomenological works also appear to support a massive gluon solution, 
as can be seen in the review of ref.~\cite{Binosi:2009qm}, and references therein, or, as very 
recent examples, in the works of refs.~\cite{Roberts:2010rn,Roberts:2011wy,Iddir:2011is,Oliveira:2011pg}. 
This also favours a decoupling solution.



\section{Conclusions}
\label{sec:conclu}

With this paper, we aimed to give an overview for the current state-of-the-art concerning 
the infrared properties of pure Yang-Mills QCD Green functions. Very much work has been 
reported in the last few years, modifying essentially the paradigm about this subject and 
demanding some sort of update for past reviews that can be found in literature. 
About ten years ago, the results from Landau gauge DSEs and FRGs analysis agreed with  
a solution, now dubbed ``{\it scaling}'', where the low-momentum behaviour for 
the two-point correlators appears to be an enhanced ghost propagator and 
a vanishing gluon propagator at zero-momentum ($\alpha_F < 0$ and $\alpha_G > 1$, according to 
the notation of \eq{param}). At that time, the lattice estimates for both 
correlators resulted to be compatible with this low-momentum behaviour, other approaches 
like the ones applying stochastic quantization methods and Gribov-Zwanziger lagrangian 
also pointed to the same results and all together matched the Kugo-Ojima confinement 
criterion providing a framework for the infrared solutions of Yang-Mills Green functions 
that was generally accepted. However, more recent results for the two-point correlators with 
larger lattices, some of them paying special attention to the problem of Gribov's copies, 
seemed to establish a different pattern for the low-momentum solutions: a finite ghost 
dressing and a finite non-zero gluon propagator at zero-momentum ($\alpha_F=0$ and $\alpha_G=1$), 
which did not agree with the scaling behaviour that would require   the ghost and the gluon 
infrared exponents to be related by $2 \alpha_F + \alpha_G = 0$. On the other hand,  
other authors,  applying a particular truncation scheme for the gluon propagator DSE 
which involves some angular approximation for the momentum integration,  
 proposed  a massive gluon propagator ($\alpha_G=1$),  consistently with the 
PT gluon propagator. Then, the GPDSE was recently re-analysed by exploiting the interplay of DSEs and lattice 
results and, apart from the scaling solution, new ones with $\alpha_F=0$, now dubbed decoupling, 
were proven to exist, which do not observe the scaling behaviour but are totally compatible with 
lattice results.
 Thus, both scaling and decoupling solutions have been now proven to emerge as solutions of 
the coupled system of gluon and ghost propagators DSEs for different truncation 
schemes, such as for instance the PT-BFM. The same occurs for FRGs. 

On the other hand, 
we have reviewed the plethora of lattice works on the subject, also discussing with some 
detail the role and impact of the lattice artefacts, and concluded that the current paradigm 
is a massive gluon and a free ghost, {\it i.e.} a decoupling  low-momentum behaviour for 
the two-point Green functions. Apart from the lattice results,  
the application of some refinement of the Gribov-Zwanziger approach and other new approaches
 also appear now to agree with a decoupling behaviour for the low-momentum Green functions solutions.
When solving the DSEs, the solutions have been proven to be ``dialed'' by the size of 
the coupling at the renormalization point, which can be univocally related to the zero-momentum 
value of the renormalized ghost dressing function. A family of  decoupling solutions 
corresponds to a family of sub-critical ones for finite values of the zero-momentum ghost dressing and 
for any coupling below a critical value at the renormalization point. 
As for the scaling solution,  it can be considered as  ``critical'', since it emerges for a {\em unique} (critical) value of the coupling for which the ghost dressing diverges 
at zero-momentum. A very similar 
pattern is shown to happen for the equal-time spatial gluon propagator and the ghost 
dressing function in Coulomb gauge. 

Of course, the critical value of the coupling depends on the renormalization point but, once it 
is known for one particular momentum, this value can be propagated to any other by 
applying the definition of the Taylor coupling in \eq{alphaTfirst} and the scaling 
solutions for ghost and gluon dressing functions. This is of course a consequence of the 
renormalization scaling for the coupling definition, which does not depend on either the cut-off, 
when written in terms of bare quantities, or the renormalization point, when expressed with 
renormalized ones. The critical value obtained at 
10 GeV in the PT-BFM scheme, in agreement with the analysis of the GPDSE with 
a lattice gluon propagator as input, is shown to be definitely above the lattice estimate for the 
Yang-Mills Taylor coupling, favouring again a decoupling solution in Landau gauge.

The truncation of the tower of DSEs which is necessary to obtain  a  (finite) tractable system of equations implies 
approximating the vertices. In particular, the ghost-gluon vertex plays a crucial role in the analysis  which led to 
find out the decoupling solutions, but also the three-gluon vertex is an essential ingredient 
for the gluon propagator DSE in any scheme. The ghost-gluon vertex benefits 
of Taylor's theorem that has been revised in the appendix, where we also discussed which
OPE non-perturbative corrections to the vertex the dimension-two gluon condensate induces. 
For the sake of consistency, we also overviewed the results from many lattice investigations 
about the impact of these OPE corrections on the Yang-Mills Green functions and about the possibilities of determining the 
size of the gluon condensate. Concerning the vertices, although 
some works have been devoted to investigate their properties, more 
lattice and continuum investigations would be very welcome.

To summarize, in the current state-of-the-art, although both types of solutions are compatible with 
DSEs (and FRGs too), lattice QCD and some continuum approaches, like mainly RGZ, seem to favour 
a decoupling-type of solutions which implies a free ghost and a massive gluon.


\begin{acknowledgements}

We thank M. M\"uller-Preussker and A. Sternbeck for very valuable comments and 
for providing us with some material to be published. One of us (J. R-Q) is also indebted 
to D. Dudal for very fruitful discussions and comments. This work has been partially 
supported by the research projects FPA2009-10773 form the Spanish MICINN and by 
P07FQM-02962 from ``Junta de Andaluc\'{\i}a''.

\end{acknowledgements}


\appendix


\section{A main ingredient: the non-renormalization Taylor's theorem}
\label{app:Taylor}

\subsection{What does the Taylor's theorem indeed say?}

A widely used statement, known as the ``non-renormalization theorem'', claims that, in the Landau gauge, the renormalization constant $\tilde Z_1$ of the ghost gluon vertex is exactly one. Note that there is no reference to a particular 
renormalisation scheme. Formulated in this way, this claim is wrong. Let us first state
and then explain below what is true in our opinion :

1) There is a true and very clear statement which can be extracted from Taylor's paper (the argument
is given below), ref.~\cite{Taylor}.
\bea
{\Gamma_\mu^{abc,Bare}}(-p,0;p)= -i f^{abc}p_\mu \label{taylor}
\eea
i.e. there is no radiative correction in this particular momentum configuration
(with zero momentum of the ingoing ghost)

2) This entails that ${\Gamma}_\mu^{abc,Bare}(p,k;q)$ is {\it finite} whatever the external momenta, and that 
therefore $\widetilde{Z_1}^{\overline{\mbox{\sc ms}}}=1$. In addition, we get also trivially 
$\tilde{Z_1}^{{\mbox{\scriptsize \sc MOM}_h}}=1$, where {\sc MOM}$_h$ refers to the configuration of momenta 
in equation~(\ref{taylor}). In general, in other schemes, {\it there is} a finite renormalisation, and this is why
one must be very careful when using the misleading  expression: "non-renormalization".

3) In particular, one finds in the very extensive calculations of radiative corrections at least two cases of 
{\sc MOM} schemes where there is a finite renormalisation (and certainly many more) : {\sc MOM}$_g$ in the 
notations of  ref.\cite{Chetyrkin:2000dq}, and the {\it symmetric } {\sc MOM} scheme. For the latter, we give 
the proof below.

The essence of Taylor's argument  is actually very simple. In a kinematical situation where the incoming 
ghost momentum is zero, consider any perturbative contribution to the ghost-gluon vertex. Following the 
ghost line in the direction of the flow,  the first vertex will be proportional to the outgoing ghost 
momentum $p_\mu$, i.e. to  the gluon momentum  $-p_\mu$.  In the Landau gauge this contribution will 
thus give 0 upon contraction with the gluon propagator. 
Therefore the only contribution to  remain is  the tree-level one.  In other words the bare ghost-gluon 
vertex  
is shown to be  equal to its tree-level value in these kinematics : 
${\Gamma_\mu^{abc,Bare}}(-p,0;p)= -i f^{abc}p_\mu$. 
This  result has been checked by means of  a direct evaluation  to  three loops in perturbation theory by Chetyrkin. 
In our notations :
\beq\label{H1pH2e1}
H_1(p,0) + H_2(p,0)=1
\eeq
Note that in the Schwinger-Dyson equation (\ref{SD1}), only
$H_1$ is present, and the theorem by Taylor does not tell that $H_1(p,0)=1$,
as seems assumed in many Schwinger-Dyson calculations.

\begin{figure}[h]
\begin{center}
\includegraphics{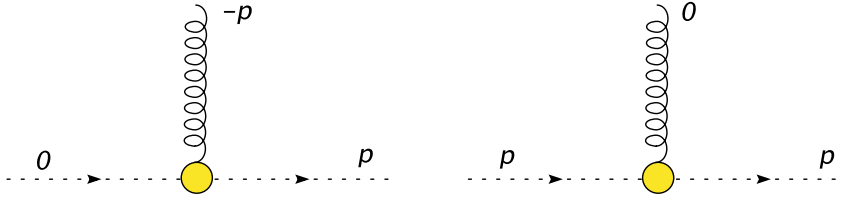}
\end{center}
\caption{The kinematical situations considered below. The left diagram (0-momentum incoming ghost) corresponds to $\Gamma_h$ below which is known to be equal to one. The right one (0-momentum gluon) corresponds to $\Gamma_g $ and  leads to a non-trivial $p^2$-dependence}
\label{cinem}
\end{figure}

As an illustration of our point 3), let us quote the formulas from the appendix of ref.~\cite{Chetyrkin:2000dq}, 
reduced to the situation we are interested in ($\xi_L = 0, n_f = 0$). 
The two dressing functions $\tilde\Gamma_h$ (resp. $\tilde\Gamma_g$) are defined 
by $\Gamma_\mu^{abc}(-p,0,p)= -i  f^{abc}\tilde\Gamma_h(p)$  
(resp. $\Gamma_\mu^{abc}(-p,p,0)= -i  f^{abc}\tilde\Gamma_g(p)$ ) and correspond 
to the kinematical situations depicted in the left (resp. right) part of fig.~(\ref{cinem}). 
We have already mentioned that $\Gamma_h$ is exactly one, but this does not hold for $\Gamma_g$ and, 
indeed, one has at three loops :
 \begin{eqnarray}
\label{GammagrenMSb}
\tilde \Gamma_{\mathrm{g}}^{\smsbar}\vert_{p^2=\mu^2} & = & 
1\;
{}
+ \,\frac{3}{4}\,  \alvps\, C_A\;
+ \, \frac{599}{96}    \alvp^{2}     C_A^{2}\;
+ \,   \left[
     \frac{43273}{432} 
    + \frac{783}{64}      \zeta_3
    - \frac{875}{64}      \zeta_5
\right]  \alvp^{3}     C_A^{3}\;\Break
+ \, \left[
     \frac{27}{4} 
    - \frac{639}{16}      \zeta_3
    + \frac{225}{8}      \zeta_5
\right]    \alvp^{3}     C_A^{2}     C_F\;
{}.\end{eqnarray}

It is then easy to find the  $p^2$-dependence  :

\begin{eqnarray}
\tilde\Gamma_{\mathrm{g}}=\widetilde \Gamma_{\mathrm{g}}^{\smsbar}\vert_{p^2=\mu^2} + \left[ \frac{11}{4} C_A^{2} \alvp^{2} +\frac{7813}{144} C_A^{3} \alvp^{3} + \cdots\right] \log(\frac{\mu^2}{-p^2}) +\cdots\nonumber
\end{eqnarray}
In ref.~\cite{Lerche:2002ep} the non-renormalization theorem is understood as the statement that the 
vertex reduces to its tree-level form at all symmetric-momenta points in a symmetric subtraction scheme. 
However this statement is not supported by a direct  evaluation. Using the one-loop results of 
Davydychev (ref.~\cite{Davydychev:1996pb}) one gets in  a symmetric configuration the value 

\begin{eqnarray}
\Gamma_\mu^{abc}(p,k;q)\vert_{p^2=k^2=q^2=\mu^2}= -i  f^{abc}\left\{p_\mu \,\left(1 +\,\alvps\, 
\frac{C_A}{12}(9+ \frac{5}{2}\phi) \right)+q_\mu\, \alvps\, \frac{C_A}{12}(3+ \frac{5}{4}\phi)\right\}
\end{eqnarray} 
with $\phi=\frac{4}{\sqrt{3}}Cl_2(\frac{\pi}{3}), Cl_2(\frac{\pi}{3})=1.049\cdots$. 

According to   ref.~\cite{Lerche:2002ep} the coefficient  of $p_\mu$ should be one. The presence of $\alpha_s$  
in the above formulas implies on the contrary that the vertex will in general depend on the momenta : 
using the results given in the appendices of ref.\cite{Chetyrkin:2000dq} one finds for the  
leading $p^2$-dependence 
$$ 
-i  f^{abc}\left\{\frac{11}{3}\frac{C_A^2}{12}\alvp^2\log(\frac{p^2}{\mu2})\left( (9+\frac{5}{2}\phi) p_\mu + (3+ \frac{5}{4}\phi) q_\mu\right)\right\} \ .
$$
This dependence is  logarithmic, as is expected   in a perturbative approach.   
Furthermore, in ref.~(\cite{Lerche:2002ep}) it is supposed that the vertex function takes the 
form $(q^2)^\ell (k^2)^m ((q-k)^2)^n$ with the restriction $\ell+m+n=0$.  
This restriction comes from the assumption that the {\emph symmetric} vertex is equal to 1 for any $p^2$, 
which, as we have just seen, is actually not the case. 
 Therefore, it should be necessary to adopt a more general point of view 
and keep open  the possibility of a non perturbative effect on $H_1$.
We should mention that, actually, the problem of the $p^2$-dependence of the ghost-gluon  
vertex has already been addressed in refs.~\cite{Alkofer:2004it, Schleifenbaum:2004id}. 
However these authors work under the condition $2 \alpha_F+\alpha_G=0$ 
which appears not to be satisfied by lattice data. Also in ref.~\cite{Boucaud:2011eh}, the impact of 
the OPE non-perturbative corrections, as the one we dealt with in sec.~\ref{sec:A2lat}, is studied 
for the ghost-gluon vertex in order to go beyond the approximation of taking 
the Taylor kinematics for the transverse form factor. We will briefly pay attention to this in the next 
subsection.

\subsection{Non-perturbative corrections for the ghost-gluon vertex}
\label{subsec:A2H1}

The non-perturbative effect resulting from a non-zero $\VA$ in a OPE expansion, 
shown in sec.~\ref{sec:A2lat} to have non-negligeable effects on ghost and gluon propagators 
at energies of the order of 2-3 GeV and still visible at around 7 GeV,  can be also 
advocated to have an impact on the ghost-gluon form factors introduced in \eq{DefH12},  
in particular on the transverse form factor, $H_1$, needed for the integration 
of the GPDSE in sec.~\ref{subsec:num1}. 
The procedure outlined in sec.~\ref{sec:A2lat} is also in order to compute the Wilson coefficients for 
such a non-perturbative contributions. Then, \eq{eq:fact} can be particularized to be

\beq\label{eq:expOPEH1}
H_1(q,k) \ = \ H_1^{\rm pert}(q,k) \  \left (1 + 
\frac{C^H_{\rm wilson}(q,k)}{H_1^{\rm pert}(q,k)} \;\;
\langle A^2(\mu^2)\rangle_{\rm \overline MS} \right ) \ ,
\eeq

\begin{figure}[htb]
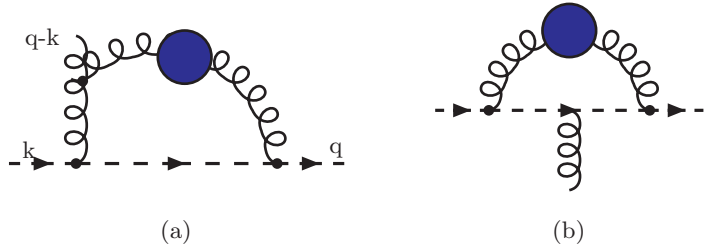

\begin{center}
\vspace*{0.5cm}
\begin{tabular}{cc}
$\ghThreeOneRS$ & $\ghThreeTwo$ \\
\\
\\
(a) & (b) 
\end{tabular}
\end{center}
\caption{\small Diagrams contributing (altogether with their appropriate permutations) to the ghost-gluon form factors for the proper ghost-gluon vertex~\cite{Boucaud:2011eh}.}
\label{fig:diagOPEH1}
\end{figure}

\noindent 
where, although it is not divergent, one can apply a finite renormalization to require 
$H_1$ to take its tree-level value, $H_1=1$, at a given momentum scale, $\mu^2$, for 
a particular kinematical configuration of $p$ and $q$ 
(for instance, $p^2=q^2=\mu^2$). This dependence on $\mu$ should be understood 
for the Wilson coefficient and the form factors. 
The Wilson coefficient can be obtained~\cite{Boucaud:2011eh} at tree-level by evaluating the diagrams in 
Fig.~\ref{fig:diagOPEH1} and reads:

\beq\label{eq:OPEH1-1}
\frac{C^H_{\rm wilson}(q,k)}{H_1^{\rm pert}(q,k)} 
\ = \ \frac 3 {64} g^2(\mu^2) \
\left( 2 \ \frac{(q-k)\cdot q}{q^2 (q-k)^2} + 2 \ \frac{(k-q)\cdot k}{k^2 (q-k)^2} + \frac{k\cdot q}{k^2 q^2} \right) \ .
\eeq
Thus, after modeling the ghost-gluon form factor all over the range of their momenta
by the insertion of some infrared mass scale to saturate the powers of momenta in the denominators of 
Eqs.~(\ref{eq:OPEH1-1}) (as a simple way to avoid the non-physical divergence coming 
from these inverse powers), one obtains the model for the ghost-gluon transverse form factor 
that was mentioned in sec.~\ref{subsec:comp-latt}. This model provides us with a first correction 
to the usual hypothesis for the GPDSE integration: $H_1=1$ that is encoded by the OPE expansion through $\VA$.

It is interesting to notice that, had we considered the kinematic configuration for the Taylor scheme, $k = 0$, 
one would obtain a non-vanishing OPE power correction for the transverse form factor; although no deviation from 
the Taylor result shown by \eq{H1pH2e1} would result, because 

\beq
H_1(q,0) + H_2(q,0) \ = \ H_1^{\rm pert}(q,0) + H_2^{\rm pert}(q,0) \ = \ 1 \ ,
\eeq
as it is proved in ref.~\cite{Boucaud:2011eh}.



\section{The Dyson-Schwinger equation as a Ward-Slavnov-Taylor identity}
\label{App}

A very general method to derive Slavnov-Taylor identities consists in taking advantage of the transformation properties of 

\beq\label{action}
e^{G(J)} = \int {\cal D} (A) \det {\cal M} \exp\left [ i \int d^4x
\left({\cal L} - \frac 1 {2\alpha} (\partial_\mu A_\mu^a)(\partial_\mu A_\mu^a) + 
J_\mu^a A_a^\mu\right)\right]\label{pathInt}
\eeq 
under gauge transformations (cf. \cite{IZ}).

${\cal M}$ is the Faddeev-Popov operator and the notation   $<$, $>_J$  indicates that the source term $J$ has to be kept, although it will eventually be set to 0 (this is denoted in the following by the suppression of the $J$ subscript). Taking the derivative of the gauge transformed of eq.~(\ref{pathInt}) with respect to the gauge parameters leads to the general Slavnov-Taylor equation
\beq\label{ST}
\frac 1 \alpha <(\partial_\mu A_\mu^a(x))>_J = <\int d^4y 
J_\mu^c(y) \,D_\mu^{cb}(y) F^{(2)ba}(y,x) >_J \ .
\eeq
$F^{(2)ba}(y,x)$ is the ghost propagator and its presence here is simply due to its very definition as the inverse of the Faddeev-Popov operator.
If one derives eq.~(\ref{ST}) with respect to $J_\rho^d(z)$  one gets  : 
 \bea\label{dJ1}
 \frac 1 \alpha <(\partial_\mu A_\mu^a(x)) A_\rho^d(z)>_J
 &=& <D_\rho^{db}(z)F^{(2)ba}(z,x)>_J \nonumber \\
 &+&  <\int d^4y  J_\mu^c(y)\,D_\mu^{cb}(y) F^{(2)ba}(y,x) A_\rho^d(z)>_J \ .
\eea
A first consequence of this relation is the triviality of the longitudinal gluon propagator. To see this, it suffices to derive both its sides with respect to $z_\rho$ and to set $J$ to zero.
The result is
  \bea
  \frac 1 \alpha <(\partial_\mu A_\mu^a(x))(\partial_\rho A_\rho^d(z))> 
  &=& <\partial_\rho D_\rho^{db}(z)\,F^{(2)ba}(z,x)> \nonumber \\
  &=& \delta_{ad}\,\delta_4(z-x) \ .
  \eea
In order to derive the second line we have invoked the fact that  $\partial_\rho D_\rho^{db}(z)$, the Faddeev-Popov operator, is the inverse of the ghost propagator $F^{(2)}$.
Thus,  in momentum space, the general form of the gluon  propagator for an arbitrary covariant gauge reads 
 \beq\label{G2}
 G^{(2)ab}_{\mu\nu}{(q)} =\delta^{ab}\left[ G^{(2)}(q^2)
 \left(\delta_{\mu\nu}-\frac{q_\mu q_\nu}{q^2}
 \right) + \alpha \frac{q_\mu q_\nu}{(q^2)^2}\right] \ .
 \eeq

Turning now back to eq.~(\ref{dJ1}) and letting $J$ go to zero we obtain
\bea\label{dJ10}
  \frac 1  \alpha <(\partial_\mu A_\mu^a(x)) A_\rho^d(z)>
 = <D_\rho^{db}(z)F^{(2)ba}(z,x)> 
 \eea
which  is nothing else than the  GPDSE.
Actually  its l.h.s. involves only the longitudinal part of the gluon propagator, that we have just seen to be  trivial :
\bea\label{proj}
 \frac 1 \alpha <(\partial_\mu A_\mu^a(x)) A_\rho^d(z)>  \partial_\rho \square^{-1}(x,z) 
\eea 

\noindent where the $\square$ symbol stands as usual for the d'Alembertian. 
 As for the r.h.s it can be rewritten as :
 \bea\label{drho}
<D_\rho^{db}(z)F^{(2)ba}(z,x)> =<\partial_\rho F^{(2)da}(z,x)> + 
i<g f^{deb} A_\rho^{e}(z) F^{(2)ba}(z,x)> \ .
 \eea
 
 The 3-point gluon-ghost Green's function can be expressed in terms of vertex functions and propagators through
 \bea\label{G3}
 &\widetilde G^{(3)fgh}_\rho(p,q,r) \equiv \displaystyle -i\int d^4x \ d^4t \ d^4z  e^{ipx}e^{irz}e^{iqt} 
 <A_\rho^{f}(t) F^{(2)gh}(z,x)> \\
 &= \displaystyle g \frac {F(p^2)}{p^2} \frac{F(r^2)}{r^2}  
 \left[ \frac{G(q^2)}{q^2}
 \left(\delta_{\rho\nu}-\frac{q_\rho q_\nu}{q^2}
 \right) + \alpha \frac{q_\rho q_\nu}{(q^2)^2}\right]f^{fgh}
 \widetilde\Gamma_\nu(p,r;q)(2\pi)^4\delta_4(p+q+r) \nonumber
 \eea
Now, we Fourier-transform \eq{dJ10}, 
use  eqs.~(\ref{proj}
-\ref{G3}) and obtain
 \bea\label{dJ10f}
 \frac{k_\rho}{k^2} &=&  \frac{k_\rho}{k^2} F(k^2) - g 
 f^{deb}f^{eba}\int \frac{d^4q}{(2\pi)^4}
 \frac{F(k^2)}{k^2} \frac{F((k+q)^2)}{(k+q)^2} \nonumber\\
& &  \left[ \frac{G(q^2)}{q^2}
 \left(\delta_{\rho\nu}-\frac{q_\rho q_\nu}{q^2}
 \right) +  \alpha \frac{q_\rho q_\nu}{(q^2)^2}\right]
 \widetilde\Gamma_{\nu}(k,-k-q;q) \ ,
 \eea
where the usual form of GPDSE can be recovered from by multiplying with  $k_\rho$
and dividing by $F(k^2)$, which leads to  
  \bea\label{GPDSE}
 F^{-1}(k^2) &=& 1 - g f^{deb}f^{eba}\int \frac{d^4q}{(2\pi)^4}
 \frac{F((k+q)^2)}{(k+q)^2}\nonumber\\
 &\,&\left[ \frac{G(q^2)}{q^2}\left(k_\nu-\frac{(q \cdot k) q_\nu}{q^2}\right) +  \alpha \frac{(q \cdot k) q_\nu}{(q^2)^2}\right]
 \widetilde\Gamma_{\nu}(k,-k-q;q) \ .
 \eea
This is a  general result, valid in any covariant gauge. Of course  the $\alpha$-depending (longitudinal) term disappears  in Landau gauge.
$\widetilde\Gamma_{\nu}(k,-k-q;q)$ is related to the quantity, $\widetilde\Gamma_{\mu\nu}$, 
that was previously introduced in 
section~\ref{sec:Intro} 
through
$$ \widetilde\Gamma_{\nu}(k,-k-q;q)=-i g k_\mu\widetilde\Gamma_{\mu\nu}(k,-k-q;q)$$
and it is usually decomposed into
$\widetilde\Gamma_{\nu}(k,-k-q;q)= g\left[ k_\nu H_1(k,q)+   q_\nu H_2(k,q)\right]$.
After inser\-ting this in eq.(\ref{GPDSE}) and restricting to the Landau gauge case one 
finally obtains 
\bea\label{GPDSEL}
 F^{-1}(k^2) = 1 +g^2 N_c \int \frac{d^4q}{(2\pi)^4}
 \frac{F((k+q)^2)}{(k+q)^2}\left[ \frac{G(q^2)}{q^2}\left(\frac{(q \cdot k)^2}{q^2} -k^2\right) 
\right]H_1(k,q) \ .
 \eea


\end{document}